\documentclass[aps,prb,preprint,groupedaddress,
amsmath,amssymb,showkeys]{revtex4-2}

\usepackage{graphicx}
\usepackage{dcolumn}
\usepackage{enumitem}
\usepackage{bm}
\usepackage{physics}
\usepackage{tensor}
\usepackage{slashed}
\usepackage[linktocpage=true]{hyperref}
\hypersetup{
    colorlinks,
    linkcolor={blue},
    urlcolor={blue},
    citecolor={blue},
}
\usepackage{color}
\numberwithin{equation}{section}
\renewcommand{\theequation}{\arabic{section}.\arabic{equation}}

\newcommand{\del}{\partial}
\newcommand{\nn}{\nonumber}
\newcommand{\mathZ}{\mathcal{Z}}
\newcommand{\mathD}{\mathcal{D}}
\newcommand{\frakh}{\mathfrak{h}}
\newcommand{\frakz}{\mathfrak{z}}
\newcommand{\frakw}{\mathfrak{w}}

\newcommand{\nocontentsline}[3]{}
\newcommand{\tocless}[2]{\bgroup\let\addcontentsline=\nocontentsline#1{#2}\egroup}

\def\C60{A$_x$C$_{60}$}

\def\HgCu3{HgCa$_2$Cu$_3$O$_{8+y}$}
\def\HgCu4{HgBa$_2$Ca$_3$Cu$_4$O$_{10+y}$}
\def\TlCu{Tl$_2$Ba$_2$CuO$_{6+\delta}$}
\def\TlCu3{Tl$_2$Ba$_2$Ca$_2$Cu$_3$O$_{10+y}$}
\def\TlCu4{Tl$_2$Ba$_2$Ca$_3$Cu$_4$O$_{12+y}$}

\def\BiCu3{Bi$_2$Sr$_2$Ca$_{2}$Cu$_3$O$_y$}

\def\8LSCO{La$_{1.88}$Sr$_{.12}$CuO$_4$}

\def\110LNSCO{La$_{1.5}$Nd$_{0.4}$Sr$_{0.1}$CuO$_{4}$}
\def\stage4LCO{La$_{2}$CuO$_{4+\delta}$}
\def\Y248{YBa$_2$Cu$_4$O$_8$}

\def\NbSe2{NbSe$_2$}
\def\TaSe2{TaSe$_2$}
\def\TiSe2{TiSe$_2$}
\def\NaCoOH2O{Na$_{0.3}$CoO$_{2y}$H$_2$O}
\def\MgB2{MgB${}_2$}

\def\URu2Si2{URu$_2$Si$_2$}

\def\Ba122{Ba(Fe$_{1-x}$Co$_x$)$_2$As$_2$}

\begin{document}

\title{An Exactly Solvable Model of Randomly Pinned \texorpdfstring{\\}{} Charge Density Waves in Two Dimensions}

\author{Matthew C. O'Brien}
\email{mco5@illinois.edu}
\author{Eduardo Fradkin}%
\email{efradkin@illinois.edu}
\affiliation{%
Department of Physics and Institute for Condensed Matter Theory,
University of Illinois Urbana-Champaign, 1110 West Green Street, Urbana, IL 61801, USA
}%

\date{\today}

\begin{abstract}
The nature of the interplay between fluctuations and quenched random disorder is a long-standing open problem, particularly in systems with a continuous order parameter. This lack of a full theoretical treatment has been underscored by recent advances in experiments on charge density wave materials. To address this problem, we formulate an exactly solvable model of a two-dimensional randomly pinned incommensurate charge density wave, and use the large-$N$ technique to map out the phase diagram and order parameter correlations. Our approach captures the physics of the Berezinskii-Kosterlitz-Thouless phase transition in the clean limit at large $N$. We pay particular attention to the roles of thermal fluctuations and quenched random field disorder in destroying long-range order, finding a novel crossover between weakly- and strongly-disordered regimes.
\end{abstract}

\maketitle

\clearpage

\tableofcontents

\clearpage

\section{Introduction \label{sec:intro}}

It has been well-understood for many decades that below four dimensions random impurities in a material will ``pin'' the $U(1)$ phase of an incommensurate charge density wave (ICDW) \cite{Efetov1977}. The pinning leads to a frustration that prevents a well-defined CDW state with a uniform phase $\theta$ from forming \cite{Imry1975}. In fact, this destruction of long-range symmetry breaking order by random fields (when not forbidden by gauge invariance) is a general feature in systems with a global $U(1)$ symmetry. The special case of two spatial dimensions is particularly subtle since a state with long-range order is already forbidden at finite temperature by the Mermin-Wagner theorem \cite{Mermin1966}. In the absence of quenched random field disorder, the Berezinskii-Kosterlitz-Thouless (BKT) transition, which involves the proliferation of topological defects (vortices in magnets with planar anisotropy, and dislocations in the case of 2D ICDWs), is special since it separates a (critical) phase with power law correlations and a disordered phase with unbroken $U(1)$ symmetry \cite{Berezinskii1971,Kosterlitz1973}. The question of how the critical phase is destroyed by a quenched random field has been understood only qualitatively.

Over the last four decades, a significant amount of work has been done to attempt to answer this question. Initially, advances were made using perturbative field-theoretic renormalization group (RG) techniques \cite{Houghton1981,Cardy1982,Goldschmidt1982}. Those authors studied the random field $XY$ model as an idealization of, among other systems, two-dimensional CDWs in systems with charge disorder. They found that the original BKT transition separating the high temperature neutral vortex plasma and low temperature power law correlated phases was destroyed by disorder due to impurities becoming (RG) relevant at an intermediate temperature. However, the nature of the low temperature impurity-dominated state remained undetermined, spurring the application of non-perturbative techniques. For example, the functional RG was used to disprove the long-standing dimensional reduction hypothesis \cite{Young1977,Fisher1985,Tarjus2004}. To this day, the nature of the low temperature state remains hotly debated, with some proposing the ground state to be a so-called ``Bragg glass'' of dislocations \cite{Giamarchi1994,LeDoussal2000}, while others claim that this state  is featureless and lacks any form of order \cite{Zeng1999}. Despite the disagreements, it is now evident that non-perturbative techniques are necessary for shedding light on the physics of the disordered system \cite{Andreanov2014,Nie2014,Nie2015,Tarjus2020,LeDoussal2006}.

Interest in disordered ICDWs has not purely been driven by the theoretical challenges described above. The cuprate high-temperature superconductors, which have focused the attention of much of the condensed matter physics community, are strongly layered compounds which display quasi-two-dimensional charge order proximate to the superconducting phase \cite{Tranquada1995,Abbamonte2005,Kivelson2003,Ghiringhelli2012,Mesaros2016}. The possibility of intertwined \cite{Berg2009,Fradkin2015} charge and superconducting orders makes understanding the role of CDW order crucial to the physics of high-$T_c$ superconductors. Recently, significant experimental advances have been made in x-ray scattering \cite{Jang2016,Mitrano2019,Lee2021}, scanning electron microscopy and spectroscopy \cite{Fujita2014,Mesaros2016} and momentum-resolved electron energy loss spectroscopy (M-EELS) \cite{Vig2017,Kogar2017}. The increased resolution of measurements has allowed for precise determination of dynamic charge correlations in CDW materials. 

Nevertheless, despite the intense theoretical and experimental interest, a full theoretical treatment of the interplay of thermal and quantum fluctuations in the presence of random disorder is an open problem. In this paper, we introduce a model which is exactly solvable in a suitable large-$N$ limit even in the presence of a quenched random field. Our results apply to the specific case of an ICDW in two dimensions which we will take to be unidirectional. Although our main motivation stems from the experimentally observed ICDWs in the lanthanum cuprates \cite{Li2007,Hucker2011,Lee2022}, the problem is of much broader interest since such CDWs are seen in many systems, including the dichalcogenides and many other quasi-2D materials. In this paper we will focus on the classical theory, leaving the quantum theory to another publication. 

The order parameter of a unidirectional CDW is the Fourier component of the local charge density $\rho(\vb{x})$ at the ordering wave vector $\vb{Q}$:
\begin{equation}
    \rho(\vb{x})=\rho_0(\vb{x})+ \rho_{\vb{Q}}(\vb{x}) \, e^{i \vb{Q} \cdot \vb{x}}+\rho_{-\vb{Q}}(\vb{x}) \, e^{-i \vb{Q} \cdot \vb{x}}+\text{higher  harmonics},
    \label{eq:cdw-op}
\end{equation}
where $\rho_0(\vb{x})$ is the slowly-varying uniform component and the also slowly-varying field $\rho_{\vb{Q}}(\vb{x})=\rho_{-\vb{Q}}^*(\vb{x})$ is the CDW order parameter. Here we will assume that the ordering wave vector is incommensurate with the underlying lattice and that we have an ICDW. A CDW, commensurate or not and regardless of which microscopic mechanism is responsible for it, is a phase of an electronic system in which $\langle \rho_{\vb{Q}}(\vb{x})\rangle\neq 0$. This state breaks translation invariance and the point group symmetry of the underlying lattice. States of this type are inherently fragile to disorder since a local (effectively random) electrostatic potential $V_{\rm imp}(\vb{x})$ due to charged impurities couples linearly to the local charge density $\rho(\vb{x})$ and, consequently, disorder couples linearly to the CDW order parameter. In what follows, we will denote the complex field of the ICDW order parameter as $\rho_{\vb{Q}}(\vb{x})\equiv \sigma(\vb{x})$ and the ordering wave vector $\vb{Q}$ will be left implicit.

In this work, we present a non-perturbative approach to treating the physics of a $U(1)$ order parameter coupled to quenched disorder. Our method is based on the well-known large-$N$ technique (see, for example, Refs. \cite{Stanley1968,Amit-book,Zinn-book,fradkin_2021}). The large-$N$ approach has been used to investigate the $O(N)$ model in a random field by Nie, Tarjus and Kivelson \cite{Nie2014}, who applied their results to the case of randomly pinned ICDWs. Taken literally, their results apply to dimensions $d>2$, since in the absence of a random field the 2D $O(N)$ model does not have long-range order. For this reason, we were motivated to explore other large-$N$ models that are, in principle, able to capture unique aspects of 2D physics, including the BKT transition (in the absence of disorder), even in the regime where $N$ is large.

To address this problem, in this paper we consider a theory of a two-component generalization of the $\mathbb{C}P^N$ model with a global $U(1)$ symmetry between the two components. This global symmetry characterizes the order parameter manifold of interest. Our generalized model includes an interaction term between the two components which is solvable in the large-$N$ limit. Here we use the fact that the large-$N$ limit of the $\mathbb{C}P^N$ model is well understood \cite{Witten1979,Coleman-1985}, and show that this coupled theory is also solvable in the large-$N$ limit.  In the absence of disorder, the $N=\infty$ ground state we find appears to spontaneously break the $U(1)$ symmetry, but we show that the inclusion of the leading order fluctuations about the $N=\infty$ ground state is sufficient for a BKT phase transition to emerge. In fact, we show explicitly that the BKT transition appears as a $1/N$ correction to the $N=\infty$ transition. 

Having confirmed that the clean theory behaves as expected, we then include quenched disorder and solve this model exactly in the $N=\infty$ limit as a function of disorder and of the coupling strength between the two components. Similarly to the clean theory, the disordered theory appears to have a broken symmetry phase. However, we again demonstrate that allowing for fluctuations about the large-$N$  state necessarily restores the symmetry. The large-$N$ limit of the model predicts a complex phase diagram in the presence of quenched random fields. While in the strong disorder regime we find a phase with short-range order (as expected), in the weak disorder regime the naive large-$N$ limit is the same as in the clean theory. We also derive explicit results for the order parameter correlator in the strong disorder regime (to leading order in the $1/N$ expansion) and show that it has the same functional form as found in earlier theories of the random field Ising model \cite{Pytte1981} and $O(N)$ model with a random field \cite{Nie2014,Nie2015}. Finally, an analysis of the $1/N$ corrections reveals that the correlation functions have essentially the same behavior in the weak and strong disorder regimes, resolving the naive disagreement with the Imry-Ma argument \cite{Imry1975}, and implying the existence of a subtle crossover between the two regimes. 

The rest of this paper is structured as follows: In Sec. \ref{sec:parton} we present the model we will be using in this work. We demonstrate our method for studying a $U(1)$ order parameter with the large-$N$ technique by solving the model exactly in the $N=\infty$ limit (\ref{sec:cleanNinfty}) and then showing how the BKT transition emerges at order $1/N$ (\ref{sec:BKT}). In Sec. \ref{sec:disorder} we begin by coupling our model to quenched disorder and then derive the phase diagram of the theory at $N=\infty$ (\ref{sec:dirtyNinfty}). Finally, we resolve the apparent inconsistencies of the mean field limit with the Imry-Ma condition by incorporating the effects of fluctuations (\ref{sec:strongDisorder} \& \ref{sec:weakDisorder}). Section \ref{sec:disc} presents our conclusions. Technical details and pedagogical reviews of known material are presented in a number of appendices. Appendix \ref{app:vorticity} supports the results of Sec. \ref{sec:BKT}, Appendix \ref{app:ON} reviews the effects of quenched disorder in $O(N)$ models, Appendix \ref{app:CPN} does the same for the conventional $\mathbb{C}P^N$ model, and Appendix \ref{app:fluctuations} supports tht results of Sec. \ref{sec:strongDisorder}.

\section{Two-Component \texorpdfstring{$\mathbb{C}P^N$}{CPN} Model for a \texorpdfstring{$U(1)$}{U(1)} Order Parameter \label{sec:parton}}

\subsection{The Model and its Symmetries}

To apply the non-perturbative large-$N$ technique to a theory with a global $U(1)$ symmetry, we wish to construct a model with an internal symmetry group $G$ with dimension scaling as $N$ and a $U(1)$ subgroup which can subsequently be spontaneously broken. A simple approach is a theory of two $N$-component complex scalar fields $\vb*{z}$ and $\vb*{w}$, each of which transforms under the fundamental representation of $U(N)$; the larger symmetry group is then $G = U(N) \times U(N)$ and each of the $U(N)$ products contributes a $U(1)$ subgroup. The particular model we have chosen to study is a two flavor generalization of the $\mathbb{C}P^N$ model, and is described by the action
\begin{equation}
    S = \frac{1}{g} \int \dd^d \vb{x} \, \left( \abs{D^\mu[a] \vb*{z}}^2 + \abs{D^\mu[a] \vb*{w}}^2 - \frac{K}{g} \abs{\vb*{z}^* \cdot \vb*{w}}^2 \right), \kern3em \abs{\vb*{z}}^2 = \abs{\vb*{w}}^2 = 1, \label{eq:multicompaction}
\end{equation}
where $a^\mu$ is a fluctuating $U(1)$ gauge field and $D^\mu[a] = \del^\mu + i a^\mu$. Importantly, we take both $\vb*{z}$ and $\vb*{w}$ to be coupled to $a^\mu$, as opposed to two independent fluctuating gauge fields. This model has a formal similarity with the chiral Gross-Neveu model which has a well-understood large-$N$ limit \cite{Gross1974,Witten1978}. It will be convenient to work in a representation where the $U(1) \times U(1) \subset G$ subgroup is generated by two types of transformations:
\begin{subequations} \label{eq:u1transforms}
\begin{alignat}{3}
    &\text{(i) diagonal (local)} \kern3em &&\vb*{z}(\vb{x}) \longrightarrow e^{i\phi(\vb{x})} \vb*{z}(\vb{x}), \kern3em &&\vb*{w}(\vb{x}) \longrightarrow e^{i\phi(\vb{x})} \vb*{w}(\vb{x}), \\
    & && a^\mu(\vb{x}) \longrightarrow a^\mu(\vb{x}) - \del^\mu \phi(\vb{x}), && \nn \\
    &\text{(ii) relative (global)} \kern3em &&\vb*{z}(\vb{x}) \longrightarrow e^{i\phi} \vb*{z}(\vb{x}), \kern3em &&\vb*{w}(\vb{x}) \longrightarrow e^{-i\phi} \vb*{w}(\vb{x}),
\end{alignat}
\end{subequations}
Note that the quartic interaction term is invariant under the global relative $U(1)$ symmetry (ii) which can be spontaneously broken (with the usual caveats on the restrictions imposed by the Mermin-Wagner theorem).

\subsection{Large-\texorpdfstring{$N$}{N} Solution \label{sec:cleanNinfty}}

We will now demonstrate that this model has the desired properties. The partition function for the theory is
\begin{equation}
    \mathcal{Z} = \int \mathD \lambda_1 \mathD \lambda_2 \mathD a^\mu  \mathcal{D} \vb*{z} \mathD \vb*{w} \exp\left(-S - \int \dd^d \vb{x} \left[ \frac{\lambda_1}{g}( \abs{\vb*{z}}^2 - 1) + \frac{\lambda_2}{g}( \abs{\vb*{w}}^2 - 1) \right] \right),
\end{equation}
where $\lambda_1$ and $\lambda_2$ are Lagrange multiplier fields imposing the constraints $\abs{\vb*{z}}^2 = \abs{\vb*{w}}^2 = 1$. Additionally, the quartic interaction term $\abs{\vb*{z}^* \cdot \vb*{w}}^2$ can be decoupled using a Hubbard-Stratonovich (HS) transformation:
\begin{align}
    \exp\left(\int \dd^d \vb{x}\, \frac{K}{g} \abs{\vb*{z}^* \cdot \vb*{w}}^2 \right) &= \int \mathD \sigma \exp\left(-\int \dd^d \vb{x}\, \left[ \frac{g}{K}  \sigma^* \sigma - \sigma \vb*{z}^* \cdot \vb*{w} - \sigma^* \vb*{z} \cdot \vb*{w}^*  \right] \right). \label{eq:HS_trans}
\end{align}
Since the Gaussian integral over $\sigma$ is peaked around the value $\sigma_{\mathrm{max}} = (K/g) \vb*{z} \cdot \vb*{w}^*$, we identify the complex field $\sigma$ as an emergent $U(1)$ order parameter. In fact, invariance of the right hand side of Eq. \eqref{eq:HS_trans} under a relative $U(1)$ transformation imposes the transformation rule $\sigma \rightarrow e^{2i\phi} \sigma$. After the HS transformation, the functional integrals over $\vb*{z}$ and $\vb*{w}$ become Gaussian, and hence, can be performed exactly to obtain
\begin{subequations}\label{eq:effact}
    \begin{align}
        \mathcal{Z} &= \int \mathD \lambda_1 \mathD \lambda_2 \mathD a^\mu \mathD \sigma\, e^{-S_{\mathrm{eff}}}, \\
        S_{\mathrm{eff}}/N &= \tr \ln \begin{pmatrix} -D_\mu^2[a] + \lambda_1 & -\sigma \\ -\sigma^* & -D_\mu^2[a] + \lambda_2 \end{pmatrix} + \int \dd^d \vb{x} \left[ \frac{\sigma^* \sigma}{K_0} - \frac{\lambda_1 + \lambda_2}{g_0} \right], 
    \end{align} 
\end{subequations}
where we have defined $g = g_0/N$ and $K = K_0/N$. As $N \rightarrow \infty$, the partition function can be evaluated exactly within mean field theory. We then take the spatially uniform ansatz $\sigma(\vb{x}) = \sigma_0$, $\lambda_1(\vb{x}) = \lambda_2(\vb{x}) = m^2$ (since the theory also has a $\mathbb{Z}_2$ symmetry which exchanges the $\vb*{z}$ and $\vb*{w}$ fields) and $a^\mu(\vb{x}) = 0$ (by gauge invariance). Then, 
\begin{subequations}
    \begin{align}
        \mathZ &= e^{-N V U_{\mathrm{eff}}}, \\
        U_{\mathrm{eff}} &= \int^\Lambda \frac{\dd^d \vb{q}}{(2\pi)^d} \ln \left( [\vb{q}^2 + m^2]^2 - \abs{\sigma_0}^2 \right) + \frac{\abs{\sigma_0}^2}{K_0} - \frac{2m^2}{g_0}, \label{eq:effpot}
    \end{align}
\end{subequations}
where $V$ is the volume of $d$-dimensional space and $\Lambda$ denotes some regulator; e.g., a UV cutoff.

We will now specialize to the case $d=2$. The effective potential \eqref{eq:effpot} is logarithmically ultraviolet divergent, but renormalization of the coupling constant $g_0$ suffices to cure this. Defining
\begin{equation}
    \frac{1}{g_0} = \frac{1}{g_R} \left(1 + g_R  \int^\Lambda \frac{\dd^d \vb{q}}{(2\pi)^d} \frac{1}{\vb{q}^2 + \mu^2} \right), \label{eq:gRenorm}
\end{equation}
for some renormalization scale $\mu$, the regulator $\Lambda$ can then be removed from Eq. \eqref{eq:effpot}, yielding a renormalized effective potential:
\begin{equation}
    U_R = \frac{m^2}{4\pi} \left[2 - \ln\left(\frac{m^4 - \abs{\sigma_0}^2}{\mu^4}\right)\right] + \frac{\abs{\sigma_0}}{4\pi} \ln\left(\frac{m^2 - \abs{\sigma_0}}{m^2 + \abs{\sigma_0}}\right) + \frac{\abs{\sigma_0}^2}{K_0} - \frac{2m^2}{g_R} . \label{eq:effpotclean}
\end{equation}
The ground state values of $m^2$ and $\abs{\sigma_0}$ are then obtained by minimizing the potential. First,
\begin{equation}
    m^4 - \abs{\sigma_0}^2 = \mu^4 e^{-8\pi/g_R}.
\end{equation}
Next, it is straightforward to check that for $K_0 < K_c = 4\pi \mu^2 e^{-4\pi/g_R}$, the potential is minimized by a vanishing order parameter $\sigma_0 = 0$. However, for $K_0 > K_c$, the magnitude of $\sigma_0$ is determined by a Curie-Weiss type transcendental equation
\begin{equation}
    \abs{\Tilde{\sigma}} = \frac{K_0}{K_c} \sinh^{-1}\left(\abs{\Tilde{\sigma}}\right) , \label{eq:weissMFT}
\end{equation}
where $\abs{\Tilde{\sigma}} = \abs{\sigma_0}/(\mu^2 e^{-4\pi/g_R})$, which is well-known to yield a $\beta = 1/2$ critical exponent in the limit $\abs{\Tilde{\sigma}} \simeq 0$. Since the effective potential Eq. \eqref{eq:effpotclean} is invariant under the relative $U(1)$ symmetry, a mean-field solution $\sigma_0$ spontaneously breaks this symmetry. However, it is evident that in $d=2$ this spontaneous symmetry breaking (SSB) is an artifact of the large-$N$ limit, since the Mermin-Wagner theorem forbids SSB \cite{Mermin1966}. This signals the need to include fluctuations around the mean-field limit, for which the large-$N$ technique provides a systematic method.

\subsection{Fluctuations \& BKT Transition \label{sec:BKT}}

 Since we are primarily interested in the infrared physics of this model, we can focus on the Goldstone manifold of degenerate ground states parameterized by the $U(1)$ phase $\theta$ of the complex order parameter field; $\sigma(\vb{x}) \equiv \rho(\vb{x}) e^{i\theta(\vb{x})}$, where $\rho(\vb{x})$ is the (real) amplitude of the order parameter (not to be confused with the charge density). By freezing $\rho(\vb{x}) = \rho_0$ and allowing $\theta(\vb{x})$ to vary slowly in space, the leading corrections to the effective action will be determined by a gradient expansion. However, $\theta \in [0,2\pi)$ is a compactified boson, and its periodicity allows the existence of vortices \cite{Berezinskii1971,Kosterlitz1973}. 

Following the usual analysis (see, for example, Ref. \cite{kardar_2013}), the periodicity is imposed by a non-fluctuating source of vorticity $A^\mu$ which satisfies
\begin{equation}
    \varepsilon^{\mu\nu}\del^\mu A^\nu(\vb{x}) = 2\pi \sum_j m_j \delta^{(2)}(\vb{x} - \vb{x}_j),
\end{equation}
for some configuration $\lbrace m_j \rbrace$ of vortices with topological charge $m_j \in \mathbb{Z}$. This source is then minimally coupled to the relative $U(1)$ symmetry, so that the partition function Eq. \eqref{eq:effact} becomes
\begin{subequations}
    \begin{align}
        \mathcal{Z} &= \sum_{\lbrace m_j \rbrace}\int \mathD \lambda_1 \mathD \lambda_2 \mathD a^\mu \mathD \rho \mathD \theta\, e^{-S_{\mathrm{eff}}[A]}, \\
        S_{\mathrm{eff}}[A]/N &= \tr \ln \begin{pmatrix} -D_\mu^2[a+A] + \lambda_1 & -\rho e^{i\theta} \\ -\rho e^{-i\theta} & -D_\mu^2[a-A] + \lambda_2 \end{pmatrix} + \int \dd^d \vb{x} \left[ \frac{\rho^2}{K_0} - \frac{\lambda_1 + \lambda_2}{g_0} \right] ,
    \end{align}
\end{subequations}
where the sum is over all possible configurations of vortices. Expanding to quadratic order in $\theta$, $A^\mu$, and their derivatives, yields
\begin{figure}[!t]
    \centering
    \includegraphics[scale=0.5]{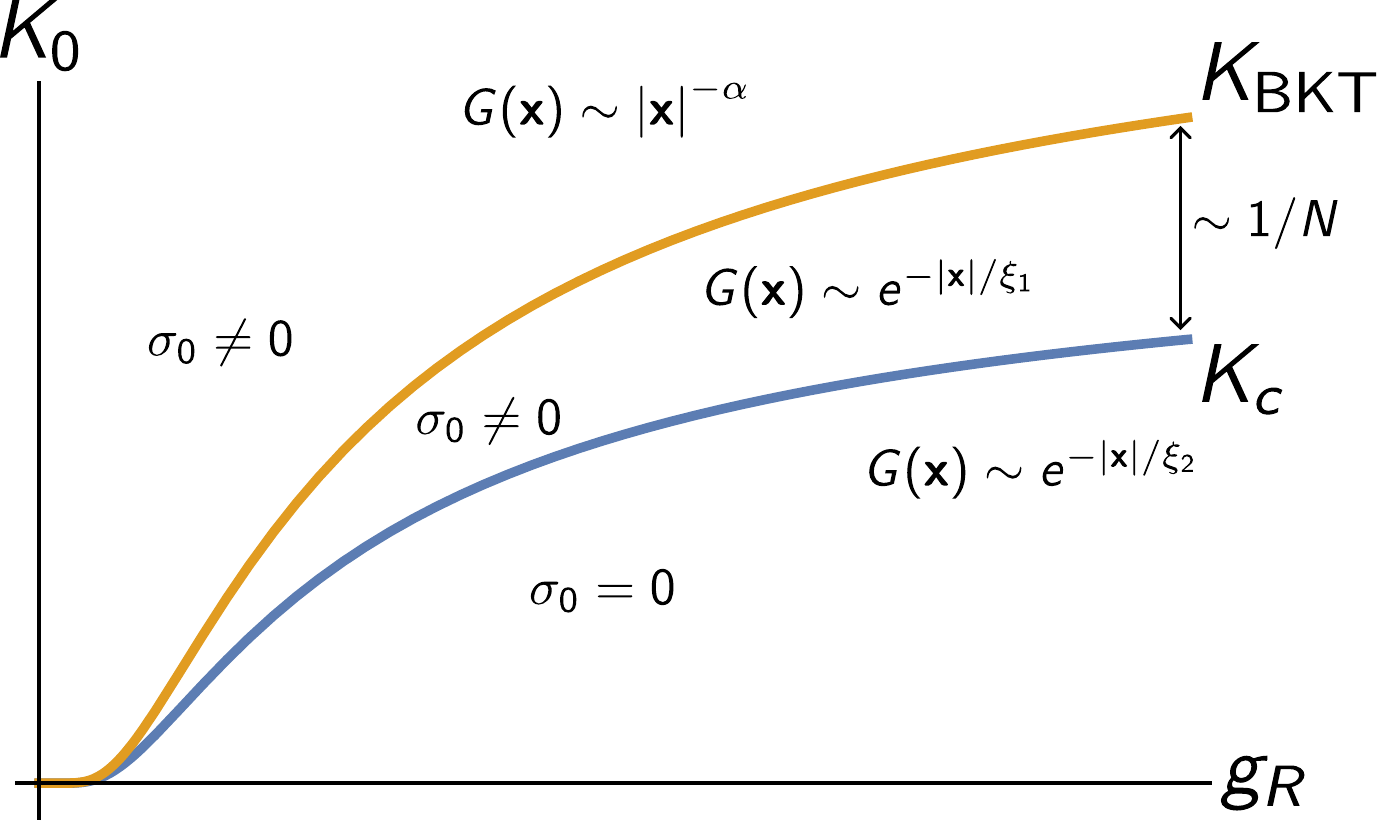}
    \caption{Phase diagram corresponding to the action Eq. \eqref{eq:multicompaction}. For $K_0 > K_{\mathrm{BKT}}$ dislocations are confined, leading to quasi-long-range order and power law correlations with exponent $\alpha > 0$. For $K_c < K_0 < K_{\mathrm{BKT}}$ there is a condensate but dislocations proliferate; correlations of $\sigma$ decay exponentially with length $\xi_1$. For $K_0 < K_c$ no $U(1)$ condensate forms, so correlations of $\sigma$ decay exponentially with different length $\xi_2$. $K_{\mathrm{BKT}}$ is split from $K_c$ to order $1/N$.}
    \label{fig:phase_diagram_clean}
\end{figure}
\begin{align}
    S_{\mathrm{eff}}[A] &\simeq \frac{\gamma_\theta}{2} \int \dd^2 \vb{x} (\del^\mu \theta + 2A^\mu)^2 + \frac{1}{4 e^2} \int \dd^2 \vb{x}  (F^{\mu\nu})^2,
\end{align}
where $F^{\mu\nu} = \del^\mu A^\nu - \del^\nu A^\mu$, $e^2 = \rho_0^2/(2m^2 \gamma_\theta)$ is the effective coupling constant for the ``electrodynamic'' response of $A^\mu$, $\gamma_\theta$ is the phase stiffness
\begin{equation}
    \gamma_\theta = \frac{N}{4\pi \rho_0} \left[m^2 \tanh^{-1}\left(\frac{\rho_0}{m^2}\right) - \rho_0\right] = \frac{N}{4\pi \tilde{\rho}} \left[\sqrt{1+\Tilde{\rho}^2} \sinh^{-1}(\Tilde{\rho}) - \Tilde{\rho} \right],
\end{equation}
and $\Tilde{\rho} = \rho_0/(\mu^2 e^{-4\pi/g_R})$; see Appendix \ref{app:vorticity} for the derivation of these quantities. Note that $\sigma$ has $U(1)$ charge $q=2$ (not to be confused with the topological charge of a vortex) since it is a charged composite of $q = \pm 1$ fields. 

This result implies a critical value of the phase stiffness $\gamma_{\mathrm{BKT}} = 2/(\pi q^2) = 1/2\pi$ such that vortices/dislocations proliferate for $\gamma_\theta < \gamma_{\mathrm{BKT}}$ (see Appendix \ref{app:vorticity}). Since $ \gamma_{\mathrm{BKT}} $ is of order $1/N$ relative to generic values of $\gamma_\theta(\Tilde{\rho})$, for large enough $N$ we can safely approximate $\gamma_{\mathrm{BKT}} \simeq N \Tilde{\rho}_{\mathrm{BKT}}^2/12\pi$, where $\Tilde{\rho}_{\mathrm{BKT}}$ is the value of $\Tilde{\rho}$ which solves the equation $\gamma_{\mathrm{BKT}} = 1/2\pi$. Substituting $\Tilde{\rho}_{\mathrm{BKT}} \simeq 6/\sqrt{N}$ into the saddle point equation Eq. \eqref{eq:weissMFT} and expanding to first order in $1/N$ yields one of our first main results:
\begin{equation}
    \frac{K_{\mathrm{BKT}}}{K_c} = \frac{\Tilde{\rho}_{\mathrm{BKT}}}{\sinh^{-1}(\Tilde{\rho}_{\mathrm{BKT}})} = 1 + \frac{1}{N} + \mathcal{O}(N^{-2}).
\end{equation}
That is, we have explicitly shown how the BKT transition emerges at order $1/N$ relative to the mean-field transition; $K_c$ remains the point at which the amplitude of the condensate forms, but only exponentially-short range order exists for $K_c < K < K_{\mathrm{BKT}}$. The fluctuation-dominated regime between $K_c$ and $K_{\rm BKT}$ is narrow in the large-$N$ regime and it is expected to become broad for smaller values of $N$. The existence of this fluctuational regime is a feature of two-dimensional physics; this behavior is analogous to the situation in (quasi-)two-dimensional superconductors, where the condensate develops at a higher temperature than the onset of zero resistivity. The results of this section are summarized in the phase diagram in Fig. \ref{fig:phase_diagram_clean}. Finally, we note that the apparent contradiction with the Mermin-Wagner theorem is resolved at this order in the fluctuations, since the critical Goldstone phase $K > K_{\mathrm{BKT}}$ also has no long-range order, though with power-law decaying correlations. This is unsurprising given that the Mermin-Wagner theorem is fundamentally a statement of the importance of fluctuations in low dimensions. This behavior is closely analogous to what was found long ago in the chiral Gross-Neveu model, though, importantly, the phase stiffness in that model is a pure number that cannot be tuned \cite{Witten1978}.

\section{The Role of Disorder \label{sec:disorder}}

\subsection{Coupling to Quenched Disorder}

Having established that our model reproduces the salient features of a $U(1)$ order parameter within the large-$N$ limit, we turn to the primary focus of this work: disorder. Any quenched disorder must couple only to gauge invariant combinations of fluctuating fields. With our model, we are spoilt for choice with possibilities, not all of which are physically interesting. For example, the simplest option, a complex scalar disorder field coupled linearly to the emergent $U(1)$ order parameter $\vb*{z}^* \cdot \vb*{w}$, can be easily shown to produce a trivial large-$N$ limit since the dimension of the disorder field does not scale with $N$. As such, it is useful to draw inspiration from the conventional $\mathbb{C}P^N$ model, the large-$N$ solution of which is presented in Appendix \ref{app:CPN}. We find that it is natural to consider ``adjoint disorder''
\begin{equation}
    \mathcal{L}_{\mathrm{dis}}^{(1)} = \mathfrak{z}^a(\vb{x}) z^*_\alpha(\vb{x}) \tau^a_{\alpha\beta} z_\beta(\vb{x}) + \mathfrak{w}^a(\vb{x}) w^*_\alpha(\vb{x}) \tau^a_{\alpha\beta} w_\beta(\vb{x}),
\end{equation}
where, for simplicity and without loss of generality, we take the disorder $\mathfrak{z}^a(\vb{x})$ and $\mathfrak{w}^a(\vb{x})$ to be $N^2$ component real vectors in the adjoint representation of $U(N)$, with generators $\tau^a$ satisfying $\tau^a_{\alpha\beta} \tau^a_{\gamma\delta} = N \delta_{\alpha\delta} \delta_{\beta\gamma}$ (implied summation over repeated indices), and distributed with variance $\eta_1^2$ according to
\begin{equation}
    \overline{\mathfrak{z}^a(\vb{x})} = \overline{\mathfrak{w}^a(\vb{x})} = 0, \kern 2em \overline{\frakz^a(\vb{x}) \frakz^b(\vb{y})} = \overline{\frakw^a(\vb{x}) \frakw^b(\vb{y})} = \eta_1^2 \delta^{ab} \delta^{(d)}(\vb{x}-\vb{y}) ,
\end{equation}
where overlines denote averaging over disorder configurations. Note that cross-correlations between $\frakz^a(\vb{x})$ and $\frakw^a(\vb{x})$ are generally allowed by symmetry, but we do not observe any qualitative impact as a result of this added model complexity. A general approach, however, should also include the new gauge invariant bilinear
\begin{equation}
    \mathcal{L}_{\mathrm{dis}}^{(2)} = \frakh^a(\vb{x}) z^*_\alpha(\vb{x}) \tau^a_{\alpha\beta} w_\beta(\vb{x}) + \frakh^{a*}(\vb{x}) w^*_\alpha(\vb{x}) \tau^a_{\alpha\beta} z_\beta(\vb{x}) ,
    \label{eq:disorder2}
\end{equation}
where $\frakh^a(\vb{x})$ is a $N^2$ component complex random vector with variance $\eta_2^2$:
\begin{equation}
    \overline{\frakh^a(\vb{x})} = 0, \kern2em \overline{\frakh^{a}(\vb{x}) \frakh^b(\vb{y})}  = 0, \kern2em \overline{\frakh^{a*}(\vb{x}) \frakh^b(\vb{y})} = \eta_2^2 \delta^{ab} \delta^{(d)}(\vb{x}-\vb{y}).
    \label{eq:ensemble-relative}
\end{equation}
The coupling to the disorder shown in Eq. \eqref{eq:disorder2} is manifestly invariant under the local symmetry of diagonal $U(1)$ transformations and transforms non-trivially under the global symmetry of the relative $U(1)$ transformations [see Eqs. \eqref{eq:u1transforms}]. Thus, for each realization of the disorder, the random fields explicitly break the relative global symmetry, but it remains unbroken in the ensemble of the distribution of Eq. \eqref{eq:ensemble-relative}. In this way, this second form of disorder has the same $U(1)$ symmetry properties as a disorder field coupled linearly to $\vb*{z}^*\cdot\vb*{w}$. We will see in a later section that symmetry-breaking disorder is vital to the physics of fluctuations, while the neutral disorder is important for stabilizing the ground state of the theory. Other forms of disorder are certainly possible, but the adjoint disorder presented here provides useful mathematical simplifications, and, we believe, is the most physically motivated and natural choice leading to a non-trivial large-$N$ limit.

Since we are interested in thermodynamic observables that are independent of any specific realization of the disorder, we use the replica trick:
\begin{align}
    \overline{\mathZ^n} &= \int \mathD \mathfrak{z} \mathD \mathfrak{w} \mathD \mathfrak{h}\,  \exp\left(- \int \dd^d \vb{x} \left[ \frac{\mathfrak{z}^2 + \mathfrak{w}^2}{2\eta_1^2} + \frac{\abs{\frakh}^2}{\eta_2^2} \right] \right) \mathZ[\frakz,\frakw,\frakh]^n \nn \\
\begin{split}
    &= \int \mathcal{D} \lambda_{1,j} \mathcal{D} \lambda_{2,j}  \mathcal{D} a^\mu_j \mathcal{D} \vb*{z}_j \mathcal{D} \vb*{w}_j \prod_{j=1}^n e^{-S_j} \\
    &\times  \exp\left(  \frac{N}{2} \int \dd^d \vb{x} \sum_{i\neq j=1}^n \left[ \eta_1^2  \left( \abs{\vb*{z}^*_i \cdot \vb*{z}_j}^2 + \abs{\vb*{w}^*_i \cdot \vb*{w}_j}^2 \right) + 2\eta_2^2 (\vb*{z}^*_i \cdot \vb*{z}_j) (\vb*{w}_i \cdot \vb*{w}^*_j) \right] \right), \label{eq:disavgedact}
\end{split}
\end{align}
where $S_j$ are the replicas of the original action Eq. \eqref{eq:multicompaction}. We are excluding the terms with $i=j$ from the sum in the last line because the unit vector constraints $\abs{\vb*{z}_i} = \abs{\vb*{w}_i} = 1$ render them trivial constants. Note that the novelty of quenched disorder in $\mathbb{C}P^N$-type models is the generation of \textit{quartic} inter-replica interaction terms, reminiscent of the Sherrington-Kirkpatrick model of spin glasses \cite{Sherrington1975}. We will see that this produces fundamentally different large-$N$ solutions compared to $O(N)$ models (see Appendix \ref{app:ON}).

\subsection{Large-\texorpdfstring{$N$}{N} Solution \label{sec:dirtyNinfty}}

Next, we must take some care with the Hubbard-Stratonovich transformation to avoid over-counting the effective degrees of freedom in the theory. We start by introducing the collective coordinates $\zeta_{ij}$ and $\omega_{ij}$ for $i\neq j$
\begin{align}
    &\int \mathD \zeta_{ij} \mathD \omega_{ij} \, \delta(\zeta_{ij} - \vb*{z}_i \cdot \vb*{z}^*_j ) \delta( \omega_{ij} - \vb*{w}_i \cdot \vb*{w}^*_j) \nn \\
    &\times \exp(\frac{N}{2} \int \dd^d \vb{x} \left[ \eta_1^2 (\abs{\vb*{z}^*_i \cdot \vb*{z}_j}^2 + \abs{\vb*{w}^*_i \cdot \vb*{w}_j}^2) + 2 \eta_2^2 (\vb*{z}^*_i \cdot \vb*{z}_j) (\vb*{w}_i \cdot \vb*{w}^*_j) \right] ) \nn  \\
    &\kern4em = \int \mathD \zeta_{ij} \mathD \omega_{ij} \mathD \kappa_{ij} \mathD \psi_{ij} \, \exp(-N \int \dd^d \vb{x} \left[ \kappa_{ji}(\zeta_{ij} - \vb*{z}_i \cdot \vb*{z}^*_j ) + \psi_{ji}( \omega_{ij} - \vb*{w}_i \cdot \vb*{w}^*_j)  \right]) \nn \\
    &\kern4em \times  \exp(\frac{N}{2} \int \dd^d \vb{x} \left[ \eta_1^2 ( \zeta_{ij} \zeta_{ji} + \omega_{ij} \omega_{ji}) + 2 \eta_2^2 \omega_{ij} \zeta_{ji} \right] ) ,
\end{align}
and fix $\zeta_{ii} = \omega_{ii} = 0$ for the reason discussed just above. Since this expression is quadratic in $\zeta_{ij}$ and $\omega_{ij}$, the Lagrange multipliers $\kappa_{ij}$ and $\psi_{ij}$ can be eliminated exactly by the saddle point equations
\begin{equation}
    \kappa_{ij} = \eta_1^2 \zeta_{ij} + \eta_2^2 \omega_{ij}, \kern3em \psi_{ij} = \eta_1^2 \omega_{ij} + \eta_2^2 \zeta_{ij} . \label{eq:disorder_lagrange}
\end{equation}
Therefore, the interaction term can be expressed as
\begin{align}
&\exp\left(  \frac{N}{2} \int \dd^d \vb{x} \sum_{i\neq j=1}^n \left[ \eta_1^2  \left( \abs{\vb*{z}^*_i \cdot \vb*{z}_j}^2 + \abs{\vb*{w}^*_i \cdot \vb*{w}_j}^2 \right) + 2\eta_2^2 (\vb*{z}^*_i \cdot \vb*{z}_j) (\vb*{w}_i \cdot \vb*{w}^*_j) \right] \right) \nn \\
    &\kern4em=\int \mathD \zeta_{ij} \mathD \omega_{ij}\, \exp\left(-\frac{1}{2N} \int \dd^d \vb{x} \sum_{i\neq j=1}^n  \left[ \eta_1^2 (\zeta_{ij} \zeta_{ji} + \omega_{ij} \omega_{ji}) +2 \eta_2^2 \omega_{ij} \zeta_{ji}  \right]\right) \\
    &\kern4em\times \exp\left( \int \dd^d \vb{x} \sum_{i\neq j=1}^n  \left[ \vb*{z}^*_i \cdot \vb*{z}_j (\eta_1^2 \zeta_{ij} + \eta_2^2 \omega_{ij}) + \vb*{w}^*_i \cdot \vb*{w}_j (\eta_1^2 \omega_{ij} + \eta_2^2 \zeta_{ij})  \right]  \right) \nn ,
\end{align}
after rescaling $\zeta_{ij}$ and $\omega_{ij}$ by $1/N$. The $\vb*{z}$ and $\vb*{w}$ fields can then be integrated out of the partition function to obtain the effective action
\begin{align}
\begin{split}
    S_{\mathrm{eff}}/N = \Tr \ln \left[ \mathrm{diag}\begin{pmatrix}
    -D_\mu^2[a] + \lambda_1 & - \sigma \\
    -\sigma^* & -D_\mu^2[a] + \lambda_2
    \end{pmatrix}  - \begin{pmatrix} \tilde{\eta}_1^2 \hat{\zeta} + \tilde{\eta}_2^2 \hat{\omega} & 0 \\ 0 & \tilde{\eta}_1^2 \hat{\omega} + \tilde{\eta}_2^2 \hat{\zeta} \end{pmatrix} \right] \\
    + \tr \int \dd^d \vb{x} \left[ \frac{\tilde{\eta}_1^2}{2} (\hat{\zeta}^2 + \hat{\omega}^2) + \tilde{\eta}_2^2  \hat{\zeta} \hat{\omega} + \frac{1}{K_0} \mathrm{diag}(\abs{\sigma}^2) - \frac{1}{g_0} \mathrm{diag}(\lambda_1 + \lambda_2) \right] , \label{eq:disorderedEffAct}
\end{split}
\end{align}
where we have rescaled $(\zeta,\omega) \rightarrow g_0 (\zeta,\omega)$, and defined $g = g_0/N$, $K = K_0/N$, and $\eta_{1,2} = \tilde{\eta}_{1,2}/g_0$ to obtain a well-defined large-$N$ limit. $\mathrm{Tr}(\,\cdot\,)$ includes the functional operator trace as well as the trace over replica indices, $\mathrm{diag}(\,\cdot\,)$ denotes a matrix which is diagonal in replica indices, and $\hat{\zeta}$ and $\hat{\omega}$ are the matrices with elements $\zeta_{ij}$ and $\omega_{ij}$, respectively. We note that this effective action is positive definite as long as $\Tilde{\eta}_1 > \Tilde{\eta}_2$. While the disorder averaged action Eq. \eqref{eq:disavgedact} is bounded from below regardless of the disorder strengths, maintaining $\Tilde{\eta}_1 > \Tilde{\eta}_2$ is necessary if one wishes to avoid spontaneously breaking the replica permutation symmetry, which is beyond the scope of this work. Then, observe that the HS disorder fields inherit the transformation rules
\begin{subequations}
\begin{alignat}{3}
    &\text{(i) diagonal} \kern2em && \zeta_{jk}(\vb{x}) \longrightarrow e^{i(\phi_j(\vb{x}) - \phi_k(\vb{x}))} \zeta_{jk}(\vb{x}), \kern2em && \omega_{jk}(\vb{x}) \longrightarrow e^{i(\phi_j(\vb{x}) - \phi_k(\vb{x}))} \omega_{jk}(\vb{x}),   \\
    &\text{(ii) relative} \kern2em && \zeta_{jk}(\vb{x}) \longrightarrow e^{i(\phi_j - \phi_k)} \zeta_{jk}(\vb{x}), \kern2em && \omega_{jk}(\vb{x}) \longrightarrow e^{-i(\phi_j - \phi_k)} \omega_{jk}(\vb{x}) .
\end{alignat}
\end{subequations}
Therefore, the $\eta_1$ disorder is neutral under all $U(1)$ transformations, while the $\eta_2$ disorder \textit{explicitly} breaks the replicated relative global $U(1)^n$ symmetry down to its replica-diagonal $U(1)$ subgroup; a non-zero expectation value of $\zeta_{ij}$ and $\omega_{ij}$ will also \textit{spontaneously} break the entire $U(1)^n\times U(1)^n$ symmetry group (see Appendix  \ref{app:CPN} for a discussion of the symmetries of the conventional $\mathbb{C}P^N$ model). This means that the natural replica-symmetric and $\mathbb{Z}_2$ exchange-symmetric saddle point $\zeta_{ij}(\vb{x}) = \omega_{ij}(\vb{x}) = \omega_0 (1 - \delta_{ij})$ is actually a privileged choice and expanding around this specific configuration loses information about the global phase diagram. However, this assumption allows us to obtain a closed-form expression for the disorder-averaged effective potential. We will first derive this effective potential and then show that the replica-symmetric configuration is stable in the regime of interest $\Tilde{\eta}_1 > \Tilde{\eta}_2$.

We then take the spatially uniform and replica-diagonal ansatz $\sigma_i(\vb{x}) = \sigma_0$, $\lambda_{1,i}(\vb{x}) = \lambda_{2,i}(\vb{x}) = m^2$, $a^\mu_i(\vb{x}) = 0$, and for convenience, $\zeta_{ij}(\vb{x}) = \omega_{ij}(\vb{x}) = \omega_0(1 - \delta_{ij})/(\eta_1^2 + \eta_2^2)$ (suppressing tildes from hereon for notational clarity). Next, one can simplify the block matrix structure in the effective action to obtain the replicated effective potential
\begin{align}
\begin{split}
    \overline{U_{\mathrm{eff}}^n} = \int \frac{\dd^d \vb{q}}{(2\pi)^d} \ln \det \left(\left[ (\vb{q}^2 + m^2 + \omega_0) \hat{I} - \omega_0 \hat{M} \right]^2 - \abs{\sigma_0}^2 \hat{I} \right) \\
    + n (n - 1) \frac{\omega_0^2}{\eta_1^2 + \eta_2^2} + n \frac{\abs{\sigma_0}^2}{K_0} - n \frac{2 m^2}{g_0},
\end{split}
\end{align}
where $\hat{I}$ is the $n\times n$ identity matrix and $\hat{M}$ is the $n\times n$ matrix of ones. After the same renormalization procedure as in all the previous cases, it follows that the disorder-averaged physical effective potential is
\begin{align}
\begin{split}
    \overline{U_R} =  \int \frac{\dd^d \vb{q}}{(2\pi)^d}  \left[ \ln \left(\frac{(\vb{q}^2 + m^2 + \omega_0)^2 - \abs{\sigma_0}^2}{\vb{q}^4} \right) - \frac{2 \omega_0 (\vb{q}^2 + m^2 + \omega_0)}{(\vb{q}^2 + m^2 + \omega_0)^2 - \abs{\sigma_0}^2} \right. \\ - \left. \frac{2m^2}{\vb{q}^2 + \mu^2} \right]
    - \frac{\omega_0^2}{\eta_1^2 + \eta_2^2} + \frac{\abs{\sigma_0}^2}{K_0} - \frac{2 m^2}{g_R} .
\end{split}
\end{align}
In $d=2$ this evaluates to
\begin{equation}
\begin{split}
    \overline{U_R} = \frac{m^2}{4\pi} \left[2 - \ln\left( \frac{(m^2+\omega_0)^2 - \abs{\sigma_0}^2}{\mu^4}\right) \right] + \frac{\abs{\sigma_0}}{4\pi} \ln \left( \frac{m^2 + \omega_0 - \abs{\sigma_0}}{m^2 + \omega_0  + \abs{\sigma_0}} \right) \\ + \frac{ \omega_0}{2\pi} - \frac{\omega_0^2}{\eta_1^2 + \eta_2^2} + \frac{\abs{\sigma_0}^2}{K_0} - \frac{2m^2}{g_R}. \label{eq:multicompEffPot}
\end{split}
\end{equation}

The full multivariate structure is complicated, so we first consider the case $\sigma_0 = 0$. This reduces the problem to two decoupled $\mathbb{C}P^N$ models with disorder strength $\eta_1^2 + \eta_2^2 \equiv \eta_{\mathrm{tot}}^2$ (see Appendix \ref{app:CPN}); the lack of any dependence on the relative magnitudes of $\eta_1$ and $\eta_2$ is a result of the replica-symmetric ansatz. The saddle point equations for $\omega_0$ and $m^2$ are, respectively,
\begin{align}
    m^2 + \omega_0 = \frac{\eta_{\mathrm{tot}}^2}{4\pi}, \kern3em
    \frac{\omega_0}{m^2 + \omega_0} - \ln\left( \frac{m^2 + \omega_0}{\mu^2}\right)  = \frac{4\pi}{g_R}.
\end{align}
However, there is an apparent problem with these equations, as they seem to imply that $\omega_0 < 0$ for $\eta_{\mathrm{tot}}^2 < \eta_{c,0}^2 = 4\pi \mu^2 e^{-4\pi/g_R}$; if this were the case, the scalar field propagator in Eq. \eqref{eq:disorderedEffAct} would not be positive-definite. However, in this problem regime, the configuration with $\omega_0 = 0$ is more energetically favorable, being the global extremum (maximum, because of the effectively negative number of degrees of freedom of $\zeta_{ij}$ and $\omega_{ij}$ when $n \rightarrow 0$) \cite{Parisi-1982}. This implies that $\eta_{c,0}$ is a crossover scale which divides a ``clean'' or weakly-disordered regime from a strongly-disordered regime:
\begin{equation}
    m^2 + \omega_0 = \begin{cases}
        \eta_{c,0}^2/4\pi, \kern1em & \eta_{\mathrm{tot}} < \eta_{c,0}, \\
        \eta_{\mathrm{tot}}^2/4\pi, \kern1em & \eta_{\mathrm{tot}} \geq \eta_{c,0} .
    \end{cases}
    \kern3em
    \omega_0 = \begin{cases}
        0, & \eta_{\mathrm{tot}} < \eta_{c,0}, \\
        \dfrac{\eta_{\mathrm{tot}}^2}{2\pi} \ln\left( \dfrac{\eta_{\mathrm{tot}}}{\eta_{c,0}} \right), \kern1em & \eta_{\mathrm{tot}} \geq \eta_{c,0} ,
    \end{cases}
\end{equation}
Observe that $m^2$ and $\omega_0$ are continuous at $\eta_{c,0}$, and that for $\eta_0 < \eta_{c,0}$, $m = \mu e^{-2\pi/g_R}$ has the same value as in the theory without disorder.

\begin{figure}[!b]
    \centering
    \includegraphics[scale=0.5]{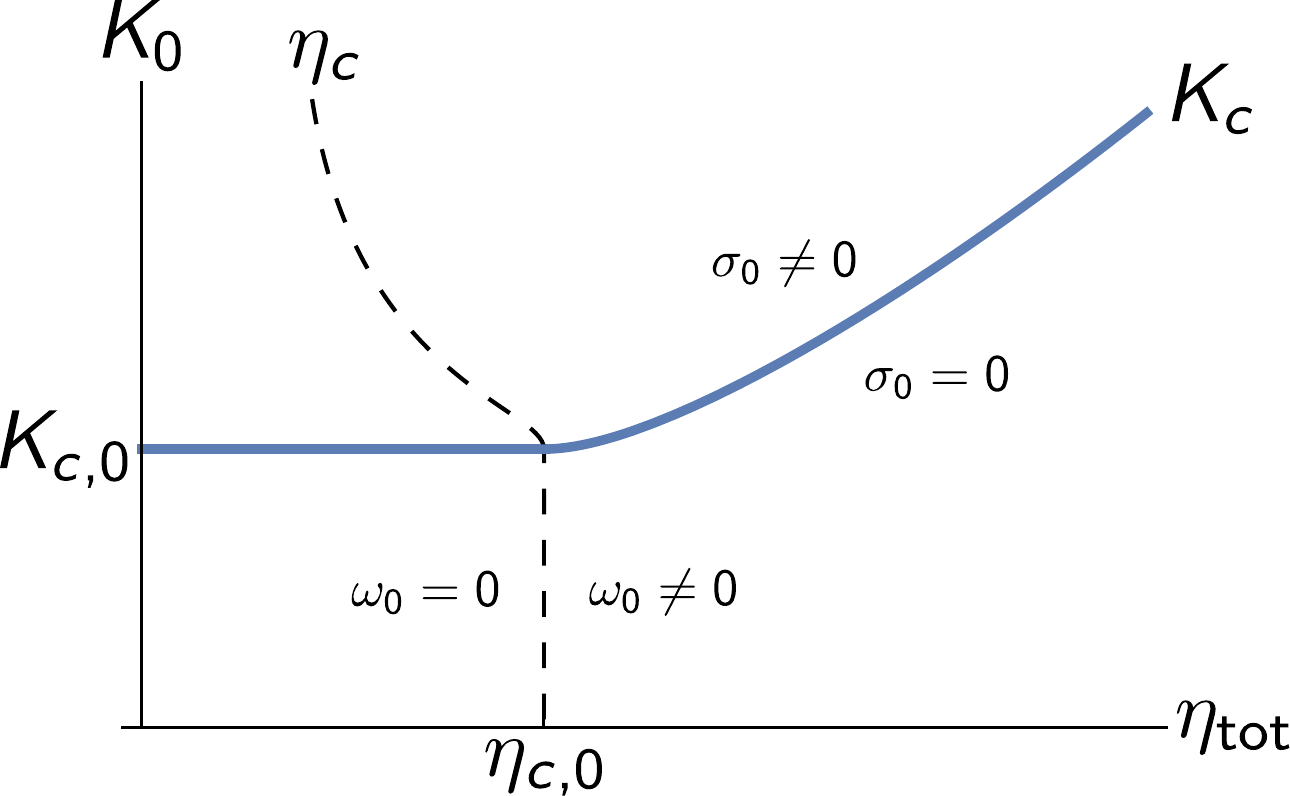}
    \caption{$N=\infty$ phase diagram of the theory with quenched disorder. $\omega_0 = 0$ everywhere to the left of the dashed line. $\sigma_0 = 0$ everywhere below the solid blue line. $K_{c,0}$ is the critical coupling in the clean limit and $\eta_{c,0}$ is the critical disorder in the absence of condensate.}
    \label{fig:phase_diagram_dirty}
\end{figure}

For $\sigma_0 \neq 0$, the full saddle point equations cannot be solved explicitly, so we simply state them here for completeness:
\begin{subequations} \label{eq:saddle_disordered}
\begin{align}
    m^2 + \omega_0 = \frac{1}{8\pi} \left( \eta_{\mathrm{tot}}^2 + \sqrt{\eta_{\mathrm{tot}}^4 + 64\pi^2 \abs{\sigma_0}^2} \right) ,\\
    \frac{\omega_0}{(m^2 + \omega_0)^2 - \abs{\sigma_0}^2} - \frac{1}{2} \ln\left(\frac{(m^2 + \omega_0)^2 -  \abs{\sigma_0}^2}{\mu^4}\right) = \frac{4\pi}{g_R} , \\
    \abs{\sigma_0} \left(\frac{4\pi}{K_0} - \frac{\omega_0}{(m^2 + \omega_0)^2 - \abs{\sigma_0}^2}\right) = \tanh^{-1}\left(\frac{\abs{\sigma_0}}{m^2 + \omega_0} \right). \label{eq:saddle_sigma_disordered}
\end{align}
\end{subequations}
The phase boundary for the onset of a condensate density $\abs{\sigma_0}$ follows from the non-trivial solution to Eq. \eqref{eq:saddle_sigma_disordered}:
\begin{align}
    K_c(\eta_{\mathrm{tot}}) = \begin{cases}
        K_{c,0}, \kern1em & \eta_{\mathrm{tot}} < \eta_{c,0}, \\
        \dfrac{(\eta_{\mathrm{tot}}/\eta_{c,0})^2}{1 + 2\ln(\eta_{\mathrm{tot}}/\eta_{c,0})} K_{c,0}, \kern1em & \eta_{\mathrm{tot}} \geq \eta_{c,0} ,
    \end{cases}
\end{align}
where $ K_{c,0} = \eta_{c,0}^2 = 4\pi \mu^2 e^{-4\pi/g_R}$ is the critical coupling for the mean-field $U(1)$ transition in the clean limit ($\eta_{\mathrm{tot}} = 0$). Therefore, at the mean-field level, the disorder only shifts the boundary for the onset of the condensate; a larger coupling $K_0$ is required to overcome increasing amounts of disorder. Importantly, however, the disorder does not eliminate the condensate (at least at the mean-field level). Similarly, the boundary for the disorder-driven crossover is shifted at finite condensate density:
\begin{equation}
    \eta_c(K_0) = \begin{cases}
        \eta_{c,0}, \kern1em & K_0 < K_{c,0}, \\
        \eta_{c,0} (1 + \abs{\Tilde{\sigma}(K_0)}^2)^{-1/4}, \kern1em & K_0 \geq K_{c,0},
    \end{cases}
\end{equation}
where $\Tilde{\sigma} = \sigma_0/(\mu^2 e^{-4\pi/g_R})$, and since $\omega_0 = 0$ on the phase boundary, $\Tilde{\sigma}(K_0)$ is determined by the clean saddle point equation Eq. \eqref{eq:weissMFT}. The above results are summarized in the phase diagram in Fig. \ref{fig:phase_diagram_dirty}.

Just as the large-$N$ solution in the absence of disorder appeared to violate the Mermin-Wagner theorem, our present solution appears to violate the Imry-Ma condition \cite{Imry1975}; in dimensions $d<4$, any infinitesimal amount of disorder should not only destroy long-range order, but correlations must decay at least exponentially; i.e., the existence of a Goldstone phase with power law correlations must also be eliminated. Note that a condensate of $\omega_0$ does not break any physical symmetries of the non-replicated theory, and hence, is not problematic. In the following section, we will show that in the dirty regime $\omega_0 > 0$ the contradiction with Imry-Ma is directly resolved by the inclusion of fluctuations to leading order in $N$. This will mirror the resolution of the Mermin-Wagner theorem in the clean theory. Next, we will demonstrate that sub-leading order fluctuations in the clean regime $\omega_0 = 0$ must be included, and generate qualitatively the same interactions which are responsible for the destruction of long-range order in the dirty regime.

\subsection{Strong Disorder Regime \label{sec:strongDisorder}}

\tocless{\subsubsection}{Correlations in the Symmetric Phase}

We begin with the case $\sigma_0 = 0$ and $\omega_0 > 0$. The physical observable of most interest in this regime is the correlation function of the order parameter. The natural parameterization here is the Cartesian form $\sigma_j(\vb{x}) = \sigma^R_j(\vb{x}) + i \sigma^I_j(\vb{x})$. Expanding the effective action Eq. \eqref{eq:disorderedEffAct} to quadratic order in the $\sigma^R$ and $\sigma^I$ yields
\begin{equation}
    S_{\mathrm{eff}} = \frac{N}{2} \sum_{a=R,I} \sum_{i,j=1}^n \int \dd^2 \vb{x} \dd^2 \vb{y}\, \sigma^{(a)}_i(\vb{x}) \left[ \Pi^{(1)}_\sigma(\vb{x}-\vb{y};n) \hat{I} - \Pi^{(2)}_\sigma(\vb{x}-\vb{y};n) \hat{M} \right]_{ij} \sigma^{(a)}_j(\vb{y}), \label{eq:OP_effAct_disordered}
\end{equation}
where the full integral expressions for the kernels $\Pi^{(1)}(\vb{x}-\vb{y};n)$ and $\Pi^{(2)}(\vb{x}-\vb{y};n)$ are given in Appendix \ref{app:fluctuations}. It follows directly from the matrix structure of this action that the propagator for $\sigma_i$ is
\begin{equation}
    \langle \sigma_i(\vb{p}) \sigma_j^*(-\vb{p}) \rangle = 2\left( \frac{\delta_{ij}}{\Pi^{(1)}_\sigma(\vb{p};n)} + \frac{\Pi^{(2)}_\sigma(\vb{p};n)}{\Pi^{(1)}_\sigma(\vb{p};n)[\Pi^{(1)}_\sigma(\vb{p};n) - n \Pi^{(2)}_\sigma(\vb{p};n)]} \right),
\end{equation}
and hence, that the disorder-averaged propagator is
\begin{equation}
    \overline{\langle \sigma(\vb{p}) \sigma^*(-\vb{p}) \rangle} = \lim_{n\rightarrow0} \frac{1}{n} \tr\langle \sigma_i(\vb{p}) \sigma_j^*(-\vb{p}) \rangle  = 2\left(\frac{1}{\Pi^{(1)}_\sigma(\vb{p};0)} + \frac{\Pi^{(2)}_\sigma(\vb{p};0)}{[\Pi^{(1)}_\sigma(\vb{p};0)]^2} \right).
\end{equation}
Immediately, we see that the Imry-Ma condition holds: if $\Pi^{(2)}_\sigma(0;0) \neq 0$, the replica-diagonal kernel $\Pi^{(1)}_\sigma(\vb{p};0)$ must be gapped for all $d < 4$, barring any pathological momentum dependence. Specifically, in $d = 2$ we find for small momentum
\begin{subequations}
\begin{align}
    \Pi^{(1)}_\sigma(\vb{p};0) &= \left[\frac{2}{K_0} - \frac{m^2 + 2\omega_0}{2\pi(m^2 + \omega_0)^2}\right] + \frac{m^2 + 3\omega_0}{12\pi(m^2 + \omega_0)^3}  \vb{p}^2 + \mathcal{O}(\vb{p}^4) , \label{eq:dirty_propA} \\
    \Pi^{(2)}_\sigma(\vb{p};0) &= \frac{\omega_0^2}{6\pi(m^2 + \omega_0)^3} - \frac{\omega_0^2}{10\pi(m^2 + \omega_0)^4} \vb{p}^2 + \mathcal{O}(\vb{p}^4) .
\end{align}
\end{subequations}
Note that the pole mass of the order parameter is proportional to the square-bracketed term in Eq. \eqref{eq:dirty_propA}; substituting the saddle-point values for $m^2$ and $\omega_0$ recovers the critical value $K_c(\eta_{\mathrm{tot}})$ above which the $\sigma_0 = 0$ state becomes unstable. To compare this propagator with the disorder averaged propagator of an $O(N)$ model (see Appendix \ref{app:ON}), we can put the propagator into canonical form by rescaling $\sigma$ by the coefficient of $\vb{p}^2$. Then, using the small momentum expansion of the kernels,
\begin{equation}
    \overline{\langle \sigma(\vb{p}) \sigma^*(-\vb{p}) \rangle} \simeq \frac{1}{\vb{p}^2 + m^2_\sigma} + \frac{\eta_\sigma^2}{(\vb{p}^2 + m^2_\sigma)^2}, \label{eq:doubleLorentzian}
\end{equation}
where the effective mass and (static) disorder strength are
\begin{subequations}
\begin{align}
    m^2_\sigma &= \frac{12\pi(m^2 + \omega_0)^3}{m^2 + 3\omega_0}\left[\frac{2}{K_0} - \frac{m^2 + 2\omega_0}{2\pi(m^2 + \omega_0)^2}\right], \\
    \eta^2_\sigma &= \frac{2 \omega_0^2}{m^2 + 3\omega_0}. \label{eq:sigma_disorder}
\end{align}
\end{subequations}
Therefore, the propagator has the well-known \cite{Pytte1981,Yoshizawa1982} double-Lorentzian form; any approximation of the double-Lorentzian term in powers of momentum should at least respect the double-pole structure. A potentially surprising feature of this result, compared to the momentum-independent numerator of the propagator in the $O(N)$ model, is the ``disorder kernel'' $\Pi^{(2)}(\vb{p})$. We note that this momentum dependence is simply a consequence of the fact that $\sigma_i$ is a bound state of the $\vb*{z}_i$ and $\vb*{w}_i$, and \textit{not} of the composite nature of the disorder fields $\zeta_{ij}$ and $\omega_{ij}$, since to this order in $N$ fluctuations of $\sigma$ do not couple to fluctuations of the disorder fields.

\tocless{\subsubsection}{Nature of the Symmetry-Breaking Phase}

Here we consider the regime $\sigma_0 \neq 0$ and $\omega_0 > 0$. In the absence of explicit symmetry breaking ($\eta_2$ disorder), the low energy degrees of freedom of the effective action Eq. \eqref{eq:disorderedEffAct} are the $n$ gapless Goldstone modes corresponding to long-wavelength distortions of the replicated relative symmetry phases $\theta_i(\vb{x})$
\begin{equation}
    \sigma_j(\vb{x}) \simeq \rho_0 e^{i\theta_j(\vb{x})}, \kern2em \zeta_{jk}(\vb{x}) \simeq \frac{\omega_0}{\eta^2_{\mathrm{tot}}} e^{i(\theta_j(\vb{x}) - \theta_k(\vb{x}))/2}, \kern2em \omega_{jk}(\vb{x}) \simeq \frac{\omega_0}{\eta^2_{\mathrm{tot}}} e^{-i(\theta_j(\vb{x}) - \theta_k(\vb{x}))/2},
\end{equation}
where the phases $\theta_i$ are scaled so as to correspond to the replicated order parameter phase. With the $U(1)^n$ global symmetry explicitly broken down to the replica-diagonal $U(1)$ subgroup, $n-1$ of these modes will become gapped. However, as long as the symmetry breaking disorder is sufficiently weak, other gapped modes will remain frozen out at comparatively higher energies. Given these considerations, the infrared sector of the effective action in $d=2$ is
\begin{align}
    S_{\mathrm{eff}} \simeq \frac{1}{2} \sum_{i,j=1}^n \int \dd^2 \vb{x}\, \theta_i\left[ -\gamma_\theta \del^2 \hat{I} +  \left( - \Gamma_\theta \del^2 + m^2_{\theta} \right) \left( \hat{I} - \frac{1}{n} \hat{M}\right) \right]_{ij} \theta_j, \label{eq:IRdisEffAct}
\end{align}
where the three parameters $\gamma_\theta(n)$, $\Gamma_\theta(n)$ and $m_\theta(n)$ depend explicitly on the number $n$ of replicas. The derivation of this effective action and the full (rather uninstructive) expressions for the three parameters are given in Appendix \ref{app:fluctuations}. Note that the $m_\theta$ term also naively appears to break the periodicity of the $\theta_j$. This is forbidden, since Eq. \eqref{eq:disorderedEffAct} is invariant under shifts $\theta_j \rightarrow \theta_j + 2\pi$, so the constant term must actually be the total contribution to quadratic order from terms in the ``true'' IR effective theory of the form
\begin{equation}
    \mathcal{L} = \mathcal{L}_0 + \sum_{p=1}^\infty \mathcal{L}_p, \kern3em
    \mathcal{L}_p = -\alpha_p \sum_{i,j=1}^n \cos(p(\theta_i - \theta_j)), \label{eq:anistropy}
\end{equation}
where $p\in\mathbb{N}$ are the degrees of $p$-fold anisotropy and $\mathcal{L}_0$ is the part of the theory which remains fully $U(1)^n$ invariant [the gradient terms in Eq. \eqref{eq:IRdisEffAct}]. Our conventional $1/N$ expansion does not give us direct access to the coefficients $\alpha_p$, but it is clear that the full effective theory should respect the periodicity of the $\theta_i$.

We will now unpack this result. The spectrum of the inverse propagator is
\begin{subequations}
\begin{alignat}{3}
    e_0 &= \gamma_\theta \vb{p}^2, \kern2em &&\text{multiplicity 1}, \kern2em &&\varphi_0 = \frac{1}{n} \begin{pmatrix}
        1 & 1 & \dots & 1
    \end{pmatrix}^T, \\
    e_i &= (\gamma_\theta + \Gamma_\theta) \vb{p}^2 + m_\theta^2, \kern2em &&\text{multiplicity $n-1$}, \kern2em &&\varphi_1 = \frac{1}{2} \begin{pmatrix}
        1 & -1 & 0 & \dots & 0
    \end{pmatrix}^T,\\
    & && &&\varphi_2 = \frac{1}{4} \begin{pmatrix}
        1 & 1 & -2 & 0 \dots & 0
    \end{pmatrix}^T, \nn \\
    & && &&\dots, \nn
\end{alignat}
\end{subequations}
where the $\varphi_i$ are orthogonal eigenvectors of the kernel in the basis of the $\theta_i$, and their normalization is chosen to preserve the compactification radius ($L^1$ norm). The single remaining gapless mode $\varphi_0$ corresponds to in-phase fluctuations of all the replicas; the gapped modes correspond to the $n-1$ linearly independent out-of-phase motions, where, to leading order in the replica number $n$,
\begin{equation}
    m_\theta^2 \simeq n N  \frac{(1 - \Delta) \omega_0^2}{4\pi} \left[\frac{m^2 + \omega_0}{(m^2 + \omega_0)^2 - \rho_0^2} - \frac{1 - \Delta}{\rho_0} \tanh^{-1}\left(\frac{\rho_0}{m^2 + \omega_0}\right)\right], \label{eq:phase_gap}
\end{equation}
and $\Delta = (\eta_1^2 - \eta_2^2)/(\eta_1^2 + \eta_2^2)$. Since $m_\theta^2 > 0$ for $\eta_1 > \eta_2$, this result confirms that the ground state obtained in the large-$N$ limit is stable.

Next, we must address the nature of physical observables in this dirty regime and compatibility with Imry-Ma. It is useful to consider an effective disorder-averaged phase stiffness. To this end, we first project Eq. \eqref{eq:IRdisEffAct} into the ``deep IR'' scale below the mass gap $m_\theta$,
\begin{align}
    S_{\mathrm{eff}} \simeq \frac{n \gamma_\theta(n)}{2} \int \dd^2 \vb{x} (\del^\mu \varphi_0)^2 .
\end{align}
We again emphasize that the factor of $n$ is needed to preserve the compactification radius on changing basis from $\theta_i$ to $\varphi_0$. Next, we impose a uniform static twist $\varphi_0(\vb{x}) = \vb{Q} \cdot \vb{x}$. Finally, the phase stiffness should be identified with the usual thermodynamic helicity modulus, determined from the disorder-averaged effective potential in the presence of the twist:
\begin{align}
    \overline{\gamma_\theta} &= \nabla_{\vb{Q}}^2 \overline{U_R(\vb{Q})} \Big\vert_{\vb{Q}=0} \nn \\
    &= \lim_{n\rightarrow 0} \gamma_\theta(n) \nn \\
    &= \frac{N}{4\pi \rho_0} \left[m^2 \tanh^{-1}\left(\frac{\rho_0}{m^2 + \omega_0}\right) - \rho_0 \frac{m^2 (m^2 + \omega_0) - \rho_0^2}{(m^2 + \omega_0)^2 - \rho_0^2}\right]. \label{eq:disorder_stiffness}
\end{align}
We could repeat our analysis from the clean theory to determine what this expression predicts for the evolution of the BKT transition as a function of disorder. However, it is well-understood that $\overline{\gamma_\theta}$ controls the relevance (in the renormalization group sense) of the disorder-induced $p$-fold anisotropy Eq. \eqref{eq:anistropy} \cite{Houghton1981,Goldschmidt1982}. Since the theory can be tuned to make $\omega_0$ small, we can treat disorder as a perturbation to the sine-Gordon representation of the clean theory [see Eq. \eqref{eq:sineGordon}]. Then, the results of Ref. \cite{Houghton1981} imply that order-$p$ anisotropy is a relevant perturbation when $\overline{\gamma_\theta} > p^2/16\pi$. Therefore, the BKT transition at $\overline{\gamma_\theta} = 1/2\pi$ is actually preempted by random field ($p=1$) and random bond ($p=2$) anisotropy. Unfortunately, our large-$N$ analysis does not reveal the nature of the disorder-dominated phase. However, it is clear that the conventional Goldstone phase with power law correlations cannot survive, and the BKT vortex plasma is known to have exponentially-decaying correlations.

To determine the boundary of the disorder dominated phase $K_{\mathrm{dis}}(\eta_{\mathrm{tot}})$, we begin by using the saddle point equations Eqs. \eqref{eq:saddle_disordered}, to write $\overline{\gamma_\theta}$ as a function of only the dimensionless parameters $\Tilde{\rho} = \rho_0/(\mu^2 e^{-4\pi/g_R})$ and $\eta_{\mathrm{tot}}/\eta_{c,0}$. Expanding the resulting expression to quadratic order in $\Tilde{\rho}$ yields
\begin{equation}
    \overline{\gamma_\theta} \simeq \left[\frac{1 + 4 \ln(\eta_{\mathrm{tot}}/\eta_{c,0}) }{(\eta_{\mathrm{tot}}/\eta_{c,0})^4} \right] \frac{N \Tilde{\rho}^2}{12 \pi}.
\end{equation}
Similarly, the saddle point equation Eq. \eqref{eq:saddle_sigma_disordered} can be expressed entirely in terms of $\Tilde{\rho}$, $\eta_{\mathrm{tot}}/\eta_{c,0}$ and $K_0/K_{c,0}$. Expanding the result to quadratic order in $\Tilde{\rho}$ yields
\begin{equation}
    \frac{K_0}{K_{c,0}} \simeq \frac{(\eta_{\mathrm{tot}}/\eta_{c,0})^2}{1 + 2\ln(\eta_{\mathrm{tot}}/\eta_{c,0})} + \frac{1 + 12\ln(\eta_{\mathrm{tot}}/\eta_{c,0})}{6(\eta_{\mathrm{tot}}/\eta_{c,0})^2 [1 + 2\ln(\eta_{\mathrm{tot}}/\eta_{c,0})]^2} \Tilde{\rho}^2 .
\end{equation}
Solving $\overline{\gamma_{\mathrm{dis}}} = p^2/16\pi$ for $\Tilde{\rho}_{\mathrm{dis}}$ and substituting into the saddle point equation yields
\begin{equation}
    \frac{K_{\mathrm{dis}}(\eta_{\mathrm{tot}})}{K_c(\eta_{\mathrm{tot}})} \simeq 1 +  \frac{1 + 12\ln(\eta_{\mathrm{tot}}/\eta_{c,0})}{1 + 6 \ln(\eta_{\mathrm{tot}}/\eta_{c,0}) + 8 \ln^2(\eta_{\mathrm{tot}}/\eta_{c,0})} \frac{p^2}{8N}. \label{eq:BKT_disordered}
\end{equation}
Predictably, for $\eta_{\mathrm{tot}} \gg \eta_{c,0}$ the vortex-dominated BKT phase eventually vanishes, though only logarithmically slowly. The surprising feature of this result is its non-monotonicity; the vortex phase is actually slightly \textit{extended} to a maximum value $K_{\mathrm{dis}}(\eta_{\mathrm{max}})/K_c(\eta_{\mathrm{max}}) \approx (1 + 0.17 p^2/N)$ at $\eta_{\mathrm{max}}/\eta_{c,0} \approx 1.2$; see Fig. \ref{fig:phase_diagram_final}. Note that this result is not an artifact of the expansion in powers of $\Tilde{\rho}$ as it can be confirmed by solving the equation $\overline{\gamma_{\mathrm{dis}}} = p^2/16\pi$ numerically. It may seem counter-intuitive for small amounts of disorder to extend the BKT phase relative to the impurity-dominated phase. However, this simply reflects the fact that the disorder initially has a stronger effect on the condensate density than on the phase stiffness.

\subsection{Weak Disorder Regime \label{sec:weakDisorder}}

In the previous section, the primary role of disorder was to mediate interactions between replicated fields via the $N=\infty$ $\vb*{z}$ and $\vb*{w}$ field propagator
\begin{equation}
    \hat{G}_0^{-1}(\vb{p}) = \begin{pmatrix}
    (\vb{p}^2 + m^2 + \omega_0) \hat{I} - \omega_0 \hat{M} & - \sigma_0 \hat{I} \\
    -\sigma_0^* \hat{I} & (\vb{p}^2 + m^2 + \omega_0) \hat{I} - \omega_0 \hat{M}
    \end{pmatrix}. \label{eq:zwprop}
\end{equation}
Naively, it might appear as though the system is blind to disorder in the regimes where $\omega_0 = 0$. In this section, we will show that this is not the case and that fluctuations which are sub-leading in $1/N$ generate qualitatively the same inter-replica couplings as in the strong disorder regime. To demonstrate this, it will suffice to consider the case where $\sigma_0 = 0$.

\begin{figure}[!t]
    \centering
    \includegraphics[scale=0.75]{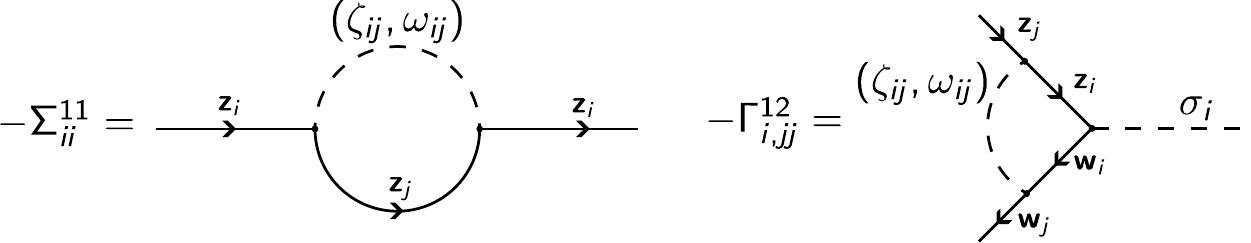}
    \caption{One-loop self energy and three-point vertex contributions from fluctuations of $\zeta_{ij}$ and $\omega_{ij}$.}
    \label{fig:vertex_correction}
\end{figure}

In the strong disorder regime, the propagator had the double-Lorentzian form characteristic of disordered systems due to the replica-mixing kernel $\Pi_\sigma^{(2)}(\vb{p},n)$. If we can show that this kernel is actually non-zero even when $\omega_0 = 0$, then the same conclusions as before will hold. We can write the kernel extremely generally in the form
\begin{equation}
    -\Pi_{\sigma,ij}^{(2)}(\vb{p}) = \int \frac{\dd^2 \vb{q}}{(2\pi)^2} \Gamma^{ab}_{i,k_1k_2}(\vb{p},-\vb{p}-\vb{q},\vb{q}) \Gamma^{cd}_{j,\ell_1\ell_2}(-\vb{p},\vb{p} + \vb{q},-\vb{q}) G^{ac}_{k_1\ell_1}(\vb{p} + \vb{q}) G^{bd}_{k_2\ell_2}(\vb{q}) \label{eq:kernelgeneralform}
\end{equation}
where $a,b,c,d=1,2$ denote either $\vb*{z}$ or $\vb*{w}$, respectively, and $i,k_1,k_2,\ell_1,\ell_2$ are replica indices, with implied summation over repeated indices. $G^{ab}_{ij}(\vb{p})$ is the \textit{exact} $\vb*{z}$, $\vb*{w}$ propagator, and $\Gamma^{ab}_{i,jk}(\vb{p},\vb{q},\vb{k})$ is, similarly, the exact three-point vertex between $\sigma_i$ and the $\vb*{z}$ and $\vb*{w}$ fields with replica indices $j$ and $k$; within the large-$N$ method, both these quantities have a perturbative expansion in powers of $1/N$. In the $N=\infty$ limit, that is, at tree level, the propagator is simply the bare propagator Eq. \eqref{eq:zwprop}, and the three-point vertex is
\begin{equation}
    (\Gamma_{i,jk})^{ab} = \begin{pmatrix}
        0 & 1 \\
        1 & 0
    \end{pmatrix}_{ab} \delta_{ij} \delta_{ik},
\end{equation}
independent of momentum and completely replica-diagonal. At this level, the necessary replica off-diagonal terms only appear when $\omega_0 > 0$. However, within the $1/N$ expansion, both the propagator and the vertex functions receive corrections due to fluctuations of all the collective fields. For example, when $\sigma_0 = 0$, the propagator can acquire off-diagonal components which couple the $\vb*{z}_i$ and $\vb*{w}_i$ fields together due to fluctuations of the $\sigma_i$. Similarly, when $\omega_0 = 0$, the three-point vertex can acquire contributions which are off-diagonal in the replica indices due to fluctuations of the $\zeta_{ij}$ and $\omega_{ij}$. At least at one-loop level, the propagator cannot acquire replica off-diagonal components when $\omega_0 = 0$ (see Fig. \ref{fig:vertex_correction}). 

A quantitatively correct calculation of the $1/N$ corrections to physical observables, such as the order parameter correlation function, requires the inclusion of self energy and vertex corrections from \textit{every} fluctuating field. This is not a particularly instructive calculation, so we will content ourselves with examining the replica-mixing contribution to the three-point vertex, as shown in Fig. \ref{fig:vertex_correction}. To leading order, this involves the order-$N$ contribution to the propagator for the $\zeta_{ij}$ and $\omega_{ij}$ fields. Expanding Eq. \eqref{eq:disorderedEffAct} to quadratic order in these fields yields
\begin{equation}
    S_{\mathrm{eff}} \simeq \frac{N}{2} \sum_{i,j=1}^n \int \dd^2 \vb{x} \dd^2 \vb{y} \begin{pmatrix}
        \zeta_{ij}(\vb{x}) & \omega_{ij}(\vb{x}) \end{pmatrix} \begin{pmatrix}
            \Pi^{(1)}_\omega(\vb{x}-\vb{y}) & \Pi^{(2)}_\omega(\vb{x}-\vb{y}) \\ \Pi^{(2)}_\omega(\vb{x}-\vb{y}) & \Pi^{(1)}_\omega(\vb{x}-\vb{y})
        \end{pmatrix} \begin{pmatrix}
            \zeta_{ji}(\vb{y}) \\ \omega_{ji}(\vb{y})
        \end{pmatrix},
\end{equation}
where the matrix of kernels in momentum space is (recall that we are setting $\sigma_0 = 0$),
\begin{equation}
    \hat{\Pi}_\omega (\vb{p}) = \begin{pmatrix}
            \eta_1^2 & \eta_2^2 \\
            \eta_2^2 & \eta_1^2
        \end{pmatrix} - \begin{pmatrix}
            \eta_1^4 + \eta_2^4 & 2\eta_1^2\eta_2^2 \\
            2\eta_1^2\eta_2^2 & \eta_1^4 + \eta_2^4
        \end{pmatrix} \int \frac{\dd^2 \vb{q}}{(2\pi)^2} \frac{1}{(\vb{q}^2 + m^2)[(\vb{p} + \vb{q})^2 + m^2]} ,
\end{equation}
and there is no dependence on $n$ when $\omega_0 = 0$. This yields a contribution to the three-point function
\begin{align}
    \Gamma^{12}_{i,jk}(\vb{p}_1,\vb{p}_2,-\vb{p}_1 - \vb{p}_2) &= \Gamma^{21}_{i,jk}(\vb{p}_1,\vb{p}_2,-\vb{p}_1 - \vb{p}_2) \label{eq:3pointIntegral}\\
    &\kern-9em= \frac{1}{N} \int \frac{\dd^2 \vb{k}}{(2\pi)^2} \frac{(1 - \delta_{ij}) \delta_{jk}}{[(\vb{k} + \vb{p_1} + \vb{p}_2)^2 + m^2][(\vb{k} + \vb{p}_2)^2 + m^2]} \frac{(\eta_1^4 + \eta_2^4) \Pi^{(2)}_\omega(\vb{k}) - 2 \eta_1^2 \eta_2^2 \Pi^{(1)}_\omega(\vb{k})}{[\Pi^{(1)}_\omega(\vb{k})]^2 - [\Pi^{(2)}_\omega(\vb{k})]^2}.  \nn 
\end{align}
Next, observe that the leading $1/N$ contribution in an expansion of Eq. \eqref{eq:kernelgeneralform} comes from the two ways of inserting a single copy of the replica-mixing contribution above,
\begin{equation}
    \Pi^{(2)}_{\sigma,ij}(\vb{p}) \simeq -2 \int \frac{\dd^2 \vb{q}}{(2\pi)^2}  \frac{\Gamma^{12}_{i,jj}(\vb{p},-\vb{p}-\vb{q},\vb{q})}{(\vb{q}^2 + m^2)[(\vb{p} + \vb{q})^2 + m^2]}.
\end{equation}
As noted in the previous section, when deriving the double-Lorentzian propagator we are primarily interested in $\Pi^{(2)}_{\sigma,ij}(\vb{p}=0)$, and since $\Gamma^{12}_{i,jj}(0,-\vb{q},\vb{q}) \rightarrow 0$ as $\abs{\vb{q}} \rightarrow \infty$, the kernel is largely determined by the value of the vertex function at $\vb{q} = 0$, which we find to be
\begin{figure}[!b]
    \centering
    \includegraphics[scale=0.5]{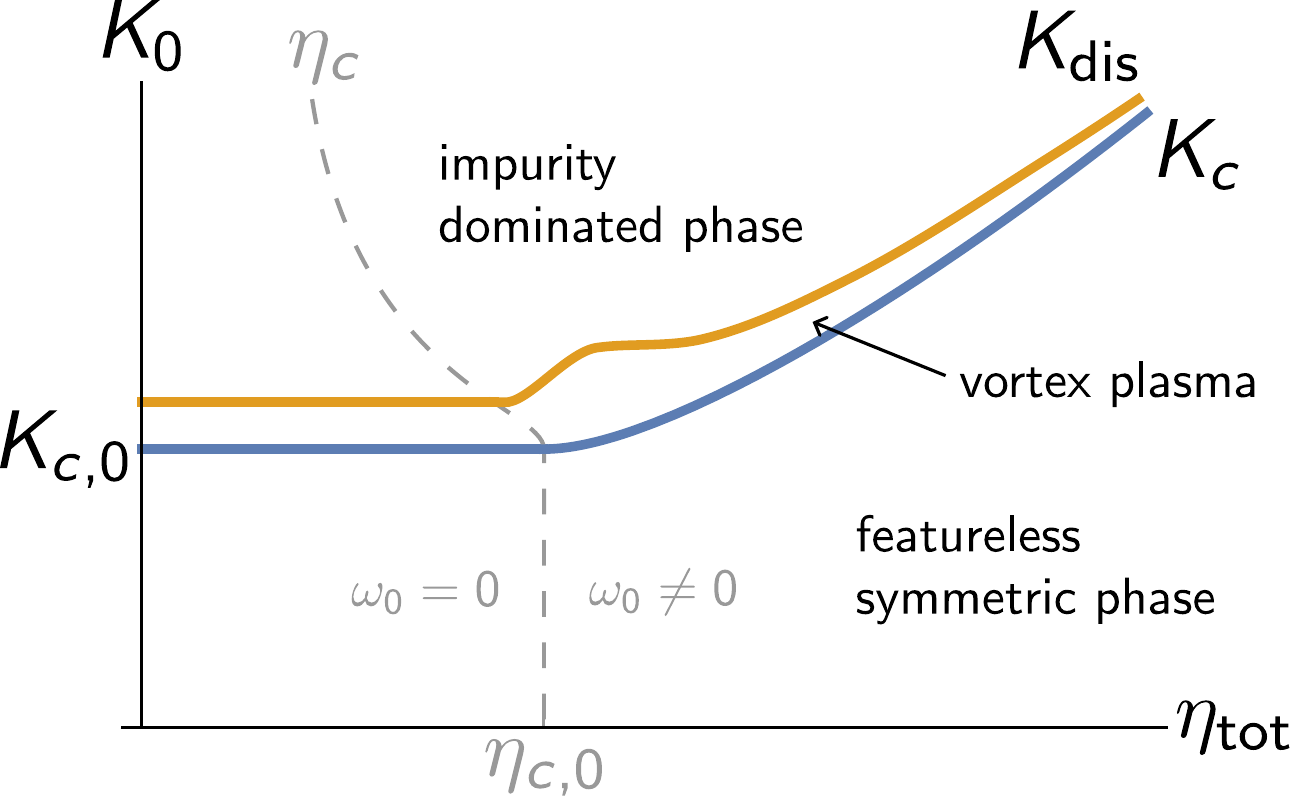}
    \caption{Schematic phase diagram of the theory with quenched disorder, including the effects of fluctuations. For $K_0 < K_c$ there is a featureless symmetric phase ($\sigma_0 = 0$) with double-Lorentzian correlations Eq. \eqref{eq:doubleLorentzian}. For $K_c < K_0 < K_{\mathrm{dis}}$ the ground state is a BKT-type vortex plasma ($\sigma_0 \neq 0$); disorder is RG irrelevant; this region is narrow when $N$ is large. For $K_0 > K_{\mathrm{dis}}$ disorder is RG relevant and the (unknown) ground state is dominated by impurities. The light gray dashed line represents the mean-field disorder-driven transition, which does not affect the nature of the ground state.}
    \label{fig:phase_diagram_final}
\end{figure}
\begin{align}
    \Gamma^{12}_{i,jj}(0,0,0) &\simeq \frac{1}{N}\frac{(1 - \delta_{ij}) 3\pi m^2 (\Delta - 1) \eta^2_{\mathrm{tot}}}{[ 24 \pi^2 m^4 - 7\pi m^2(1 + \Delta) \eta^2_{\mathrm{tot}} + 2 \Delta \eta^4_{\mathrm{tot}}]^2 } \left[  24 \pi^2 m^4 - 7 \pi m^2 (1 + \Delta) \eta^2_{\mathrm{tot}}  \vphantom{\ln\left( \frac{2\pi m^2 (1 + \Delta) \eta^2_{\mathrm{tot}} - \Delta \eta^4_{\mathrm{tot}}}{3(4\pi m^2 - \eta^2_{\mathrm{tot}})(4\pi m^2 - \Delta \eta^2_{\mathrm{tot}})} \right)} \right. \nn \\
    & + \left. 2 \Delta \eta^4_{\mathrm{tot}}+ \frac{1}{2} (2 \pi m^2(1 + \Delta)\eta^2_{\mathrm{tot}} - \Delta \eta^4_{\mathrm{tot}}) \ln\left( \frac{2\pi m^2 (1 + \Delta) \eta^2_{\mathrm{tot}} - \Delta \eta^4_{\mathrm{tot}}}{3(4\pi m^2 - \eta^2_{\mathrm{tot}})(4\pi m^2 - \Delta \eta^2_{\mathrm{tot}})} \right) \right].
\end{align}
This expression was obtained by using a Pad\'e approximant of order (0,2) for the $(\zeta,\omega)$ propagator in Eq. \eqref{eq:3pointIntegral}. This result has two important features: i) It vanishes for $\Delta=1$, that is, $\eta_2 = 0$. Therefore, some symmetry breaking disorder is necessary to generate replica-mixing interactions. ii) It diverges at the phase boundary $\eta^2_{c,0} = 4\pi m^2$ (the dependence of $m^2$ on model parameters will be shifted to order $1/N$ even if the phase boundary equation appears unchanged in this form). This divergence originates from the infrared divergence in Eq. \eqref{eq:3pointIntegral} due to gapless $\zeta_{ij}$ and $\omega_{ij}$ fluctuations. However, a similar IR divergence must occur in the calculation of the replica-diagonal kernel $\Pi^{(1)}_\sigma(\vb{p})$. Therefore, the effective disorder strength will not be infinite.

Finally, the result of this section shows that the order parameter correlation function will still have a double-Lorentzian form in the weak disorder regime $\omega_0 = 0$,
\begin{equation}
    \overline{\langle \sigma(\vb{p}) \sigma^*(-\vb{p}) \rangle} \simeq \frac{1}{\vb{p}^2 + m^{\prime\,2}_\sigma} + \frac{\eta_\sigma^{\prime\,2}}{(\vb{p}^2 + m^{\prime\,2}_\sigma)^2}, \label{eq:doubleLorentzianWeak}
\end{equation}
albeit with modified parameters $m^{\prime\,2}_\sigma$ and $\eta_\sigma^{\prime\,2}$. It is also clear that this analysis carries over to the $\sigma_0 \neq 0$ regime. Therefore, whether $\omega_0$ is zero or not does not fundamentally alter the nature of the ground state, but simply determines whether the effect of disorder is suppressed by an additional factor of $1/N$. All of the previous analysis is summarized in the phase diagram shown in Fig. \ref{fig:phase_diagram_final}. In closing this section we re-emphasize that the presence of the double-Lorentzian term in the disorder-averaged two-point correlator of Eq. \eqref{eq:doubleLorentzianWeak} implies a stronger infrared singularity in the averaged susceptibility. In turn, this behavior implies the absence of long-range order below four dimensions even in the weak disorder regime, as expected from the Imry-Ma argument \cite{Imry1975}.

\section{Discussion \label{sec:disc}}

In this paper we have introduced  a new model, solvable in the large-$N$ limit, to understand the interplay of thermal fluctuations and quenched random disorder in two-dimensional systems with a phase transition in the clean limit. While we were motivated by the existence of charge-ordered states in the cuprate high-$T_c$ superconductors, the formalism we have developed is completely general and can, in principle, be applied to any system described by a $U(1)$ order parameter coupled to random field disorder. Previous studies \cite{Nie2014} using the large-$N$ technique did not capture the physics of the Berezinskii-Kosterlitz-Thouless phase transition because the order parameter was encoded on a manifold with dimension that scaled with $N$, which, for any $N>2$, does not have a phase transition in 2D in the clean limit. In contrast, we held the $U(1)$ manifold fixed as a subgroup of the larger symmetry group $U(N)\times U(N)$. Naively, this produced an inconsistency with the Mermin-Wagner theorem \cite{Mermin1966} in the absence of disorder, and the Imry-Ma theorem \cite{Imry1975} in the presence of random field disorder. Throughout this paper, we have shown that including fluctuations about the $N=\infty$ ground state resolves any apparent contradictions. 

Our first main result demonstrated how these fluctuations produce a BKT transition split from the mean-field transition in the clean theory at order $1/N$; the vortex plasma phase is narrow when $N$ is large. Next, we derived the $N=\infty$ phase diagram of the model as a function of the disorder strength and the coupling between the two $\mathbb{C}P^N$ components. This revealed a novel disorder-driven transition between a weakly- and a strongly-disordered regime. We then derived an explicit expression for the order parameter correlation function in the strongly-disordered symmetric phase of the theory, finding agreement with the well-known double-Lorentzian distribution \cite{Pytte1981,Yoshizawa1982}. In the strongly-disordered phase with a $U(1)$ condensate, we derived an infrared effective theory which had the form of a random field $XY$ model. By mapping our expression to known results \cite{Houghton1981,Goldschmidt1982}, we were able to derive the phase boundary between a vortex-dominated BKT-like phase and the contentious \cite{Giamarchi1994,LeDoussal2000,Zeng1999} impurity-dominated phase. 

Finally, we proved that fluctuations are sufficient for ensuring agreement with the Imry-Ma condition in the weakly-disordered regime by showing that the order parameter correlation function again has the double-Lorentzian form. We note that non-analytic contributions that are non-perturbative in the $1/N$ expansion are not included in our analysis. These include rare configurations of the disorder \cite{Griffiths1969}, which may also play a role, for example, by rounding the disorder-driven transition into a broad crossover.

In this paper, we have considered a purely classical theory. This suitably describes static correlations in materials when quantum effects are weak, such as when electron-electron interactions are strong enough to form an insulating CDW state. A complete theory of dynamic correlations in randomly pinned ICDWs necessarily requires the inclusion of quantum fluctuations. Luckily, the approach used in this work lends itself well to such a generalization. Investigating the $N=\infty$ ground state properties at $T=0$ will require little more than changing the dimension from $d=2$ to $d=3$ in all of the calculations presented, due to the usual quantum-classical correspondence. However, being quenched, disorder also introduces significant temporal non-locality into the system, which will lead to richer behavior of dynamic order parameter correlations. It would also be interesting to study conducting systems, in which case the CDW order parameter will experience Landau damping \cite{Sun2008}. This would provide a unified theoretical picture of the role of disorder in a wide range of physical systems, and help answer many of the questions raised by modern experiments.

\begin{acknowledgments}
MCO thanks J. Gliozzi for useful discussions. This work was supported in part by the US National Science Foundation through the grants DMR 1725401 and DMR 2225920 at the University of Illinois.
\end{acknowledgments}

\renewcommand{\theequation}{\Alph{section}.\arabic{equation}}

\appendix

\addcontentsline{toc}{section}{Appendix \ref*{app:vorticity}: Response of Two-Component \texorpdfstring{$\mathbb{C}P^N$}{CPN} Model without Disorder to a Background \texorpdfstring{$U(1)$}{U(1)} Gauge Field}
\tocless{\section}{Response of Two-Component \texorpdfstring{$\mathbb{C}P^N$}{CPN} Model without Disorder to a Background \texorpdfstring{$U(1)$}{U(1)} Gauge Field\label{app:vorticity}}

In this appendix, we sketch the calculation of the response of the two-component $\mathbb{C}P^N$ model to a $U(1)$ background gauge field.
We start from the effective action with a background field $A^\mu$ minimally coupled to the (previously global) relative $U(1)$ symmetry
\begin{equation}
    S_{\mathrm{eff}}[A]/N = \tr \ln \begin{pmatrix} -D_\mu^2[a+A] + \lambda_1 & -\rho e^{i\theta} \\ -\rho e^{-i\theta} & -D_\mu^2[a-A] + \lambda_2 \end{pmatrix} + \int \dd^d \vb{x} \left[ \frac{\rho^2}{K_0} - \frac{\lambda_1 + \lambda_2}{g_0} \right].
\end{equation}
The leading order behavior of the sector corresponding to the relative $U(1)$ symmetry involves only $\theta$ and $A^\mu$. All other couplings are either forbidden by symmetry or sub-leading in $1/N$. Expanding the $\tr\ln$ to quadratic order in these fields yields the following two-point kernels:
\begin{subequations}
\begin{align}
    \Pi_{\theta\theta}(\vb{p}) &= 2N \rho_0^2 \int \frac{\dd^d \vb{q}}{(2\pi)^d} \frac{(\vb{q}^2 + m^2)[\vb{q}^2 - (\vb{q} + \vb{p})^2]}{[(\vb{q}^2 + m^2)^2 - \rho_0^2]([(\vb{q} + \vb{p})^2 + m^2]^2 - \rho_0^2)} , \\
    \Pi_{A\theta}^\mu(\vb{p}) &= -2 i N \rho_0^2 \int \frac{\dd^d \vb{q}}{(2\pi)^d} \frac{(2q^\mu + p^\mu)[\vb{q}^2 - (\vb{q} + \vb{p})^2]}{[(\vb{q}^2 + m^2)^2 - \rho_0^2]([(\vb{q} + \vb{p})^2 + m^2]^2 - \rho_0^2)} ,\\
    \Pi_{AA}^{\mu\nu}(\vb{p}) &= 2N \int \frac{\dd^d \vb{q}}{(2\pi)^d} \frac{(2q^\mu + p^\mu)(2q^\nu + p^\nu)[\rho_0^2 - (\vb{q}^2 + m^2)[(\vb{q} + \vb{p})^2 + m^2]^2]}{[(\vb{q}^2 + m^2)^2 - \rho_0^2]([(\vb{q} + \vb{p})^2 + m^2]^2 - \rho_0^2)} \nn \\
    &\kern5em + 4N \delta^{\mu\nu} \int \frac{\dd^d \vb{q}}{(2\pi)^d} \frac{\vb{q}^2 + m^2}{(\vb{q}^2 + m^2)^2 - \rho_0^2} \nn \\
    &= \Pi_{a a}^{\mu\nu}(\vb{p}) + 4N \rho_0^2 \int \frac{\dd^d \vb{q}}{(2\pi)^d} \frac{(2q^\mu + p^\mu)(2q^\nu + p^\nu)}{[(\vb{q}^2 + m^2)^2 - \rho_0^2]([(\vb{q} + \vb{p})^2 + m^2]^2 - \rho_0^2)} ,
\end{align}
\end{subequations}
where $\Pi_{a a}^{\mu\nu}(\vb{p})$ is (two times) the well-known electrodynamic response of the $\mathbb{C}P^N$ model \cite{Witten1979,Coleman-1985}, which in $d = 2$ is
\begin{equation}
    \Pi_{a a}^{\mu\nu}(\vb{p}) =  \big( \delta^{\mu\nu} \vb{p}^2 - p^\mu p^\nu \big) \left[ \frac{N m^2}{6\pi(m^4 - \rho_0^2)} + \mathcal{O}(\vb{p}^2) \right].
\end{equation}
Since $A^\mu$ couples to matter ($\theta$), the kernel $\Pi^{\mu\nu}_{AA}$ will have both transverse $\Pi^{\mu\nu}_{T}$ and longitudinal $\Pi^{\mu\nu}_{L}$ responses. One finds that
\begin{equation}
    \Pi_{\theta\theta}(\vb{p}) = \vb{p}^2 \Lambda(\vb{p}^2), \kern3em \Pi_{A\theta}^\mu(\vb{p}) = -2i p^\mu \Lambda(\vb{p}^2), \kern3em \Pi^{\mu\nu}_{L}(\vb{p}) = 4 \delta^{\mu\nu} \Lambda(\vb{p}^2),
\end{equation}
where the kernel $\Lambda(\vb{p}^2)$ can be evaluated exactly in $d=2$,
\begin{align}
    \Lambda(\vb{p}^2) &= \frac{N\rho_0^2}{2\pi \vb{p}^2} \left[ \tanh^{-1}\left(\frac{\rho_0}{m^2}\right) \right. \nn \\
    &\kern5em - \left. \frac{2\rho_0}{\sqrt{(\vb{p}^2)^2 + 4(m^2 \vb{p}^2 + \rho_0^2)}} \tanh^{-1}\left(\frac{\sqrt{(\vb{p}^2)^2 + 4(m^2 \vb{p}^2 + \rho_0^2)}}{\vb{p}^2 + 2m^2}\right) \right] \nn \\
    &=\gamma_\theta + \mathcal{O}(\vb{p}^2),
\end{align}
where $\gamma_\theta$ is the phase stiffness
\begin{equation}
    \gamma_\theta = \frac{N}{4\pi \rho_0} \left[m^2 \tanh^{-1}\left(\frac{\rho_0}{m^2}\right) - \rho_0\right] .
\end{equation}
The transverse vorticity response is
\begin{equation}
    \Pi^{\mu\nu}_{T} = \big( \delta^{\mu\nu} \vb{p}^2 - p^\mu p^\nu \big) \left[ \frac{2 m^2 \gamma_\theta}{\rho_0^2} + \mathcal{O}(\vb{p}^2) \right].
\end{equation}
Note that since $\gamma_\theta \propto \rho_0^2$ in the vicinity of the critical point $K_c$, the transverse response does not vanish. Instead, it simply becomes equal to $\Pi_{a a}^{\mu\nu}(\vb{p})$ for $K_0 \leq K_c$. Therefore, in the long-wavelength limit, we have
\begin{equation}
    S_{\mathrm{eff}}[A] \simeq \frac{\gamma_\theta}{2} \int \dd^2 \vb{x}\, \big( \del^\mu \theta + 2 A^\mu \big)^2 + \frac{1}{4 e^2} \int \dd^2 \vb{x}\, (F^{\mu\nu})^2 ,
\end{equation}
where $e^2 = \rho_0^2/(2m^2 \gamma_\theta)$ is the effective coupling constant of the probe field and $F^{\mu\nu} = \del^\mu A^\nu - \del^\nu A^\mu$ is the Maxwell tensor for $A^\mu$.

In the above discussion we ignored the fact that the phase field is actually defined mod $2\pi$ and that the full computation of the partition function is dominated by the contributions of vortices and anti-vortices. Such contributions can be computed by regarding the local flux of the gauge field $A^\mu$ as representing vortices and anti-vortices (for a recent discussion see Ref. \cite{Fradkin2023}). It is well understood that the leading contributions to the partition function come from dilute configurations of  vortices and anti-vortices, and that in terms of the Cauchy-Riemann dual $\vartheta$ of the phase field, which satisfies $\del^\mu \vartheta = \varepsilon^{\mu\nu} \del^\nu \theta$, the effective action is mapped to the sine-Gordon theory \cite{Wiegmann1978,Kadanoff1978,Amit1980}. In the present case, 
\begin{equation}
    S_{\mathrm{eff}} = \int \dd^2 \vb{x} \left[ \frac{1}{2} (\del^\mu \vartheta)^2 - v_0 \cos(\beta_0 \vartheta) \right], \label{eq:sineGordon}
\end{equation}
where $\beta_0^2 = (4\pi)^2 \gamma_\theta $, $v_0 = 2 e^{-\mathcal{E}_{\mathrm{core}}}/a^2$, $\mathcal{E}_{\mathrm{core}}$ is the core energy of a vortex, $a \sim \mu^{-1}$ is an ultraviolet cutoff, and $\vartheta$ has been rescaled by $2\sqrt{\gamma_\theta}$ to bring the action into canonical form.

\addcontentsline{toc}{section}{Appendix \ref*{app:ON}: Review of  \texorpdfstring{$O(N)$}{O(N)} Model with a Quenched Random Field}
\tocless{\section}{Review of  \texorpdfstring{$O(N)$}{O(N)} Model with a Quenched Random Field\label{app:ON}}

In this appendix, we review the large-$N$ analysis of $O(N)$ models in a quenched random field for the purpose of aiding comparison with the main results in this paper. The $O(N)$ nonlinear sigma model (NLSM) is described by the action
\begin{equation}
    S = \int \dd^d \vb{x} \left[ \frac{1}{2g} (\del_\mu \vb*{n})^2 - \vb*{h} \cdot \vb*{n}\right],\kern3em \vb*{n}^2(x) = 1,
\end{equation}
where $\vb*{h}(\vb{x})$ is a fixed external source. Suppose now that $\vb*{h}(\vb{x}) = \mathfrak{h}(\vb{x})$ is a random field drawn locally from a Gaussian distribution such that
\begin{equation}
    \overline{\mathfrak{h}^a(\vb{x})} = 0, \kern3em \overline{\mathfrak{h}^a(\vb{x}) \mathfrak{h}^b(\vb{y})} = \eta^2 \delta_{ab} \delta^{(d)}(\vb{x} - \vb{y}),
\end{equation}
where the overline denotes an average over the disorder configurations, and the variance $\eta^2$ represents the strength of the disorder. Using the replica trick formalism, we consider, for $n\in \mathbb{Z}$,
\begin{align}
    \overline{\mathcal{Z}^n} &= \int \mathcal{D} \mathfrak{h}\, \exp\left(- \int \dd^d \vb{x} \frac{\mathfrak{h}^2}{2\eta^2} \right) \mathcal{Z}[\mathfrak{h}]^n \nn \\
    &= \int \prod_{j=1}^n \mathcal{D} \vb*{n}_j \mathcal{D} \lambda_j \, \exp\left(- \sum_{i,j=1}^n \int \dd^d \vb{x} \frac{1}{2g} \Big[ (\del_\mu \vb*{n}_j)^2 + \lambda_j (\vb*{n}^2 - 1) \Big] \delta_{ij}  - \frac{\eta^2}{2} \vb*{n}_i \cdot \vb*{n}_j \right) , \label{eq:disavgON}
\end{align}
where $\vb*{n}_j$ and $\lambda_j$ (the Lagrange multiplier imposing the unit vector constraint), for $j=1,\dots,n$ are the replicated fields corresponding to each factor of $\mathcal{Z}[\frakh]$. Because the theory remains quadratic in the replicated $O(N)$ fields $\vb*{n}_j$, they can be integrated out exactly:
\begin{align}
    S_{\mathrm{eff}} &= \frac{N}{2} \Tr\ln\left[ -\del^2 \hat{I} + \mathrm{diag}(\lambda) - g \eta^2 \hat{M}  \right] - \sum_{j=1}^n \int \dd^d \vb{x} \frac{\lambda_j}{2g}  ,
\end{align}
where $\Tr(\,\cdot\,)$ denotes the functional operator trace as well as the trace over replica indices, and $\hat{I}$ is the $n\times n$ identity matrix, $\hat{M}$ is the matrix with a $1$ in every entry and $\mathrm{diag}(\lambda) = \mathrm{diag}(\lambda_1,\dots,\lambda_n)$ is the diagonal matrix of Lagrange multipliers. The remaining functional integrals over each $\lambda_j$ can be performed using steepest descent, which becomes exact in the limit $N\rightarrow \infty$. To make this limit precise, we define $g = g_0/N$ and $\eta^2 = N \eta_0^2$, keeping $g_0$ and $\eta_0$ fixed as $N \rightarrow \infty$. We then look for a replica-symmetric saddle point where $\lambda_1 = \dots = \lambda_n = m^2$, which yields the self-consistent equation
\begin{equation}
     \tr \int \frac{\dd^d \vb{p}}{(2\pi)^d} \frac{1}{(\vb{p}^2 + m^2) \hat{I} - g_0 \eta_0^2 \hat{M}} = \frac{n}{g_0},
\end{equation}
where $\tr(\,\cdot\,)$ denotes a trace over only the replica indices. Since $\hat{M}$ is an idempotent matrix ($\hat{M}^2 = n \hat{M}$), finding the inverse matrix in the integrand is straightforward:
\begin{equation}
    \frac{1}{n} \tr \int \frac{\dd^d \vb{p}}{(2\pi)^d} \left[  \frac{\hat{I}}{\vb{p}^2 + m^2} + \frac{g_0 \eta_0^2 \hat{M}}{(\vb{p}^2 + m^2)(\vb{p}^2 + m^2 - n g_0 \eta_0^2)} \right] = \frac{1}{g_0},
\end{equation}
and hence, taking the replica limit $n \rightarrow 0$,
\begin{equation}
    \int \frac{\dd^d \vb{p}}{(2\pi)^d} \left[  \frac{1}{\vb{p}^2 + m^2} + \frac{g_0 \eta_0^2}{(\vb{p}^2 + m^2)^2} \right] = \frac{1}{g_0} .
\end{equation}
While explicitly solving this equation for $m^2$ requires renormalization of the coupling constant $g_0$, it will suffice for our purposes to simply make the following remarks: (i) For any amount of disorder ($\eta_0^2 > 0$) in $d\leq 4$ there is an infrared singularity in the above integral equation unless $m^2 > 0$. This implies the absence of any broken symmetry phase, providing an exact, non-perturbative realization of the Imry-Ma argument \cite{Imry1975}. (ii) The disorder-averaged propagator of the $\vb*{n}$ field has the double-Lorentzian form
\begin{equation}
    \overline{G(\vb{p})} = \frac{1}{\vb{p}^2 + m^2} + \frac{g_0 \eta_0^2}{(\vb{p}^2 + m^2)^2}.
\end{equation}

\addcontentsline{toc}{section}{Appendix \ref*{app:CPN}: Review of \texorpdfstring{$\mathbb{C}P^N$}{CPN} Model with a Quenched Random Field}
\tocless{\section}{\texorpdfstring{Review of $\mathbb{C}P^N$}{CPN} Model with a Quenched Random Field \label{app:CPN}}

In the main body of this work, we investigate a two-component generalization of the $\mathbb{C}P^N$ model. Here, we present a summary of the physics of the simpler model in a quenched random field for pedagogical purposes.

In the $C\mathbb{P}^N$ model, quenched disorder can only couple to gauge invariant combinations of the scalar field. The natural coupling originates in the Hopf map from $O(3)$ unit vectors to elements of $\mathbb{C}P^1$, $\vb*{h} \cdot \vb*{n} \rightarrow h^a z^*_\alpha \tau^a_{\alpha\beta} z_\beta$, where $\tau^a$ are the generators of $SU(2)$. Therefore, we consider the partition function
\begin{equation}
    \mathZ[\mathfrak{h}] = \int \mathD \lambda \mathD a^\mu  \mathcal{D} \vb*{z}\, \exp\left(- \frac{1}{g} \int \dd^d \vb{x} \left[ \abs{D^\mu[a]\vb*{z}}^2 + \lambda(\abs{\vb*{z}}^2 - 1) \right] + \int \dd^d \vb{x}\, \mathfrak{h}^a z^*_\alpha \tau^a_{\alpha\beta} z_\beta \right),
\end{equation}
where $\tau^a$ are the generators of $SU(N)$, $\frakh^a$ is a real $(N^2-1)$-component random field transforming under the adjoint (vector) representation of $SU(N)$ and drawn from the locally Gaussian distribution
\begin{equation}
    \overline{\frakh^a(\vb{x})} = 0, \kern3em \overline{\frakh^a(\vb{x}) \frakh^b(\vb{y})} = \eta^2 \delta^{ab} \delta^{(d)}(\vb{x} - \vb{y}),
\end{equation}
$\eta^2$ is the variance (strength) of the disorder and overlines denote averaging with respect to disorder configurations. For quenched disorder averages, we use the replica trick formalism,
\begin{align}
    \overline{\mathZ^n} &= \int \mathD \frakh \, \exp\left(- \int \dd^d \vb{x} \frac{\frakh^2}{2\eta^2} \right) \mathZ[\frakh]^n \nn\\
\begin{split}
    &= \int \mathcal{D} \lambda_j \mathcal{D} a^\mu_j \mathcal{D} \vb*{z}_j \, \exp\left(- \sum_{i,j = 1}^n \int \dd^d \vb{x} \frac{1}{g}\left[  \abs{D^\mu[a_j] \vb*{z}_j}^2 + \lambda_j \left( \abs{\vb*{z}_j}^2 - 1 \right) \right] \delta_{ij}\right. \\
    &\kern23em- \left. \frac{N \eta^2}{2} \vert \vb*{z}^*_i \cdot \vb*{z}_j \vert^2 \right) ,
\end{split}
\end{align}
where we have used the identity $\tau^a_{\alpha\beta} \tau^a_{\gamma \delta} =  N \delta_{\alpha \delta} \delta_{\beta\gamma} - \delta_{\alpha\beta} \delta_{\gamma\delta}$ [implied summation over repeated indices; working in the convention $\tr( \tau^a \tau^b ) = N \delta^{ab}$], and dropped the unimportant constant terms proportional to $\abs{\vb*{z}_i}\abs{\vb*{z}_j} = 1$. The quartic interaction between the replicated $\vb*{z}_j$ fields can then be decoupled using a Hermitian Hubbard-Stratonovich field $\omega_{ij} = \omega^*_{ji}$; since $\vb*{z}$ is a unit vector, we also have $\omega_{ii} = 0$. Therefore, 
\begin{align}
\begin{split}
    \overline{\mathcal{Z}^n} = \int \mathcal{D} \omega_{ij} \mathcal{D} \lambda_j \mathcal{D} a^\mu_j  \mathcal{D} \vb*{z}_j \, \exp\left(- \sum_{i,j = 1}^n \int \dd^d \vb{x} \frac{1}{g} \left[  \abs{D^\mu[a_j] \vb*{z}_j}^2 \delta_{ij} + \lambda_j \left( \abs{\vb*{z}_j}^2 - 1\right) \delta_{ij} \right. \right.\\
    - \left. \left. g \omega_{ij} \vb*{z}^*_{i} \cdot \vb*{z}_j \right] - \frac{1}{2 N \eta^2} \sum_{i,j = 1}^n \int \dd^d \vb{x} \, \omega_{ij}^* \omega_{ij} \right) .
\end{split}
\end{align}
The field $\omega_{ij}$ has a simple physical interpretation by comparison with the $O(N)$ nonlinear sigma model with quenched disorder, Eq. \eqref{eq:disavgON}. Evidently the amplitude of $\omega_{ij}$ plays the role of an effective disorder strength for the $SU(N)$ scalars $\vb*{z}$. The crucial difference is that $\omega_{ij}$ must be allowed to fluctuate to ensure gauge invariance is respected. After rescaling $\omega_{ij} \rightarrow \omega_{ij}/g$, defining $g = g_0/N$ and $\eta = \eta_0/g_0$, and integrating out the $\vb*{z}_j$, one obtains an effective action
\begin{equation}
    S_{\mathrm{eff}}/N = \Tr \ln \left[ - \mathrm{diag}(D_\mu^2[a]) + \mathrm{diag}(\lambda) - \hat{\omega} \right] + \tr \int \dd^d \vb{x} \left[ \frac{\hat{\omega}^2}{2\eta_0^2} - \frac{\mathrm{diag}(\lambda)}{g_0} \right],
\end{equation}
where $\mathrm{diag}(\,\cdot\,)$ denotes a matrix which is diagonal in replica indices, $\hat{\omega}$ is the matrix with elements $\omega_{ij}$, and $\Tr$ includes a trace over functional configurations and replica indices.

Before proceeding, we briefly comment on the symmetries of the replicated theory. The original non-replicated theory has a local $U(1)$ gauge symmetry with the transformation rules
\begin{equation}
    \vb*{z}(\vb{x}) \longrightarrow e^{i\theta(\vb{x})} \vb*{z}(\vb{x}), \kern3em a^\mu(\vb{x}) \longrightarrow a^\mu(\vb{x}) - \del^\mu \theta(\vb{x}).
\end{equation}
The replicated theory has an enlarged $U(1)^n$ gauge symmetry which allows for independent gauge transformations on each replicated field
\begin{equation}
    \vb*{z}_j(\vb{x}) \longrightarrow e^{i\theta_j(\vb{x})} \vb*{z}_j(\vb{x}), \kern3em a^\mu_j(\vb{x}) \longrightarrow a^\mu_j(\vb{x}) - \del^\mu \theta_j(\vb{x}).
\end{equation}
As a consequence, the disorder Hubbard-Stratonovich field inherits the transformation
\begin{equation}
    \omega_{jk}(\vb{x}) \longrightarrow e^{i(\theta_j(\vb{x}) - \theta_k(\vb{x}))} \omega_{jk}(\vb{x}),
\end{equation}
transforming as a tensor under $U(1)^n$. Since a replica-symmetric ground state should be invariant under the permutation group $S_n$, we might have assumed this would uniquely constrain $\omega_{ij} = \omega_0$, independent of $i,j$. However, the manifold of gauge-equivalent ground states is $U(1)^n/S_n$; i.e., $\omega_{ij} = \omega_0$ is simply a particular representative of the equivalence class of configurations obtained from $\omega_0$ by gauge transformations.

Given the above considerations, we are free to expand around the replica-symmetric saddle point $\omega_{ij}(\vb{x}) = \omega_0$. Also taking the replica-symmetric ansatz $\lambda_j(\vb{x}) = m^2$, and fixing $a^\mu_j(\vb{x}) = 0$, the replicated effective potential in the $N = \infty$ limit is
\begin{equation}
    \overline{U_{\mathrm{eff}}^n} = \int \frac{\dd^d \vb{q}}{(2\pi)^d} \ln \det\left[ (\vb{q}^2 + m^2 + \omega_0) \hat{I} - \omega_0 \hat{M} \right] + n(n-1) \frac{\omega_0^2}{2\eta_0^2} - n \frac{m^2}{g_0},
\end{equation}
where $\hat{I}$ is the $n\times n$ identity matrix and $\hat{M}$ is the matrix with a $1$ in every entry. Using the same coupling constant renormalization as in Eq. \eqref{eq:gRenorm} to cure the UV divergence, we recover the physical (renormalized) effective potential by taking the replica limit
\begin{align}
    \overline{U_R} &= \lim_{n\rightarrow 0} \frac{\overline{U_R^n}}{n} \nn \\
    &= \int \frac{\dd^d \vb{q}}{(2\pi)^d}  \left[ \ln \left(1 + \frac{m^2 + \omega_0}{\vb{q}^2} \right) - \frac{\omega_0}{\vb{q}^2 + m^2 + \omega_0} - \frac{m^2}{\vb{q}^2 + \mu^2} \right] - \frac{\omega_0^2}{2\eta_0^2} - \frac{m^2}{g_R} . \label{eq:cpndirtypot}
\end{align}
In $d=2$ this evaluates to
\begin{equation}
    \overline{U_R} = \frac{m^2}{4\pi} \left[1 - \ln \left( \frac{m^2 + \omega_0}{\mu^2}\right) \right] + \frac{\omega_0}{4\pi} - \frac{\omega_0^2}{2\eta_0^2} - \frac{m^2}{g_R}.
\end{equation}
The values of $m^2$ and $\omega_0$ are then obtained from the saddle point equations, though a subtlety occurs for $\omega_0$, as the corresponding equation has two solutions. Note that since $\omega_{ij}$ has $n(n-1)$ complex degrees of freedom, in the replica limit $n\rightarrow0$, the negative number of effective degrees of freedom require us to \textit{maximize} the effective potential \cite{Parisi-1982}. This yields a crossover scale $\eta_c^2 = 4\pi \mu^2 e^{-4\pi/g_R}$ such that 
\begin{equation}
    m^2 + \omega_0 = \begin{cases}
        \eta_{c,0}^2/4\pi, \kern1em & \eta_0 < \eta_{c,0}, \\
        \eta_0^2/4\pi, \kern1em & \eta_0 \geq \eta_{c,0} .
    \end{cases}
    \kern3em
    \omega_0 = \begin{cases}
        0, & \eta_0 < \eta_{c,0}, \\
        \dfrac{\eta_0^2}{2\pi} \ln\left( \dfrac{\eta_0}{\eta_{c,0}} \right), \kern1em & \eta_0 \geq \eta_{c,0} ,
    \end{cases}
\end{equation}
We emphasize that this crossover does not break any ``physical'' symmetries; i.e., those of the non-replicated theory. It is also likely that this apparent mean-field transition would be rounded by [possibly non-perturbative $\mathcal{O}(e^{-N})$] corrections to the saddle point equations. Note that this behavior is not an artifact of the absence of a phase transition in the clean $d=2$ theory; consider the case of $d=3$, where Eq. \eqref{eq:cpndirtypot} evaluates to
\begin{equation}
    \overline{U_R} = \frac{1}{12\pi} \left[ (\omega_0 - 2m^2) \sqrt{m^2 + \omega_0}  + 3\mu m^2 \right] - \frac{\omega_0^2}{2\eta_0^2} - \frac{m^2}{g_R} .
\end{equation}
It is simple to check that the ``clean'' symmetry breaking state $m^2 = \omega_0 = 0$ is also never energetically favorable.

For the purposes of this paper, it suffices to note that the $1/N$ expansion reveals that the Higgs mechanism takes place for $\eta_0 > \eta_c$, with $n-1$ of the replicated gauge fields $a^\mu_j$ becoming massive; as expected from the form of the disorder, only the replica-diagonal $U(1)$ subgroup is unaffected, with the corresponding gauge field remaining gapless.

\addcontentsline{toc}{section}{Appendix \ref*{app:fluctuations}: Fluctuation Kernels in Disordered Two-Component \texorpdfstring{$\mathbb{C}P^N$}{CPN} Model}
\tocless{\section}{Fluctuation Kernels in Disordered Two-Component \texorpdfstring{$\mathbb{C}P^N$}{CPN} Model \label{app:fluctuations}}

In this appendix, we present the details of calculations referred to in Sec. \ref{sec:strongDisorder}. We start from the effective action Eq. \eqref{eq:disorderedEffAct} and derive the leading order fluctuation contributions in the strong disorder regime where $\omega_0 > 0$. The propagator for the $\vb*{z}$ and $\vb*{w}$ fields is
\begin{equation}
    \hat{G}_0^{-1}(\vb{p}) = \begin{pmatrix}
    (\vb{p}^2 + m^2 + \omega_0) \hat{I} - \omega_0 \hat{M} & - \sigma_0 \hat{I} \\
    -\sigma_0^* \hat{I} & (\vb{p}^2 + m^2 + \omega_0) \hat{I} - \omega_0 \hat{M}
    \end{pmatrix}.
\end{equation}
the components of which we will denote $G_{\alpha\beta}^{ij}(\vb{p})$. For example, $G_{zz}^{ij}(\vb{p})$ is the amplitude for the hybridization of $\vb*{z}_i$ and $\vb*{z}_j$.

First, the case $\sigma_0 = 0$: The kernels in Eq. \eqref{eq:OP_effAct_disordered} are given by the integrals
\begin{subequations}
\begin{align}
    \Pi_\sigma^{(1)}(\vb{p};n) - \Pi_\sigma^{(2)}(\vb{p};n) &= \frac{2}{K_0} - 2\int \frac{\dd^2 \vb{q}}{(2\pi)^2} G_{zz}^{ii}(\vb{q}) G_{zz}^{ii}(\vb{p} + \vb{q}) \\
\begin{split}
    \Pi_\sigma^{(2)}(\vb{p};n) &= 2 \int \frac{\dd^2 \vb{q}}{(2\pi)^2} G_{zz}^{ij}(\vb{q}) G_{zz}^{ij}(\vb{p} + \vb{q}) .
\end{split}
\end{align}
\end{subequations}
In practice, we only need to evaluate these kernels for $n=0$, so we have
\begin{subequations}
\begin{align}
\begin{split}
    \Pi_\sigma^{(1)}(\vb{p};0) - \Pi_\sigma^{(2)}(\vb{p};0) &= \frac{2}{K_0} - 2 \int \frac{\dd^2 \vb{q}}{(2\pi)^2} \left[ \frac{1}{\vb{q}^2 + m^2 + \omega_0} + \frac{\omega_0}{(\vb{q}^2 + m^2 + \omega_0)^2} \right] \\
    &\kern2.25em \times \left[ \frac{1}{(\vb{p} + \vb{q})^2 + m^2 + \omega_0} + \frac{\omega_0}{[(\vb{p} + \vb{q})^2 + m^2 + \omega_0]^2} \right],
\end{split}\\
    \Pi_\sigma^{(2)}(\vb{p};0) &= 2 \int \frac{\dd^2 \vb{q}}{(2\pi)^2} \frac{\omega_0^2}{(\vb{q}^2 + m^2 + \omega_0)^2 [(\vb{p} + \vb{q})^2 + m^2 + \omega_0]^2} ,
\end{align}
\end{subequations}
which are straightforward to evaluate using the usual methods (e.g., Feynman parametrization of the loop integrals).

Next, the case $\sigma_0 \neq 0$. Following the same notation as above, we write the effective action in the form
\begin{equation}
    S_{\mathrm{eff}} = \frac{N}{2} \sum_{i,j=1}^n \int \dd^2 \vb{x} \dd^2 \vb{y}\, \theta_i(\vb{x}) \left[ \Pi^{(1)}_\theta(\vb{x}-\vb{y};n) \hat{I} - \Pi^{(2)}_\theta(\vb{x}-\vb{y};n) \hat{M} \right]_{ij} \theta_j(\vb{y}).
\end{equation}
However, there are now contributions from three sources
\begin{equation}
    \sigma_j(\vb{x}) \simeq \rho_0 e^{i\theta_j(\vb{x})}, \kern2em \zeta_{jk}(\vb{x}) \simeq \frac{\omega_0}{\eta^2_{\mathrm{tot}}} e^{i(\theta_j(\vb{x}) - \theta_k(\vb{x}))/2}, \kern2em \omega_{jk}(\vb{x}) \simeq \frac{\omega_0}{\eta^2_{\mathrm{tot}}} e^{-i(\theta_j(\vb{x}) - \theta_k(\vb{x}))/2}.
\end{equation}
After some algebra, we obtain
\begin{subequations}
\begin{align} 
    &\kern-2em\Pi_\theta^{(1)}(\vb{p};n) - \Pi_\theta^{(2)}(\vb{p};n) = \\
    &= 2 \rho_0^2 \int \frac{\dd^2 \vb{q}}{(2\pi)^2} \left[ G_{zw}^{ii}(\vb{q}) G_{zw}^{ii}(\vb{p} + \vb{q}) - G_{zz}^{ii}(\vb{q}) G_{zz}^{ii}(\vb{p} + \vb{q}) \right] + 2 \rho_0 \int \frac{\dd^2 \vb{q}}{(2\pi)^2} G_{zw}^{ii}(\vb{q}) \nn \\
    &+ (n-1)\omega_0^2 \Delta^2 \int \frac{\dd^2 \vb{q}}{(2\pi)^2} \big[ G_{zw}^{ii}(\vb{q}) G_{zw}^{ii}(\vb{p} + \vb{q}) - G_{zz}^{ii}(\vb{q}) G_{zz}^{ii}(\vb{p} + \vb{q}) \nn \\    
    &\kern11em+ (n-2) G_{zw}^{ii}(\vb{q}) G_{zw}^{ij}(\vb{p} + \vb{q}) - (n-2) G_{zz}^{ii}(\vb{q}) G_{zz}^{ij}(\vb{p} + \vb{q}) \nn  \\    
    &\kern11em+ (n-1) G_{zz}^{ij}(\vb{q}) G_{zz}^{ij}(\vb{p} + \vb{q}) - (n-1) G_{zw}^{ij}(\vb{q}) G_{zw}^{ij}(\vb{p} + \vb{q}) \big] \nn \\
    & + (n-1) \omega_0 \int \frac{\dd^2 \vb{q}}{(2\pi)^2} G_{zz}^{ij}(\vb{q})  \nn \\
    &+4(n-1) \rho_0 \omega_0 \Delta \int \frac{\dd^2 \vb{q}}{(2\pi)^2} \left[ G_{zw}^{ii}(\vb{q}) G_{zz}^{ij}(\vb{p} + \vb{q}) - G_{zw}^{ij}(\vb{q}) G_{zz}^{ii}(\vb{p} + \vb{q})  \right], \nn \allowdisplaybreaks \\
    &\kern-2em \Pi_\theta^{(2)}(\vb{p};n) = \\
    &= 2 \rho_0^2 \int \frac{\dd^2 \vb{q}}{(2\pi)^2} \left[  G_{zz}^{ij}(\vb{q}) G_{zz}^{ij}(\vb{p} + \vb{q}) - G_{zw}^{ij}(\vb{q}) G_{zw}^{ij}(\vb{p} + \vb{q}) \right] \nn \\
    &- \omega_0^2 \Delta^2 \int \frac{\dd^2 \vb{q}}{(2\pi)^2} \big[ G_{zz}^{ii}(\vb{q}) G_{zz}^{ii}(\vb{p} + \vb{q}) - G_{zw}^{ii}(\vb{q}) G_{zw}^{ii}(\vb{p} + \vb{q}) \nn \\    
    &\kern8em+(n-2) G_{zz}^{ii}(\vb{q}) G_{zz}^{ij}(\vb{p} + \vb{q}) - (n-2) G_{zw}^{ii}(\vb{q}) G_{zw}^{ij}(\vb{p} + \vb{q})  \nn  \\    
    &\kern8em- (n-1) G_{zz}^{ij}(\vb{q}) G_{zz}^{ij}(\vb{p} + \vb{q}) + (n-1) G_{zw}^{ij}(\vb{q}) G_{zw}^{ij}(\vb{p} + \vb{q}) \big] \nn \\
    &+ \omega_0 \int \frac{\dd^2 \vb{q}}{(2\pi)^2} G_{zz}^{ij}(\vb{q}) \nn \\
    &- 4 \rho_0 \omega_0 \Delta \int \frac{\dd^2 \vb{q}}{(2\pi)^2} \left[ G_{zw}^{ij}(\vb{q}) G_{zz}^{ii}(\vb{p} + \vb{q}) - G_{zw}^{ii}(\vb{q}) G_{zz}^{ij}(\vb{p} + \vb{q})  \right].  \nn
\end{align}
\end{subequations}
These kernels can be shown to have the following power series expansion in $\vb{p}$,
\begin{subequations}
\begin{align}
    N[\Pi_\theta^{(1)}(\vb{p};n) - \Pi_\theta^{(2)}(\vb{p};n)] &\simeq m_\theta(n)^2 + [\gamma_\theta(n) + \Gamma_\theta(n)] \vb{p}^2, \\
    n N \Pi_\theta^{(2)}(\vb{p};n) &\simeq m_\theta(n)^2 + \Gamma_\theta(n) \vb{p}^2,
\end{align}
\end{subequations}
which yields the effective action Eq. \eqref{eq:IRdisEffAct}. The leading order-$n$ contribution to $m_\theta^2$ is given in Eq. \eqref{eq:phase_gap}, and to order $n$ the coefficients of the gradient terms are
\begin{subequations}
\begin{align}
    \gamma_\theta(n) &\simeq \overline{\gamma_\theta} + n N \frac{\rho_0^2 \omega_0^2}{4\pi[(m^2 + \omega_0)^2 - \rho_0^2]^2} , \\
    \Gamma_\theta(n) &\simeq n N \frac{\omega_0^2}{8 \pi \rho_0^2} \left( [6 - \Delta(6 - \Delta)] \left[\frac{m^2 + \omega_0}{\rho_0} \tanh^{-1}\left(\frac{\rho_0}{m^2 + \omega_0}\right) - 1\right] - \frac{2\rho_0 (1 - \Delta)}{(m^2 + \omega_0)^2 - \rho_0^2} \right),
\end{align}
\end{subequations}
where $\overline{\gamma_\theta}$ is the disorder-averaged phase stiffness given in Eq. \eqref{eq:disorder_stiffness}. Note that while $\Gamma_\theta(n)$ is not strictly positive for all ranges of parameters, the spectrum of the kernel only depends on the combination $\gamma_\theta(n) + \Gamma_\theta(n)$, which is positive for all $-1 \leq \Delta \leq 1$.

\bibliography{refs}

\providecommand{\noopsort}[1]{}\providecommand{\singleletter}[1]{#1}%
\begin{thebibliography}{54}%
\makeatletter
\providecommand \@ifxundefined [1]{%
 \@ifx{#1\undefined}
}%
\providecommand \@ifnum [1]{%
 \ifnum #1\expandafter \@firstoftwo
 \else \expandafter \@secondoftwo
 \fi
}%
\providecommand \@ifx [1]{%
 \ifx #1\expandafter \@firstoftwo
 \else \expandafter \@secondoftwo
 \fi
}%
\providecommand \natexlab [1]{#1}%
\providecommand \enquote  [1]{``#1''}%
\providecommand \bibnamefont  [1]{#1}%
\providecommand \bibfnamefont [1]{#1}%
\providecommand \citenamefont [1]{#1}%
\providecommand \href@noop [0]{\@secondoftwo}%
\providecommand \href [0]{\begingroup \@sanitize@url \@href}%
\providecommand \@href[1]{\@@startlink{#1}\@@href}%
\providecommand \@@href[1]{\endgroup#1\@@endlink}%
\providecommand \@sanitize@url [0]{\catcode `\\12\catcode `\$12\catcode
  `\&12\catcode `\#12\catcode `\^12\catcode `\_12\catcode `\%12\relax}%
\providecommand \@@startlink[1]{}%
\providecommand \@@endlink[0]{}%
\providecommand \url  [0]{\begingroup\@sanitize@url \@url }%
\providecommand \@url [1]{\endgroup\@href {#1}{\urlprefix }}%
\providecommand \urlprefix  [0]{URL }%
\providecommand \Eprint [0]{\href }%
\providecommand \doibase [0]{https://doi.org/}%
\providecommand \selectlanguage [0]{\@gobble}%
\providecommand \bibinfo  [0]{\@secondoftwo}%
\providecommand \bibfield  [0]{\@secondoftwo}%
\providecommand \translation [1]{[#1]}%
\providecommand \BibitemOpen [0]{}%
\providecommand \bibitemStop [0]{}%
\providecommand \bibitemNoStop [0]{.\EOS\space}%
\providecommand \EOS [0]{\spacefactor3000\relax}%
\providecommand \BibitemShut  [1]{\csname bibitem#1\endcsname}%
\let\auto@bib@innerbib\@empty
\bibitem [{\citenamefont {Efetov}\ and\ \citenamefont
  {Larkin}(1977)}]{Efetov1977}%
  \BibitemOpen
  \bibfield  {author} {\bibinfo {author} {\bibfnamefont {K.~B.}\ \bibnamefont
  {Efetov}}\ and\ \bibinfo {author} {\bibfnamefont {A.~I.}\ \bibnamefont
  {Larkin}},\ }\bibfield  {title} {\bibinfo {title} {{Charge-density wave in a
  random potential}},\ }\href {http://jetp.ras.ru/cgi-bin/dn/e_045_06_1236.pdf}
  {\bibfield  {journal} {\bibinfo  {journal} {Sov. Phys. JETP}\ }\textbf
  {\bibinfo {volume} {45}},\ \bibinfo {pages} {1236} (\bibinfo {year}
  {1977})}\BibitemShut {NoStop}%
\bibitem [{\citenamefont {Imry}\ and\ \citenamefont {Ma}(1975)}]{Imry1975}%
  \BibitemOpen
  \bibfield  {author} {\bibinfo {author} {\bibfnamefont {Y.}~\bibnamefont
  {Imry}}\ and\ \bibinfo {author} {\bibfnamefont {S.-k.}\ \bibnamefont {Ma}},\
  }\bibfield  {title} {\bibinfo {title} {{Random-Field Instability of the
  Ordered State of Continuous Symmetry}},\ }\href
  {https://doi.org/10.1103/PhysRevLett.35.1399} {\bibfield  {journal} {\bibinfo
   {journal} {Phys. Rev. Lett.}\ }\textbf {\bibinfo {volume} {35}},\ \bibinfo
  {pages} {1399} (\bibinfo {year} {1975})}\BibitemShut {NoStop}%
\bibitem [{\citenamefont {Mermin}\ and\ \citenamefont
  {Wagner}(1966)}]{Mermin1966}%
  \BibitemOpen
  \bibfield  {author} {\bibinfo {author} {\bibfnamefont {N.~D.}\ \bibnamefont
  {Mermin}}\ and\ \bibinfo {author} {\bibfnamefont {H.}~\bibnamefont
  {Wagner}},\ }\bibfield  {title} {\bibinfo {title} {{Absence of Ferromagnetism
  or Antiferromagnetism in One- or Two-Dimensional Isotropic Heisenberg
  Models}},\ }\href {https://doi.org/10.1103/PhysRevLett.17.1133} {\bibfield
  {journal} {\bibinfo  {journal} {Phys. Rev. Lett.}\ }\textbf {\bibinfo
  {volume} {17}},\ \bibinfo {pages} {1133} (\bibinfo {year}
  {1966})}\BibitemShut {NoStop}%
\bibitem [{\citenamefont {Berezinskii}(1971)}]{Berezinskii1971}%
  \BibitemOpen
  \bibfield  {author} {\bibinfo {author} {\bibfnamefont {V.~L.}\ \bibnamefont
  {Berezinskii}},\ }\bibfield  {title} {\bibinfo {title} {{Destruction of
  Long-Range Order in One-Dimensional and Two-Dimensional Systems Possessing a
  Continuous Symmetry Group I. Classical Systems}},\ }\href
  {http://jetp.ras.ru/cgi-bin/dn/e_032_03_0493.pdf} {\bibfield  {journal}
  {\bibinfo  {journal} {Sov. Phys. JETP}\ }\textbf {\bibinfo {volume} {32}},\
  \bibinfo {pages} {493} (\bibinfo {year} {1971})}\BibitemShut {NoStop}%
\bibitem [{\citenamefont {Kosterlitz}\ and\ \citenamefont
  {Thouless}(1973)}]{Kosterlitz1973}%
  \BibitemOpen
  \bibfield  {author} {\bibinfo {author} {\bibfnamefont {J.~M.}\ \bibnamefont
  {Kosterlitz}}\ and\ \bibinfo {author} {\bibfnamefont {D.~J.}\ \bibnamefont
  {Thouless}},\ }\bibfield  {title} {\bibinfo {title} {{Ordering, metastability
  and phase transitions in two-dimensional systems}},\ }\href
  {https://doi.org/10.1088/0022-3719/6/7/010} {\bibfield  {journal} {\bibinfo
  {journal} {J. Phys. Chem.}\ }\textbf {\bibinfo {volume} {6}},\ \bibinfo
  {pages} {1181} (\bibinfo {year} {1973})}\BibitemShut {NoStop}%
\bibitem [{\citenamefont {Houghton}\ \emph {et~al.}(1981)\citenamefont
  {Houghton}, \citenamefont {Kenway},\ and\ \citenamefont
  {Ying}}]{Houghton1981}%
  \BibitemOpen
  \bibfield  {author} {\bibinfo {author} {\bibfnamefont {A.}~\bibnamefont
  {Houghton}}, \bibinfo {author} {\bibfnamefont {R.~D.}\ \bibnamefont
  {Kenway}},\ and\ \bibinfo {author} {\bibfnamefont {S.~C.}\ \bibnamefont
  {Ying}},\ }\bibfield  {title} {\bibinfo {title} {{Effects of a random
  symmetry-breaking field on topological order in two dimensions}},\ }\href
  {https://doi.org/10.1103/PhysRevB.23.298} {\bibfield  {journal} {\bibinfo
  {journal} {Phys. Rev. B}\ }\textbf {\bibinfo {volume} {23}},\ \bibinfo
  {pages} {298} (\bibinfo {year} {1981})}\BibitemShut {NoStop}%
\bibitem [{\citenamefont {Cardy}\ and\ \citenamefont
  {Ostlund}(1982)}]{Cardy1982}%
  \BibitemOpen
  \bibfield  {author} {\bibinfo {author} {\bibfnamefont {J.~L.}\ \bibnamefont
  {Cardy}}\ and\ \bibinfo {author} {\bibfnamefont {S.}~\bibnamefont
  {Ostlund}},\ }\bibfield  {title} {\bibinfo {title} {{Random symmetry-breaking
  fields and the $XY$ model}},\ }\href
  {https://doi.org/10.1103/PhysRevB.25.6899} {\bibfield  {journal} {\bibinfo
  {journal} {Phys. Rev. B}\ }\textbf {\bibinfo {volume} {25}},\ \bibinfo
  {pages} {6899} (\bibinfo {year} {1982})}\BibitemShut {NoStop}%
\bibitem [{\citenamefont {Goldschmidt}\ and\ \citenamefont
  {Houghton}(1982)}]{Goldschmidt1982}%
  \BibitemOpen
  \bibfield  {author} {\bibinfo {author} {\bibfnamefont {Y.~Y.}\ \bibnamefont
  {Goldschmidt}}\ and\ \bibinfo {author} {\bibfnamefont {A.}~\bibnamefont
  {Houghton}},\ }\bibfield  {title} {\bibinfo {title} {{Field-theoretic
  treatment of the 2d planar model with random $p$-fold symmetry-breaking
  field}},\ }\href {https://doi.org/10.1016/0550-3213(82)90237-1} {\bibfield
  {journal} {\bibinfo  {journal} {Nucl. Phys. B}\ }\textbf {\bibinfo {volume}
  {210}},\ \bibinfo {pages} {155} (\bibinfo {year} {1982})}\BibitemShut
  {NoStop}%
\bibitem [{\citenamefont {Young}(1977)}]{Young1977}%
  \BibitemOpen
  \bibfield  {author} {\bibinfo {author} {\bibfnamefont {A.~P.}\ \bibnamefont
  {Young}},\ }\bibfield  {title} {\bibinfo {title} {{On the lowering of
  dimensionality in phase transitions with random fields}},\ }\href
  {https://doi.org/10.1088/0022-3719/10/9/007} {\bibfield  {journal} {\bibinfo
  {journal} {J. Phys. Chem.}\ }\textbf {\bibinfo {volume} {10}},\ \bibinfo
  {pages} {L257} (\bibinfo {year} {1977})}\BibitemShut {NoStop}%
\bibitem [{\citenamefont {Fisher}(1985)}]{Fisher1985}%
  \BibitemOpen
  \bibfield  {author} {\bibinfo {author} {\bibfnamefont {D.~S.}\ \bibnamefont
  {Fisher}},\ }\bibfield  {title} {\bibinfo {title} {{Random fields, random
  anisotropies, nonlinear $\sigma$ models, and dimensional reduction}},\ }\href
  {https://doi.org/10.1103/PhysRevB.31.7233} {\bibfield  {journal} {\bibinfo
  {journal} {Phys. Rev. B}\ }\textbf {\bibinfo {volume} {31}},\ \bibinfo
  {pages} {7233} (\bibinfo {year} {1985})}\BibitemShut {NoStop}%
\bibitem [{\citenamefont {Tarjus}\ and\ \citenamefont
  {Tissier}(2004)}]{Tarjus2004}%
  \BibitemOpen
  \bibfield  {author} {\bibinfo {author} {\bibfnamefont {G.}~\bibnamefont
  {Tarjus}}\ and\ \bibinfo {author} {\bibfnamefont {M.}~\bibnamefont
  {Tissier}},\ }\bibfield  {title} {\bibinfo {title} {{Nonperturbative
  Functional Renormalization Group for Random-Field Models: The Way Out of
  Dimensional Reduction}},\ }\href
  {https://doi.org/10.1103/PhysRevLett.93.267008} {\bibfield  {journal}
  {\bibinfo  {journal} {Phys. Rev. Lett.}\ }\textbf {\bibinfo {volume} {93}},\
  \bibinfo {pages} {267008} (\bibinfo {year} {2004})}\BibitemShut {NoStop}%
\bibitem [{\citenamefont {Giamarchi}\ and\ \citenamefont {{Le
  Doussal}}(1994)}]{Giamarchi1994}%
  \BibitemOpen
  \bibfield  {author} {\bibinfo {author} {\bibfnamefont {T.}~\bibnamefont
  {Giamarchi}}\ and\ \bibinfo {author} {\bibfnamefont {P.}~\bibnamefont {{Le
  Doussal}}},\ }\bibfield  {title} {\bibinfo {title} {{Elastic theory of pinned
  flux lattices}},\ }\href {https://doi.org/10.1103/PhysRevLett.72.1530}
  {\bibfield  {journal} {\bibinfo  {journal} {Phys. Rev. Lett.}\ }\textbf
  {\bibinfo {volume} {72}},\ \bibinfo {pages} {1530} (\bibinfo {year}
  {1994})}\BibitemShut {NoStop}%
\bibitem [{\citenamefont {{Le Doussal}}\ and\ \citenamefont
  {Giamarchi}(2000)}]{LeDoussal2000}%
  \BibitemOpen
  \bibfield  {author} {\bibinfo {author} {\bibfnamefont {P.}~\bibnamefont {{Le
  Doussal}}}\ and\ \bibinfo {author} {\bibfnamefont {T.}~\bibnamefont
  {Giamarchi}},\ }\bibfield  {title} {\bibinfo {title} {{Dislocations and Bragg
  glasses in two dimensions}},\ }\href
  {https://doi.org/10.1016/S0921-4534(00)00005-8} {\bibfield  {journal}
  {\bibinfo  {journal} {Physica C}\ }\textbf {\bibinfo {volume} {331}},\
  \bibinfo {pages} {233} (\bibinfo {year} {2000})}\BibitemShut {NoStop}%
\bibitem [{\citenamefont {Zeng}\ \emph {et~al.}(1999)\citenamefont {Zeng},
  \citenamefont {Leath},\ and\ \citenamefont {Fisher}}]{Zeng1999}%
  \BibitemOpen
  \bibfield  {author} {\bibinfo {author} {\bibfnamefont {C.}~\bibnamefont
  {Zeng}}, \bibinfo {author} {\bibfnamefont {P.~L.}\ \bibnamefont {Leath}},\
  and\ \bibinfo {author} {\bibfnamefont {D.~S.}\ \bibnamefont {Fisher}},\
  }\bibfield  {title} {\bibinfo {title} {{Absence of Two-Dimensional Bragg
  Glasses}},\ }\href {https://doi.org/10.1103/PhysRevLett.82.1935} {\bibfield
  {journal} {\bibinfo  {journal} {Phys. Rev. Lett.}\ }\textbf {\bibinfo
  {volume} {82}},\ \bibinfo {pages} {1935} (\bibinfo {year}
  {1999})}\BibitemShut {NoStop}%
\bibitem [{\citenamefont {Andreanov}\ and\ \citenamefont
  {Fedorenko}(2014)}]{Andreanov2014}%
  \BibitemOpen
  \bibfield  {author} {\bibinfo {author} {\bibfnamefont {A.}~\bibnamefont
  {Andreanov}}\ and\ \bibinfo {author} {\bibfnamefont {A.~A.}\ \bibnamefont
  {Fedorenko}},\ }\bibfield  {title} {\bibinfo {title} {{Localization of spin
  waves in disordered quantum rotors}},\ }\href
  {https://doi.org/10.1103/PhysRevB.90.014205} {\bibfield  {journal} {\bibinfo
  {journal} {Phys. Rev. B}\ }\textbf {\bibinfo {volume} {90}},\ \bibinfo
  {pages} {014205} (\bibinfo {year} {2014})}\BibitemShut {NoStop}%
\bibitem [{\citenamefont {Nie}\ \emph {et~al.}(2014)\citenamefont {Nie},
  \citenamefont {Tarjus},\ and\ \citenamefont {Kivelson}}]{Nie2014}%
  \BibitemOpen
  \bibfield  {author} {\bibinfo {author} {\bibfnamefont {L.}~\bibnamefont
  {Nie}}, \bibinfo {author} {\bibfnamefont {G.}~\bibnamefont {Tarjus}},\ and\
  \bibinfo {author} {\bibfnamefont {S.~A.}\ \bibnamefont {Kivelson}},\
  }\bibfield  {title} {\bibinfo {title} {{Quenched disorder and vestigial
  nematicity in the pseudogap regime of the cuprates}},\ }\href
  {https://doi.org/10.1073/pnas.1406019111} {\bibfield  {journal} {\bibinfo
  {journal} {Proc. Nat. Acad. Sci. USA}\ }\textbf {\bibinfo {volume} {111}},\
  \bibinfo {pages} {7980} (\bibinfo {year} {2014})}\BibitemShut {NoStop}%
\bibitem [{\citenamefont {Nie}\ \emph {et~al.}(2015)\citenamefont {Nie},
  \citenamefont {Sierens}, \citenamefont {Melko}, \citenamefont {Sachdev},\
  and\ \citenamefont {Kivelson}}]{Nie2015}%
  \BibitemOpen
  \bibfield  {author} {\bibinfo {author} {\bibfnamefont {L.}~\bibnamefont
  {Nie}}, \bibinfo {author} {\bibfnamefont {L.~E.~H.}\ \bibnamefont {Sierens}},
  \bibinfo {author} {\bibfnamefont {R.~G.}\ \bibnamefont {Melko}}, \bibinfo
  {author} {\bibfnamefont {S.}~\bibnamefont {Sachdev}},\ and\ \bibinfo {author}
  {\bibfnamefont {S.~A.}\ \bibnamefont {Kivelson}},\ }\bibfield  {title}
  {\bibinfo {title} {{Fluctuating orders and quenched randomness in the
  cuprates}},\ }\href {https://doi.org/10.1103/PhysRevB.92.174505} {\bibfield
  {journal} {\bibinfo  {journal} {Phys. Rev. B}\ }\textbf {\bibinfo {volume}
  {92}},\ \bibinfo {pages} {174505} (\bibinfo {year} {2015})}\BibitemShut
  {NoStop}%
\bibitem [{\citenamefont {Tarjus}\ and\ \citenamefont
  {Tissier}(2020)}]{Tarjus2020}%
  \BibitemOpen
  \bibfield  {author} {\bibinfo {author} {\bibfnamefont {G.}~\bibnamefont
  {Tarjus}}\ and\ \bibinfo {author} {\bibfnamefont {M.}~\bibnamefont
  {Tissier}},\ }\bibfield  {title} {\bibinfo {title} {{Random-field Ising and
  $O(N)$ models: theoretical description through the functional renormalization
  group}},\ }\href {https://doi.org/10.1140/epjb/e2020-100489-1} {\bibfield
  {journal} {\bibinfo  {journal} {Eur. Phys. J. B}\ }\textbf {\bibinfo {volume}
  {93}},\ \bibinfo {pages} {50} (\bibinfo {year} {2020})}\BibitemShut {NoStop}%
\bibitem [{\citenamefont {{Le Doussal}}\ and\ \citenamefont
  {Wiese}(2006)}]{LeDoussal2006}%
  \BibitemOpen
  \bibfield  {author} {\bibinfo {author} {\bibfnamefont {P.}~\bibnamefont {{Le
  Doussal}}}\ and\ \bibinfo {author} {\bibfnamefont {K.~J.}\ \bibnamefont
  {Wiese}},\ }\bibfield  {title} {\bibinfo {title} {{Random-Field Spin Models
  beyond 1 Loop: A Mechanism for Decreasing the Lower Critical Dimension}},\
  }\href {https://doi.org/10.1103/PhysRevLett.96.197202} {\bibfield  {journal}
  {\bibinfo  {journal} {Phys. Rev. Lett.}\ }\textbf {\bibinfo {volume} {96}},\
  \bibinfo {pages} {197202} (\bibinfo {year} {2006})}\BibitemShut {NoStop}%
\bibitem [{\citenamefont {Tranquada}\ \emph {et~al.}(1995)\citenamefont
  {Tranquada}, \citenamefont {Sternlieb}, \citenamefont {Axe}, \citenamefont
  {Nakamura},\ and\ \citenamefont {Uchida}}]{Tranquada1995}%
  \BibitemOpen
  \bibfield  {author} {\bibinfo {author} {\bibfnamefont {J.~M.}\ \bibnamefont
  {Tranquada}}, \bibinfo {author} {\bibfnamefont {B.~J.}\ \bibnamefont
  {Sternlieb}}, \bibinfo {author} {\bibfnamefont {J.~D.}\ \bibnamefont {Axe}},
  \bibinfo {author} {\bibfnamefont {Y.}~\bibnamefont {Nakamura}},\ and\
  \bibinfo {author} {\bibfnamefont {S.}~\bibnamefont {Uchida}},\ }\bibfield
  {title} {\bibinfo {title} {{Evidence for stripe correlations of spins and
  holes in copper oxide superconductors}},\ }\href
  {https://doi.org/10.1038/375561a0} {\bibfield  {journal} {\bibinfo  {journal}
  {Nature}\ }\textbf {\bibinfo {volume} {375}},\ \bibinfo {pages} {561}
  (\bibinfo {year} {1995})}\BibitemShut {NoStop}%
\bibitem [{\citenamefont {Abbamonte}\ \emph {et~al.}(2005)\citenamefont
  {Abbamonte}, \citenamefont {Rusydi}, \citenamefont {Smadici}, \citenamefont
  {Gu}, \citenamefont {Sawatzky},\ and\ \citenamefont {Feng}}]{Abbamonte2005}%
  \BibitemOpen
  \bibfield  {author} {\bibinfo {author} {\bibfnamefont {P.}~\bibnamefont
  {Abbamonte}}, \bibinfo {author} {\bibfnamefont {A.}~\bibnamefont {Rusydi}},
  \bibinfo {author} {\bibfnamefont {S.}~\bibnamefont {Smadici}}, \bibinfo
  {author} {\bibfnamefont {G.~D.}\ \bibnamefont {Gu}}, \bibinfo {author}
  {\bibfnamefont {G.~A.}\ \bibnamefont {Sawatzky}},\ and\ \bibinfo {author}
  {\bibfnamefont {D.~L.}\ \bibnamefont {Feng}},\ }\bibfield  {title} {\bibinfo
  {title} {{Spatially modulated `Mottness' in La$_{2-x}$Ba$_x$CuO$_4$}},\
  }\href {https://doi.org/10.1038/nphys178} {\bibfield  {journal} {\bibinfo
  {journal} {Nature Phys.}\ }\textbf {\bibinfo {volume} {1}},\ \bibinfo {pages}
  {155} (\bibinfo {year} {2005})}\BibitemShut {NoStop}%
\bibitem [{\citenamefont {Kivelson}\ \emph {et~al.}(2003)\citenamefont
  {Kivelson}, \citenamefont {Bindloss}, \citenamefont {Fradkin}, \citenamefont
  {Oganesyan}, \citenamefont {Tranquada}, \citenamefont {Kapitulnik},\ and\
  \citenamefont {Howald}}]{Kivelson2003}%
  \BibitemOpen
  \bibfield  {author} {\bibinfo {author} {\bibfnamefont {S.~A.}\ \bibnamefont
  {Kivelson}}, \bibinfo {author} {\bibfnamefont {I.~P.}\ \bibnamefont
  {Bindloss}}, \bibinfo {author} {\bibfnamefont {E.}~\bibnamefont {Fradkin}},
  \bibinfo {author} {\bibfnamefont {V.}~\bibnamefont {Oganesyan}}, \bibinfo
  {author} {\bibfnamefont {J.~M.}\ \bibnamefont {Tranquada}}, \bibinfo {author}
  {\bibfnamefont {A.}~\bibnamefont {Kapitulnik}},\ and\ \bibinfo {author}
  {\bibfnamefont {C.}~\bibnamefont {Howald}},\ }\bibfield  {title} {\bibinfo
  {title} {{How to detect fluctuating stripes in the high-temperature
  superconductors}},\ }\href {https://doi.org/10.1103/RevModPhys.75.1201}
  {\bibfield  {journal} {\bibinfo  {journal} {Rev. Mod. Phys.}\ }\textbf
  {\bibinfo {volume} {75}},\ \bibinfo {pages} {1201} (\bibinfo {year}
  {2003})}\BibitemShut {NoStop}%
\bibitem [{\citenamefont {Ghiringhelli}\ \emph {et~al.}(2012)\citenamefont
  {Ghiringhelli}, \citenamefont {{Le Tacon}}, \citenamefont {Minola},
  \citenamefont {Blanco-Canosa}, \citenamefont {Mazzoli}, \citenamefont
  {Brookes}, \citenamefont {{De Luca}}, \citenamefont {Frano}, \citenamefont
  {Hawthorn}, \citenamefont {He}, \citenamefont {Loew}, \citenamefont {Sala},
  \citenamefont {Peets}, \citenamefont {Salluzzo}, \citenamefont {Schierle},
  \citenamefont {Sutarto}, \citenamefont {Sawatzky}, \citenamefont {Weschke},
  \citenamefont {Keimer},\ and\ \citenamefont {Braicovich}}]{Ghiringhelli2012}%
  \BibitemOpen
  \bibfield  {author} {\bibinfo {author} {\bibfnamefont {G.}~\bibnamefont
  {Ghiringhelli}}, \bibinfo {author} {\bibfnamefont {M.}~\bibnamefont {{Le
  Tacon}}}, \bibinfo {author} {\bibfnamefont {M.}~\bibnamefont {Minola}},
  \bibinfo {author} {\bibfnamefont {S.}~\bibnamefont {Blanco-Canosa}}, \bibinfo
  {author} {\bibfnamefont {C.}~\bibnamefont {Mazzoli}}, \bibinfo {author}
  {\bibfnamefont {N.~B.}\ \bibnamefont {Brookes}}, \bibinfo {author}
  {\bibfnamefont {G.~M.}\ \bibnamefont {{De Luca}}}, \bibinfo {author}
  {\bibfnamefont {A.}~\bibnamefont {Frano}}, \bibinfo {author} {\bibfnamefont
  {D.~G.}\ \bibnamefont {Hawthorn}}, \bibinfo {author} {\bibfnamefont
  {F.}~\bibnamefont {He}}, \bibinfo {author} {\bibfnamefont {T.}~\bibnamefont
  {Loew}}, \bibinfo {author} {\bibfnamefont {M.~M.}\ \bibnamefont {Sala}},
  \bibinfo {author} {\bibfnamefont {D.~C.}\ \bibnamefont {Peets}}, \bibinfo
  {author} {\bibfnamefont {M.}~\bibnamefont {Salluzzo}}, \bibinfo {author}
  {\bibfnamefont {E.}~\bibnamefont {Schierle}}, \bibinfo {author}
  {\bibfnamefont {R.}~\bibnamefont {Sutarto}}, \bibinfo {author} {\bibfnamefont
  {G.~A.}\ \bibnamefont {Sawatzky}}, \bibinfo {author} {\bibfnamefont
  {E.}~\bibnamefont {Weschke}}, \bibinfo {author} {\bibfnamefont
  {B.}~\bibnamefont {Keimer}},\ and\ \bibinfo {author} {\bibfnamefont
  {L.}~\bibnamefont {Braicovich}},\ }\bibfield  {title} {\bibinfo {title}
  {{Long-Range Incommensurate Charge Fluctuations in
  (Y,Nd)Ba$_2$Cu$_3$O$_{6+x}$}},\ }\href
  {https://doi.org/10.1126/science.1223532} {\bibfield  {journal} {\bibinfo
  {journal} {Science}\ }\textbf {\bibinfo {volume} {337}},\ \bibinfo {pages}
  {821} (\bibinfo {year} {2012})}\BibitemShut {NoStop}%
\bibitem [{\citenamefont {Mesaros}\ \emph {et~al.}(2016)\citenamefont
  {Mesaros}, \citenamefont {Fujita}, \citenamefont {Edkins}, \citenamefont
  {Hamidian}, \citenamefont {Eisaki}, \citenamefont {Uchida}, \citenamefont
  {Davis}, \citenamefont {Lawler},\ and\ \citenamefont {Kim}}]{Mesaros2016}%
  \BibitemOpen
  \bibfield  {author} {\bibinfo {author} {\bibfnamefont {A.}~\bibnamefont
  {Mesaros}}, \bibinfo {author} {\bibfnamefont {K.}~\bibnamefont {Fujita}},
  \bibinfo {author} {\bibfnamefont {S.~D.}\ \bibnamefont {Edkins}}, \bibinfo
  {author} {\bibfnamefont {M.~H.}\ \bibnamefont {Hamidian}}, \bibinfo {author}
  {\bibfnamefont {H.}~\bibnamefont {Eisaki}}, \bibinfo {author} {\bibfnamefont
  {S.-i.}\ \bibnamefont {Uchida}}, \bibinfo {author} {\bibfnamefont {J.~C.~S.}\
  \bibnamefont {Davis}}, \bibinfo {author} {\bibfnamefont {M.~J.}\ \bibnamefont
  {Lawler}},\ and\ \bibinfo {author} {\bibfnamefont {E.-A.}\ \bibnamefont
  {Kim}},\ }\bibfield  {title} {\bibinfo {title} {{Commensurate 4 $a_0$-period
  charge density modulations throughout the Bi$_2$ Sr$_2$CaCu$_2$O$_{8+x}$
  pseudogap regime}},\ }\href {https://doi.org/10.1073/pnas.1614247113}
  {\bibfield  {journal} {\bibinfo  {journal} {Proc. Nat. Acad. Sci. USA}\
  }\textbf {\bibinfo {volume} {113}},\ \bibinfo {pages} {12661} (\bibinfo
  {year} {2016})}\BibitemShut {NoStop}%
\bibitem [{\citenamefont {Berg}\ \emph {et~al.}(2009)\citenamefont {Berg},
  \citenamefont {Fradkin}, \citenamefont {Kivelson},\ and\ \citenamefont
  {Tranquada}}]{Berg2009}%
  \BibitemOpen
  \bibfield  {author} {\bibinfo {author} {\bibfnamefont {E.}~\bibnamefont
  {Berg}}, \bibinfo {author} {\bibfnamefont {E.}~\bibnamefont {Fradkin}},
  \bibinfo {author} {\bibfnamefont {S.~A.}\ \bibnamefont {Kivelson}},\ and\
  \bibinfo {author} {\bibfnamefont {J.~M.}\ \bibnamefont {Tranquada}},\
  }\bibfield  {title} {\bibinfo {title} {{Striped superconductors: how spin,
  charge and superconducting orders intertwine in the cuprates}},\ }\href
  {https://doi.org/10.1088/1367-2630/11/11/115004} {\bibfield  {journal}
  {\bibinfo  {journal} {New J. Phys.}\ }\textbf {\bibinfo {volume} {11}},\
  \bibinfo {pages} {115004} (\bibinfo {year} {2009})}\BibitemShut {NoStop}%
\bibitem [{\citenamefont {Fradkin}\ \emph {et~al.}(2015)\citenamefont
  {Fradkin}, \citenamefont {Kivelson},\ and\ \citenamefont
  {Tranquada}}]{Fradkin2015}%
  \BibitemOpen
  \bibfield  {author} {\bibinfo {author} {\bibfnamefont {E.}~\bibnamefont
  {Fradkin}}, \bibinfo {author} {\bibfnamefont {S.~A.}\ \bibnamefont
  {Kivelson}},\ and\ \bibinfo {author} {\bibfnamefont {J.~M.}\ \bibnamefont
  {Tranquada}},\ }\bibfield  {title} {\bibinfo {title} {{Colloquium : Theory of
  intertwined orders in high temperature superconductors}},\ }\href
  {https://doi.org/10.1103/RevModPhys.87.457} {\bibfield  {journal} {\bibinfo
  {journal} {Rev. Mod. Phys.}\ }\textbf {\bibinfo {volume} {87}},\ \bibinfo
  {pages} {457} (\bibinfo {year} {2015})}\BibitemShut {NoStop}%
\bibitem [{\citenamefont {Jang}\ \emph {et~al.}(2016)\citenamefont {Jang},
  \citenamefont {Lee}, \citenamefont {Nojiri}, \citenamefont {Matsuzawa},
  \citenamefont {Yasumura}, \citenamefont {Nie}, \citenamefont {Maharaj},
  \citenamefont {Gerber}, \citenamefont {Liu}, \citenamefont {Mehta},
  \citenamefont {Bonn}, \citenamefont {Liang}, \citenamefont {Hardy},
  \citenamefont {Burns}, \citenamefont {Islam}, \citenamefont {Song},
  \citenamefont {Hastings}, \citenamefont {Devereaux}, \citenamefont {Shen},
  \citenamefont {Kivelson}, \citenamefont {Kao}, \citenamefont {Zhu},\ and\
  \citenamefont {Lee}}]{Jang2016}%
  \BibitemOpen
  \bibfield  {author} {\bibinfo {author} {\bibfnamefont {H.}~\bibnamefont
  {Jang}}, \bibinfo {author} {\bibfnamefont {W.-S.}\ \bibnamefont {Lee}},
  \bibinfo {author} {\bibfnamefont {H.}~\bibnamefont {Nojiri}}, \bibinfo
  {author} {\bibfnamefont {S.}~\bibnamefont {Matsuzawa}}, \bibinfo {author}
  {\bibfnamefont {H.}~\bibnamefont {Yasumura}}, \bibinfo {author}
  {\bibfnamefont {L.}~\bibnamefont {Nie}}, \bibinfo {author} {\bibfnamefont
  {A.~V.}\ \bibnamefont {Maharaj}}, \bibinfo {author} {\bibfnamefont
  {S.}~\bibnamefont {Gerber}}, \bibinfo {author} {\bibfnamefont {Y.-J.}\
  \bibnamefont {Liu}}, \bibinfo {author} {\bibfnamefont {A.}~\bibnamefont
  {Mehta}}, \bibinfo {author} {\bibfnamefont {D.~A.}\ \bibnamefont {Bonn}},
  \bibinfo {author} {\bibfnamefont {R.}~\bibnamefont {Liang}}, \bibinfo
  {author} {\bibfnamefont {W.~N.}\ \bibnamefont {Hardy}}, \bibinfo {author}
  {\bibfnamefont {C.~A.}\ \bibnamefont {Burns}}, \bibinfo {author}
  {\bibfnamefont {Z.}~\bibnamefont {Islam}}, \bibinfo {author} {\bibfnamefont
  {S.}~\bibnamefont {Song}}, \bibinfo {author} {\bibfnamefont {J.}~\bibnamefont
  {Hastings}}, \bibinfo {author} {\bibfnamefont {T.~P.}\ \bibnamefont
  {Devereaux}}, \bibinfo {author} {\bibfnamefont {Z.-X.}\ \bibnamefont {Shen}},
  \bibinfo {author} {\bibfnamefont {S.~A.}\ \bibnamefont {Kivelson}}, \bibinfo
  {author} {\bibfnamefont {C.-C.}\ \bibnamefont {Kao}}, \bibinfo {author}
  {\bibfnamefont {D.}~\bibnamefont {Zhu}},\ and\ \bibinfo {author}
  {\bibfnamefont {J.-S.}\ \bibnamefont {Lee}},\ }\bibfield  {title} {\bibinfo
  {title} {{Ideal charge-density-wave order in the high-field state of
  superconducting YBCO}},\ }\href {https://doi.org/10.1073/pnas.1612849113}
  {\bibfield  {journal} {\bibinfo  {journal} {Proc. Nat. Acad. Sci. USA}\
  }\textbf {\bibinfo {volume} {113}},\ \bibinfo {pages} {14645} (\bibinfo
  {year} {2016})}\BibitemShut {NoStop}%
\bibitem [{\citenamefont {Mitrano}\ \emph {et~al.}(2019)\citenamefont
  {Mitrano}, \citenamefont {Lee}, \citenamefont {Husain}, \citenamefont {Zhu},
  \citenamefont {{de la Pe{\~{n}}a Mu{\~{n}}oz}}, \citenamefont {Sun},
  \citenamefont {Joe}, \citenamefont {Reid}, \citenamefont {Wandel},
  \citenamefont {Coslovich}, \citenamefont {Schlotter}, \citenamefont {van
  Driel}, \citenamefont {Schneeloch}, \citenamefont {Gu}, \citenamefont
  {Goldenfeld},\ and\ \citenamefont {Abbamonte}}]{Mitrano2019}%
  \BibitemOpen
  \bibfield  {author} {\bibinfo {author} {\bibfnamefont {M.}~\bibnamefont
  {Mitrano}}, \bibinfo {author} {\bibfnamefont {S.}~\bibnamefont {Lee}},
  \bibinfo {author} {\bibfnamefont {A.~A.}\ \bibnamefont {Husain}}, \bibinfo
  {author} {\bibfnamefont {M.}~\bibnamefont {Zhu}}, \bibinfo {author}
  {\bibfnamefont {G.}~\bibnamefont {{de la Pe{\~{n}}a Mu{\~{n}}oz}}}, \bibinfo
  {author} {\bibfnamefont {S.~X.-L.}\ \bibnamefont {Sun}}, \bibinfo {author}
  {\bibfnamefont {Y.~I.}\ \bibnamefont {Joe}}, \bibinfo {author} {\bibfnamefont
  {A.~H.}\ \bibnamefont {Reid}}, \bibinfo {author} {\bibfnamefont {S.~F.}\
  \bibnamefont {Wandel}}, \bibinfo {author} {\bibfnamefont {G.}~\bibnamefont
  {Coslovich}}, \bibinfo {author} {\bibfnamefont {W.}~\bibnamefont
  {Schlotter}}, \bibinfo {author} {\bibfnamefont {T.}~\bibnamefont {van
  Driel}}, \bibinfo {author} {\bibfnamefont {J.}~\bibnamefont {Schneeloch}},
  \bibinfo {author} {\bibfnamefont {G.~D.}\ \bibnamefont {Gu}}, \bibinfo
  {author} {\bibfnamefont {N.}~\bibnamefont {Goldenfeld}},\ and\ \bibinfo
  {author} {\bibfnamefont {P.}~\bibnamefont {Abbamonte}},\ }\bibfield  {title}
  {\bibinfo {title} {{Evidence for photoinduced sliding of the charge-order
  condensate in La$_{1.875}$Ba$_{0.125}$CuO$_4$}},\ }\href
  {https://doi.org/10.1103/PhysRevB.100.205125} {\bibfield  {journal} {\bibinfo
   {journal} {Phys. Rev. B}\ }\textbf {\bibinfo {volume} {100}},\ \bibinfo
  {pages} {205125} (\bibinfo {year} {2019})}\BibitemShut {NoStop}%
\bibitem [{\citenamefont {Lee}\ \emph {et~al.}(2021)\citenamefont {Lee},
  \citenamefont {Collini}, \citenamefont {Sun}, \citenamefont {Mitrano},
  \citenamefont {Guo}, \citenamefont {Eckberg}, \citenamefont {Paglione},
  \citenamefont {Fradkin},\ and\ \citenamefont {Abbamonte}}]{Lee2021}%
  \BibitemOpen
  \bibfield  {author} {\bibinfo {author} {\bibfnamefont {S.}~\bibnamefont
  {Lee}}, \bibinfo {author} {\bibfnamefont {J.}~\bibnamefont {Collini}},
  \bibinfo {author} {\bibfnamefont {S.~X.-L.}\ \bibnamefont {Sun}}, \bibinfo
  {author} {\bibfnamefont {M.}~\bibnamefont {Mitrano}}, \bibinfo {author}
  {\bibfnamefont {X.}~\bibnamefont {Guo}}, \bibinfo {author} {\bibfnamefont
  {C.}~\bibnamefont {Eckberg}}, \bibinfo {author} {\bibfnamefont
  {J.}~\bibnamefont {Paglione}}, \bibinfo {author} {\bibfnamefont
  {E.}~\bibnamefont {Fradkin}},\ and\ \bibinfo {author} {\bibfnamefont
  {P.}~\bibnamefont {Abbamonte}},\ }\bibfield  {title} {\bibinfo {title}
  {{Multiple Charge Density Waves and Superconductivity Nucleation at Antiphase
  Domain Walls in the Nematic Pnictide Ba$_{1 - x}$Sr$_x$Ni$_2$As$_2$}},\
  }\href {https://doi.org/10.1103/PhysRevLett.127.027602} {\bibfield  {journal}
  {\bibinfo  {journal} {Phys. Rev. Lett.}\ }\textbf {\bibinfo {volume} {127}},\
  \bibinfo {pages} {027602} (\bibinfo {year} {2021})}\BibitemShut {NoStop}%
\bibitem [{\citenamefont {Fujita}\ \emph {et~al.}(2014)\citenamefont {Fujita},
  \citenamefont {Hamidian}, \citenamefont {Edkins}, \citenamefont {Kim},
  \citenamefont {Kohsaka}, \citenamefont {Azuma}, \citenamefont {Takano},
  \citenamefont {Takagi}, \citenamefont {Eisaki}, \citenamefont {Uchida},
  \citenamefont {Allais}, \citenamefont {Lawler}, \citenamefont {Kim},
  \citenamefont {Sachdev},\ and\ \citenamefont {Davis}}]{Fujita2014}%
  \BibitemOpen
  \bibfield  {author} {\bibinfo {author} {\bibfnamefont {K.}~\bibnamefont
  {Fujita}}, \bibinfo {author} {\bibfnamefont {M.~H.}\ \bibnamefont
  {Hamidian}}, \bibinfo {author} {\bibfnamefont {S.~D.}\ \bibnamefont
  {Edkins}}, \bibinfo {author} {\bibfnamefont {C.~K.}\ \bibnamefont {Kim}},
  \bibinfo {author} {\bibfnamefont {Y.}~\bibnamefont {Kohsaka}}, \bibinfo
  {author} {\bibfnamefont {M.}~\bibnamefont {Azuma}}, \bibinfo {author}
  {\bibfnamefont {M.}~\bibnamefont {Takano}}, \bibinfo {author} {\bibfnamefont
  {H.}~\bibnamefont {Takagi}}, \bibinfo {author} {\bibfnamefont
  {H.}~\bibnamefont {Eisaki}}, \bibinfo {author} {\bibfnamefont {S.-i.}\
  \bibnamefont {Uchida}}, \bibinfo {author} {\bibfnamefont {A.}~\bibnamefont
  {Allais}}, \bibinfo {author} {\bibfnamefont {M.~J.}\ \bibnamefont {Lawler}},
  \bibinfo {author} {\bibfnamefont {E.-A.}\ \bibnamefont {Kim}}, \bibinfo
  {author} {\bibfnamefont {S.}~\bibnamefont {Sachdev}},\ and\ \bibinfo {author}
  {\bibfnamefont {J.~C.~S.}\ \bibnamefont {Davis}},\ }\bibfield  {title}
  {\bibinfo {title} {{Direct phase-sensitive identification of a $d$-form
  factor density wave in underdoped cuprates}},\ }\href
  {https://doi.org/10.1073/pnas.1406297111} {\bibfield  {journal} {\bibinfo
  {journal} {Proc. Nat. Acad. Sci. USA}\ }\textbf {\bibinfo {volume} {111}},\
  \bibinfo {pages} {E3026} (\bibinfo {year} {2014})}\BibitemShut {NoStop}%
\bibitem [{\citenamefont {Vig}\ \emph {et~al.}(2017)\citenamefont {Vig},
  \citenamefont {Kogar}, \citenamefont {Mitrano}, \citenamefont {Husain},
  \citenamefont {Venema}, \citenamefont {Rak}, \citenamefont {Mishra},
  \citenamefont {Johnson}, \citenamefont {Gu}, \citenamefont {Fradkin},
  \citenamefont {Norman},\ and\ \citenamefont {Abbamonte}}]{Vig2017}%
  \BibitemOpen
  \bibfield  {author} {\bibinfo {author} {\bibfnamefont {S.}~\bibnamefont
  {Vig}}, \bibinfo {author} {\bibfnamefont {A.}~\bibnamefont {Kogar}}, \bibinfo
  {author} {\bibfnamefont {M.}~\bibnamefont {Mitrano}}, \bibinfo {author}
  {\bibfnamefont {A.}~\bibnamefont {Husain}}, \bibinfo {author} {\bibfnamefont
  {L.}~\bibnamefont {Venema}}, \bibinfo {author} {\bibfnamefont
  {M.}~\bibnamefont {Rak}}, \bibinfo {author} {\bibfnamefont {V.}~\bibnamefont
  {Mishra}}, \bibinfo {author} {\bibfnamefont {P.}~\bibnamefont {Johnson}},
  \bibinfo {author} {\bibfnamefont {G.}~\bibnamefont {Gu}}, \bibinfo {author}
  {\bibfnamefont {E.}~\bibnamefont {Fradkin}}, \bibinfo {author} {\bibfnamefont
  {M.}~\bibnamefont {Norman}},\ and\ \bibinfo {author} {\bibfnamefont
  {P.}~\bibnamefont {Abbamonte}},\ }\bibfield  {title} {\bibinfo {title}
  {{Measurement of the dynamic charge response of materials using low-energy,
  momentum-resolved electron energy-loss spectroscopy (M-EELS)}},\ }\href
  {https://doi.org/10.21468/SciPostPhys.3.4.026} {\bibfield  {journal}
  {\bibinfo  {journal} {SciPost Phys.}\ }\textbf {\bibinfo {volume} {3}},\
  \bibinfo {pages} {026} (\bibinfo {year} {2017})}\BibitemShut {NoStop}%
\bibitem [{\citenamefont {Kogar}\ \emph {et~al.}(2017)\citenamefont {Kogar},
  \citenamefont {Rak}, \citenamefont {Vig}, \citenamefont {Husain},
  \citenamefont {Flicker}, \citenamefont {Joe}, \citenamefont {Venema},
  \citenamefont {MacDougall}, \citenamefont {Chiang}, \citenamefont {Fradkin},
  \citenamefont {van Wezel},\ and\ \citenamefont {Abbamonte}}]{Kogar2017}%
  \BibitemOpen
  \bibfield  {author} {\bibinfo {author} {\bibfnamefont {A.}~\bibnamefont
  {Kogar}}, \bibinfo {author} {\bibfnamefont {M.~S.}\ \bibnamefont {Rak}},
  \bibinfo {author} {\bibfnamefont {S.}~\bibnamefont {Vig}}, \bibinfo {author}
  {\bibfnamefont {A.~A.}\ \bibnamefont {Husain}}, \bibinfo {author}
  {\bibfnamefont {F.}~\bibnamefont {Flicker}}, \bibinfo {author} {\bibfnamefont
  {Y.~I.}\ \bibnamefont {Joe}}, \bibinfo {author} {\bibfnamefont
  {L.}~\bibnamefont {Venema}}, \bibinfo {author} {\bibfnamefont {G.~J.}\
  \bibnamefont {MacDougall}}, \bibinfo {author} {\bibfnamefont {T.~C.}\
  \bibnamefont {Chiang}}, \bibinfo {author} {\bibfnamefont {E.}~\bibnamefont
  {Fradkin}}, \bibinfo {author} {\bibfnamefont {J.}~\bibnamefont {van Wezel}},\
  and\ \bibinfo {author} {\bibfnamefont {P.}~\bibnamefont {Abbamonte}},\
  }\bibfield  {title} {\bibinfo {title} {{Signatures of exciton condensation in
  a transition metal dichalcogenide}},\ }\href
  {https://doi.org/10.1126/science.aam6432} {\bibfield  {journal} {\bibinfo
  {journal} {Science}\ }\textbf {\bibinfo {volume} {358}},\ \bibinfo {pages}
  {1314} (\bibinfo {year} {2017})}\BibitemShut {NoStop}%
\bibitem [{\citenamefont {Li}\ \emph {et~al.}(2007)\citenamefont {Li},
  \citenamefont {H{\"{u}}cker}, \citenamefont {Gu}, \citenamefont {Tsvelik},\
  and\ \citenamefont {Tranquada}}]{Li2007}%
  \BibitemOpen
  \bibfield  {author} {\bibinfo {author} {\bibfnamefont {Q.}~\bibnamefont
  {Li}}, \bibinfo {author} {\bibfnamefont {M.}~\bibnamefont {H{\"{u}}cker}},
  \bibinfo {author} {\bibfnamefont {G.~D.}\ \bibnamefont {Gu}}, \bibinfo
  {author} {\bibfnamefont {A.~M.}\ \bibnamefont {Tsvelik}},\ and\ \bibinfo
  {author} {\bibfnamefont {J.~M.}\ \bibnamefont {Tranquada}},\ }\bibfield
  {title} {\bibinfo {title} {{Two-Dimensional Superconducting Fluctuations in
  Stripe-Ordered La$_{1.875}$Ba$_{0.125}$CuO$_4$}},\ }\href
  {https://doi.org/10.1103/PhysRevLett.99.067001} {\bibfield  {journal}
  {\bibinfo  {journal} {Phys. Rev. Lett.}\ }\textbf {\bibinfo {volume} {99}},\
  \bibinfo {pages} {067001} (\bibinfo {year} {2007})}\BibitemShut {NoStop}%
\bibitem [{\citenamefont {H{\"{u}}cker}\ \emph {et~al.}(2011)\citenamefont
  {H{\"{u}}cker}, \citenamefont {v.~Zimmermann}, \citenamefont {Gu},
  \citenamefont {Xu}, \citenamefont {Wen}, \citenamefont {Xu}, \citenamefont
  {Kang}, \citenamefont {Zheludev},\ and\ \citenamefont
  {Tranquada}}]{Hucker2011}%
  \BibitemOpen
  \bibfield  {author} {\bibinfo {author} {\bibfnamefont {M.}~\bibnamefont
  {H{\"{u}}cker}}, \bibinfo {author} {\bibfnamefont {M.}~\bibnamefont
  {v.~Zimmermann}}, \bibinfo {author} {\bibfnamefont {G.~D.}\ \bibnamefont
  {Gu}}, \bibinfo {author} {\bibfnamefont {Z.~J.}\ \bibnamefont {Xu}}, \bibinfo
  {author} {\bibfnamefont {J.~S.}\ \bibnamefont {Wen}}, \bibinfo {author}
  {\bibfnamefont {G.}~\bibnamefont {Xu}}, \bibinfo {author} {\bibfnamefont
  {H.~J.}\ \bibnamefont {Kang}}, \bibinfo {author} {\bibfnamefont
  {A.}~\bibnamefont {Zheludev}},\ and\ \bibinfo {author} {\bibfnamefont
  {J.~M.}\ \bibnamefont {Tranquada}},\ }\bibfield  {title} {\bibinfo {title}
  {{Stripe order in superconducting La$_{2-x}$Ba$_x$CuO$_4$ ($0.095 \leq x \leq
  0.155$)}},\ }\href {https://doi.org/10.1103/PhysRevB.83.104506} {\bibfield
  {journal} {\bibinfo  {journal} {Phys. Rev. B}\ }\textbf {\bibinfo {volume}
  {83}},\ \bibinfo {pages} {104506} (\bibinfo {year} {2011})}\BibitemShut
  {NoStop}%
\bibitem [{\citenamefont {Lee}\ \emph {et~al.}(2022)\citenamefont {Lee},
  \citenamefont {Huang}, \citenamefont {Johnson}, \citenamefont {Guo},
  \citenamefont {Husain}, \citenamefont {Mitrano}, \citenamefont {Lu},
  \citenamefont {Zakrzewski}, \citenamefont {{de la Pe{\~{n}}a}}, \citenamefont
  {Peng}, \citenamefont {Huang}, \citenamefont {Lee}, \citenamefont {Jang},
  \citenamefont {Lee}, \citenamefont {Joe}, \citenamefont {Doriese},
  \citenamefont {Szypryt}, \citenamefont {Swetz}, \citenamefont {Chi},
  \citenamefont {Aczel}, \citenamefont {MacDougall}, \citenamefont {Kivelson},
  \citenamefont {Fradkin},\ and\ \citenamefont {Abbamonte}}]{Lee2022}%
  \BibitemOpen
  \bibfield  {author} {\bibinfo {author} {\bibfnamefont {S.}~\bibnamefont
  {Lee}}, \bibinfo {author} {\bibfnamefont {E.~W.}\ \bibnamefont {Huang}},
  \bibinfo {author} {\bibfnamefont {T.~A.}\ \bibnamefont {Johnson}}, \bibinfo
  {author} {\bibfnamefont {X.}~\bibnamefont {Guo}}, \bibinfo {author}
  {\bibfnamefont {A.~A.}\ \bibnamefont {Husain}}, \bibinfo {author}
  {\bibfnamefont {M.}~\bibnamefont {Mitrano}}, \bibinfo {author} {\bibfnamefont
  {K.}~\bibnamefont {Lu}}, \bibinfo {author} {\bibfnamefont {A.~V.}\
  \bibnamefont {Zakrzewski}}, \bibinfo {author} {\bibfnamefont {G.~A.}\
  \bibnamefont {{de la Pe{\~{n}}a}}}, \bibinfo {author} {\bibfnamefont
  {Y.}~\bibnamefont {Peng}}, \bibinfo {author} {\bibfnamefont {H.}~\bibnamefont
  {Huang}}, \bibinfo {author} {\bibfnamefont {S.-J.}\ \bibnamefont {Lee}},
  \bibinfo {author} {\bibfnamefont {H.}~\bibnamefont {Jang}}, \bibinfo {author}
  {\bibfnamefont {J.-S.}\ \bibnamefont {Lee}}, \bibinfo {author} {\bibfnamefont
  {Y.~I.}\ \bibnamefont {Joe}}, \bibinfo {author} {\bibfnamefont {W.~B.}\
  \bibnamefont {Doriese}}, \bibinfo {author} {\bibfnamefont {P.}~\bibnamefont
  {Szypryt}}, \bibinfo {author} {\bibfnamefont {D.~S.}\ \bibnamefont {Swetz}},
  \bibinfo {author} {\bibfnamefont {S.}~\bibnamefont {Chi}}, \bibinfo {author}
  {\bibfnamefont {A.~A.}\ \bibnamefont {Aczel}}, \bibinfo {author}
  {\bibfnamefont {G.~J.}\ \bibnamefont {MacDougall}}, \bibinfo {author}
  {\bibfnamefont {S.~A.}\ \bibnamefont {Kivelson}}, \bibinfo {author}
  {\bibfnamefont {E.}~\bibnamefont {Fradkin}},\ and\ \bibinfo {author}
  {\bibfnamefont {P.}~\bibnamefont {Abbamonte}},\ }\bibfield  {title} {\bibinfo
  {title} {{Generic character of charge and spin density waves in
  superconducting cuprates}},\ }\href {https://doi.org/10.1073/pnas.2119429119}
  {\bibfield  {journal} {\bibinfo  {journal} {Proc. Nat. Acad. Sci. USA}\
  }\textbf {\bibinfo {volume} {119}},\ \bibinfo {pages} {e2119429119} (\bibinfo
  {year} {2022})}\BibitemShut {NoStop}%
\bibitem [{\citenamefont {Stanley}(1968)}]{Stanley1968}%
  \BibitemOpen
  \bibfield  {author} {\bibinfo {author} {\bibfnamefont {H.~E.}\ \bibnamefont
  {Stanley}},\ }\bibfield  {title} {\bibinfo {title} {{Spherical Model as the
  Limit of Infinite Spin Dimensionality}},\ }\href
  {https://doi.org/10.1103/PhysRev.176.718} {\bibfield  {journal} {\bibinfo
  {journal} {Phys. Rev.}\ }\textbf {\bibinfo {volume} {176}},\ \bibinfo {pages}
  {718} (\bibinfo {year} {1968})}\BibitemShut {NoStop}%
\bibitem [{\citenamefont {Amit}(1980)}]{Amit-book}%
  \BibitemOpen
  \bibfield  {author} {\bibinfo {author} {\bibfnamefont {D.~J.}\ \bibnamefont
  {Amit}},\ }\href@noop {} {\emph {\bibinfo {title} {{Field Theory, the
  Renormalization Group and Critical Phenomena}}}}\ (\bibinfo  {publisher}
  {McGraw Hill},\ \bibinfo {address} {New York, NY},\ \bibinfo {year}
  {1980})\BibitemShut {NoStop}%
\bibitem [{\citenamefont {Zinn-Justin}(2002)}]{Zinn-book}%
  \BibitemOpen
  \bibfield  {author} {\bibinfo {author} {\bibfnamefont {J.}~\bibnamefont
  {Zinn-Justin}},\ }\href@noop {} {\emph {\bibinfo {title} {{Quantum Field
  Theory and Critical Phenomena}}}},\ \bibinfo {edition} {4th}\ ed.,\
  International Series of Monographs in Physics\ (\bibinfo  {publisher} {Oxford
  University Press},\ \bibinfo {address} {Oxford, U.K.},\ \bibinfo {year}
  {2002})\BibitemShut {NoStop}%
\bibitem [{\citenamefont {Fradkin}(2021)}]{fradkin_2021}%
  \BibitemOpen
  \bibfield  {author} {\bibinfo {author} {\bibfnamefont {E.}~\bibnamefont
  {Fradkin}},\ }\href@noop {} {\emph {\bibinfo {title} {Quantum Field Theory:
  An Integrated Approach}}}\ (\bibinfo  {publisher} {Princeton University
  Press},\ \bibinfo {address} {Princeton, NJ},\ \bibinfo {year}
  {2021})\BibitemShut {NoStop}%
\bibitem [{\citenamefont {Witten}(1979)}]{Witten1979}%
  \BibitemOpen
  \bibfield  {author} {\bibinfo {author} {\bibfnamefont {E.}~\bibnamefont
  {Witten}},\ }\bibfield  {title} {\bibinfo {title} {{Instatons, the quark
  model, and the $1/N$ expansion}},\ }\href
  {https://doi.org/10.1016/0550-3213(79)90243-8} {\bibfield  {journal}
  {\bibinfo  {journal} {Nucl. Phys. B}\ }\textbf {\bibinfo {volume} {149}},\
  \bibinfo {pages} {285} (\bibinfo {year} {1979})}\BibitemShut {NoStop}%
\bibitem [{\citenamefont {Coleman}(1985)}]{Coleman-1985}%
  \BibitemOpen
  \bibfield  {author} {\bibinfo {author} {\bibfnamefont {S.}~\bibnamefont
  {Coleman}},\ }\href@noop {} {\emph {\bibinfo {title} {Aspects of Symmetry}}}\
  (\bibinfo  {publisher} {Cambridge University Press},\ \bibinfo {address}
  {Cambridge, U.K.},\ \bibinfo {year} {1985})\BibitemShut {NoStop}%
\bibitem [{\citenamefont {Pytte}\ \emph {et~al.}(1981)\citenamefont {Pytte},
  \citenamefont {Imry},\ and\ \citenamefont {Mukamel}}]{Pytte1981}%
  \BibitemOpen
  \bibfield  {author} {\bibinfo {author} {\bibfnamefont {E.}~\bibnamefont
  {Pytte}}, \bibinfo {author} {\bibfnamefont {Y.}~\bibnamefont {Imry}},\ and\
  \bibinfo {author} {\bibfnamefont {D.}~\bibnamefont {Mukamel}},\ }\bibfield
  {title} {\bibinfo {title} {{Lower Critical Dimension and the Roughening
  Transition of the Random-Field Ising Model}},\ }\href
  {https://doi.org/10.1103/PhysRevLett.46.1173} {\bibfield  {journal} {\bibinfo
   {journal} {Phys. Rev. Lett.}\ }\textbf {\bibinfo {volume} {46}},\ \bibinfo
  {pages} {1173} (\bibinfo {year} {1981})}\BibitemShut {NoStop}%
\bibitem [{\citenamefont {Gross}\ and\ \citenamefont
  {Neveu}(1974)}]{Gross1974}%
  \BibitemOpen
  \bibfield  {author} {\bibinfo {author} {\bibfnamefont {D.~J.}\ \bibnamefont
  {Gross}}\ and\ \bibinfo {author} {\bibfnamefont {A.}~\bibnamefont {Neveu}},\
  }\bibfield  {title} {\bibinfo {title} {{Dynamical symmetry breaking in
  asymptotically free field theories}},\ }\href
  {https://doi.org/10.1103/PhysRevD.10.3235} {\bibfield  {journal} {\bibinfo
  {journal} {Phys. Rev. D}\ }\textbf {\bibinfo {volume} {10}},\ \bibinfo
  {pages} {3235} (\bibinfo {year} {1974})}\BibitemShut {NoStop}%
\bibitem [{\citenamefont {Witten}(1978)}]{Witten1978}%
  \BibitemOpen
  \bibfield  {author} {\bibinfo {author} {\bibfnamefont {E.}~\bibnamefont
  {Witten}},\ }\bibfield  {title} {\bibinfo {title} {{Some properties of the
  $(\bar{\psi}\psi)^2$ model in two dimensions}},\ }\href
  {https://doi.org/10.1016/0550-3213(78)90204-3} {\bibfield  {journal}
  {\bibinfo  {journal} {Nucl. Phys. B}\ }\textbf {\bibinfo {volume} {142}},\
  \bibinfo {pages} {285} (\bibinfo {year} {1978})}\BibitemShut {NoStop}%
\bibitem [{\citenamefont {Kardar}(2013)}]{kardar_2013}%
  \BibitemOpen
  \bibfield  {author} {\bibinfo {author} {\bibfnamefont {M.}~\bibnamefont
  {Kardar}},\ }\href@noop {} {\emph {\bibinfo {title} {Statistical Physics of
  Fields}}}\ (\bibinfo  {publisher} {Cambridge University Press},\ \bibinfo
  {address} {Cambridge, U.K.},\ \bibinfo {year} {2013})\BibitemShut {NoStop}%
\bibitem [{\citenamefont {Sherrington}\ and\ \citenamefont
  {Kirkpatrick}(1975)}]{Sherrington1975}%
  \BibitemOpen
  \bibfield  {author} {\bibinfo {author} {\bibfnamefont {D.}~\bibnamefont
  {Sherrington}}\ and\ \bibinfo {author} {\bibfnamefont {S.}~\bibnamefont
  {Kirkpatrick}},\ }\bibfield  {title} {\bibinfo {title} {{Solvable Model of a
  Spin-Glass}},\ }\href {https://doi.org/10.1103/PhysRevLett.35.1792}
  {\bibfield  {journal} {\bibinfo  {journal} {Phys. Rev. Lett.}\ }\textbf
  {\bibinfo {volume} {35}},\ \bibinfo {pages} {1792} (\bibinfo {year}
  {1975})}\BibitemShut {NoStop}%
\bibitem [{\citenamefont {Parisi}(1984)}]{Parisi-1982}%
  \BibitemOpen
  \bibfield  {author} {\bibinfo {author} {\bibfnamefont {G.}~\bibnamefont
  {Parisi}},\ }\bibfield  {title} {\bibinfo {title} {{An Introduction to the
  Statistical Mechanics of Amorphous Systems}},\ }in\ \href@noop {} {\emph
  {\bibinfo {booktitle} {{Recent Advances in Field Theory and Statistical
  Mechanics}}}},\ \bibinfo {series} {{\'Ecole d'\'Et\'e de Les Houches}}, Vol.\
  \bibinfo {volume} {{XXXIX}},\ \bibinfo {editor} {edited by\ \bibinfo {editor}
  {\bibfnamefont {J.-B.}\ \bibnamefont {Zuber}}\ and\ \bibinfo {editor}
  {\bibfnamefont {R.}~\bibnamefont {Stora}}}\ (\bibinfo  {publisher} {Elsevier
  Science Publishers},\ \bibinfo {address} {Amsterdam, The Netherlands},\
  \bibinfo {year} {1984})\ pp.\ \bibinfo {pages} {473--524}\BibitemShut
  {NoStop}%
\bibitem [{\citenamefont {Yoshizawa}\ \emph {et~al.}(1982)\citenamefont
  {Yoshizawa}, \citenamefont {Cowley}, \citenamefont {Shirane}, \citenamefont
  {Birgeneau}, \citenamefont {Guggenheim},\ and\ \citenamefont
  {Ikeda}}]{Yoshizawa1982}%
  \BibitemOpen
  \bibfield  {author} {\bibinfo {author} {\bibfnamefont {H.}~\bibnamefont
  {Yoshizawa}}, \bibinfo {author} {\bibfnamefont {R.~A.}\ \bibnamefont
  {Cowley}}, \bibinfo {author} {\bibfnamefont {G.}~\bibnamefont {Shirane}},
  \bibinfo {author} {\bibfnamefont {R.~J.}\ \bibnamefont {Birgeneau}}, \bibinfo
  {author} {\bibfnamefont {H.~J.}\ \bibnamefont {Guggenheim}},\ and\ \bibinfo
  {author} {\bibfnamefont {H.}~\bibnamefont {Ikeda}},\ }\bibfield  {title}
  {\bibinfo {title} {{Random-Field Effects in Two- and Three-Dimensional Ising
  Antiferromagnets}},\ }\href {https://doi.org/10.1103/PhysRevLett.48.438}
  {\bibfield  {journal} {\bibinfo  {journal} {Phys. Rev. Lett.}\ }\textbf
  {\bibinfo {volume} {48}},\ \bibinfo {pages} {438} (\bibinfo {year}
  {1982})}\BibitemShut {NoStop}%
\bibitem [{\citenamefont {Griffiths}(1969)}]{Griffiths1969}%
  \BibitemOpen
  \bibfield  {author} {\bibinfo {author} {\bibfnamefont {R.~B.}\ \bibnamefont
  {Griffiths}},\ }\bibfield  {title} {\bibinfo {title} {{Nonanalytic Behavior
  Above the Critical Point in a Random Ising Ferromagnet}},\ }\href
  {https://doi.org/10.1103/PhysRevLett.23.17} {\bibfield  {journal} {\bibinfo
  {journal} {Phys. Rev. Lett.}\ }\textbf {\bibinfo {volume} {23}},\ \bibinfo
  {pages} {17} (\bibinfo {year} {1969})}\BibitemShut {NoStop}%
\bibitem [{\citenamefont {Sun}\ \emph {et~al.}(2008)\citenamefont {Sun},
  \citenamefont {Fregoso}, \citenamefont {Lawler},\ and\ \citenamefont
  {Fradkin}}]{Sun2008}%
  \BibitemOpen
  \bibfield  {author} {\bibinfo {author} {\bibfnamefont {K.}~\bibnamefont
  {Sun}}, \bibinfo {author} {\bibfnamefont {B.~M.}\ \bibnamefont {Fregoso}},
  \bibinfo {author} {\bibfnamefont {M.~J.}\ \bibnamefont {Lawler}},\ and\
  \bibinfo {author} {\bibfnamefont {E.}~\bibnamefont {Fradkin}},\ }\bibfield
  {title} {\bibinfo {title} {{Fluctuating stripes in strongly correlated
  electron systems and the nematic-smectic quantum phase transition}},\ }\href
  {https://doi.org/10.1103/PhysRevB.78.085124} {\bibfield  {journal} {\bibinfo
  {journal} {Phys. Rev. B}\ }\textbf {\bibinfo {volume} {78}},\ \bibinfo
  {pages} {085124} (\bibinfo {year} {2008})}\BibitemShut {NoStop}%
\bibitem [{\citenamefont {Fradkin}(2023)}]{Fradkin2023}%
  \BibitemOpen
  \bibfield  {author} {\bibinfo {author} {\bibfnamefont {E.}~\bibnamefont
  {Fradkin}},\ }\href@noop {} {\bibinfo {title} {Field theoretic aspects of
  condensed matter physics: An overview}} (\bibinfo {year} {2023}),\ \bibinfo
  {note} {to appear in {\sl Encyclopedia of Condensed Matter Physics 2e}},\
  \Eprint {https://arxiv.org/abs/arXiv:2301.13234} {arXiv:2301.13234}
  \BibitemShut {NoStop}%
\bibitem [{\citenamefont {Wiegmann}(1978)}]{Wiegmann1978}%
  \BibitemOpen
  \bibfield  {author} {\bibinfo {author} {\bibfnamefont {P.~B.}\ \bibnamefont
  {Wiegmann}},\ }\bibfield  {title} {\bibinfo {title} {{One-dimensional Fermi
  system and plane $xy$ model}},\ }\href
  {https://doi.org/10.1088/0022-3719/11/8/019} {\bibfield  {journal} {\bibinfo
  {journal} {J. Phys. Chem.}\ }\textbf {\bibinfo {volume} {11}},\ \bibinfo
  {pages} {1583} (\bibinfo {year} {1978})}\BibitemShut {NoStop}%
\bibitem [{\citenamefont {Kadanoff}(1978)}]{Kadanoff1978}%
  \BibitemOpen
  \bibfield  {author} {\bibinfo {author} {\bibfnamefont {L.~P.}\ \bibnamefont
  {Kadanoff}},\ }\bibfield  {title} {\bibinfo {title} {{Lattice Coulomb gas
  representations of two-dimensional problems}},\ }\href
  {https://doi.org/10.1088/0305-4470/11/7/027} {\bibfield  {journal} {\bibinfo
  {journal} {J. Phys. A}\ }\textbf {\bibinfo {volume} {11}},\ \bibinfo {pages}
  {1399} (\bibinfo {year} {1978})}\BibitemShut {NoStop}%
\bibitem [{\citenamefont {Amit}\ \emph {et~al.}(1980)\citenamefont {Amit},
  \citenamefont {Goldschmidt},\ and\ \citenamefont {Grinstein}}]{Amit1980}%
  \BibitemOpen
  \bibfield  {author} {\bibinfo {author} {\bibfnamefont {D.~J.}\ \bibnamefont
  {Amit}}, \bibinfo {author} {\bibfnamefont {Y.~Y.}\ \bibnamefont
  {Goldschmidt}},\ and\ \bibinfo {author} {\bibfnamefont {S.}~\bibnamefont
  {Grinstein}},\ }\bibfield  {title} {\bibinfo {title} {{Renormalisation group
  analysis of the phase transition in the 2D Coulomb gas, Sine-Gordon theory
  and XY-model}},\ }\href {https://doi.org/10.1088/0305-4470/13/2/024}
  {\bibfield  {journal} {\bibinfo  {journal} {J. Phys. A}\ }\textbf {\bibinfo
  {volume} {13}},\ \bibinfo {pages} {585} (\bibinfo {year} {1980})}\BibitemShut
  {NoStop}%
\end{thebibliography}%

\end{document}